\documentclass[a4paper,fleqn,usenatbib]{mnras}

\usepackage{newtxtext,newtxmath}

\usepackage[T1]{fontenc}
\usepackage{ae,aecompl}

\usepackage{graphicx}	
\usepackage{amsmath}	
\usepackage{amssymb}	
\usepackage{enumitem}
\usepackage{times}
\usepackage{url}
\usepackage[usenames]{color}
\usepackage{xcolor}
\usepackage{longtable}
\usepackage{threeparttable}

\def \simless {\mathbin{\lower 3pt\hbox{$\rlap{\raise 4pt
              \hbox{$\char'074$}}\mathchar"7218$}}}
\def \simgreat {\mathbin{\lower 3pt\hbox{$\rlap{\raise 4pt
              \hbox{$\char'076$}}\mathchar"7218$}}}
\def\ie {{\it i.e.}}
\def\cf {{\it c.f.}}
\def\eg {{\it e.g.}}
\newcommand{\angstrom}{\mbox{\normalfont\AA}}

\title[The IRX-$\beta$ relation of high-redshift galaxies]{The IRX-$\beta$ relation of high-redshift galaxies}

\author[L. Liang et al.]{Lichen Liang$^{1}$\thanks{Email: lliang@physik.uzh.ch}, Robert Feldmann$^{1}$, Christopher C. Hayward$^{2}$, Desika Narayanan$^{3,4,5}$,  \newauthor  Onur \c{C}atmabacak$^{1}$, Du\v{s}an Kere\v{s}$^{6}$, Claude-Andr\'{e} Faucher-Gigu\`{e}re$^{7}$, Philip F. Hopkins$^{8}$\\
$^{1}$Institute for Computational Science, University of Zurich, Winterthurerstrasse 190, Zurich CH-8057, Switzerland\\
$^{2}$Center for Computational Astrophysics, Flatiron Institute, 162 Fifth Avenue, New York, NY 10010, USA\\
$^{3}$Department of Astronomy, University of Florida, 211 Bryant Space Sciences Center, Gainesville, FL 32611 USA\\
$^{4}$University of Florida Informatics Institute, 432 Newell Drive, CISE Bldg E251, Gainesville, FL 32611\\
$^{5}$Cosmic Dawn Center at the Niels Bohr Institute, University of Copenhagen and DTU-Space, Technical University of Denmark\\
$^{6}$Department of Physics, Center for Astrophysics and Space Sciences, University of California at San Diego, La Jolla, CA 92093, USA\\
$^{7}$Department of Physics and Astronomy and CIERA, Northwestern University, Evanston, IL 60208, USA\\
$^{8}$TAPIR, Mailcode 350-17, California Institute of Technology, Pasadena, CA 91125, USA}
\date{Accepted 2020. Received 2020; in original form 2020}

\pubyear{2020}

\begin{document}
\label{firstpage}
\pagerange{\pageref{firstpage}--\pageref{lastpage}}
\maketitle

\begin{abstract}
\noindent The relation between infrared excess (IRX) and UV spectral slope ($\beta_{\rm UV}$) is an empirical probe of dust properties of galaxies. The shape, scatter, and redshift evolution of this relation are not well understood, however, leading to uncertainties in estimating the dust content and star formation rates (SFRs) of galaxies at high redshift. In this study, we explore the nature and properties of the IRX-$\beta_{\rm UV}$ relation with a sample of $z=2-6$ galaxies ($M_*\approx 10^9-10^{12}\,M_\odot$) extracted from high-resolution cosmological simulations (MassiveFIRE) of the Feedback in Realistic Environments (FIRE) project. The galaxies in our sample show an IRX-$\beta_{\rm UV}$ relation that is in good agreement with the observed relation in nearby galaxies. IRX is tightly coupled to the UV optical depth, and is mainly determined by the dust-to-star geometry instead of total dust mass, while $\beta_{\rm UV}$ is set both by stellar properties, UV optical depth, and the dust extinction law. Overall, much of the scatter in the IRX-$\beta_{\rm UV}$ relation of our sample is found to be driven by variations of the intrinsic UV spectral slope. We further assess how the IRX-$\beta_{\rm UV}$ relation depends on viewing direction, dust-to-metal ratio, birth-cloud structures, and the dust extinction law and we present a simple model that encapsulates most of the found dependencies. Consequently, we argue that the reported `deficit' of the infrared/sub-millimetre bright objects at $z \simgreat 5$ does not necessarily imply a non-standard dust extinction law at those epochs.
\vspace{-5 pt}
\end{abstract}

\section{Introduction}
\label{Sec:1}


Reliable estimates of the SFR of galaxies at low and high-$z$ is crucial for constraining the various physical processes associated with galaxy evolution at different epochs \citep[\eg][]{Madau_2014}. The rest-frame UV luminosities of galaxy ($L_{\rm UV}$), which are dominated by the light of the young, massive stars, are commonly used as diagnostics of the current SFR of galaxy \citep[\eg][]{Kennicutt_1998, Kennicutt_2012, Conroy_2013, Flores_2020}. However, the accuracy of this method can be hampered by the effect of dust attenuation \citep[\eg][]{Salim_2020}. It is known that a large fraction of stellar radiation in the Universe is absorbed by interstellar dust and becomes re-emitted at infrared (IR) and millimetre (mm) wavelengths in the form of thermal radiation \citep[\eg][]{Calzetti_2000, Magnelli_2009, Reddy_2010, Burgarella_2013, Gruppioni_2013, Whitaker_2017}. Therefore, it is critical to account for both the dust thermal emission as well as the UV light of stars to accurately measure the SFR of galaxies.


However, estimating the dust luminosity of distant galaxies can be very challenging. While the UV photometry of many high-$z$ galaxies can be extracted from the deep broadband imaging surveys \citep[\eg,][]{Bouwens_2009, Ellis_2013, McLure_2013, Bouwens_2014, McLeod_2015, Oesch_2015, Laporte_2016, Oesch_2018}, reliable measurement of their dust continuum is often not possible. Many single-dish IR telescopes, such as \textit{Herschel} and \textsc{\small SCUBA}, have high confusion noise due to their poor spatial resolution \citep[\eg][]{Dole_2004, Nguyen_2010, Berta_2011, Lutz_2014}, and therefore source detection is limited to the most IR-luminous objects at high redshifts \citep[\eg,][]{Casey_2014P}. Interferometric telescopes (\eg~ALMA) have significantly improved the detection limit with higher resolution and sensitivity, but these can only probe relatively small volumes. Therefore, on many occasions, one needs to infer the bolometric IR luminosity (and the obscured SFR) of high-$z$ galaxies through alternative, indirect methods.

One common alternative strategy is by using the empirical relationship between the UV spectral slope ($\beta_{\rm UV}$), defined as the index in the power-law relationship $f_\lambda\propto\lambda^{\beta_{\rm UV}}$ over the wavelength range $1200<\lambda<3200\,\angstrom$\footnote{Throughout this paper, `$\lambda$' stands for rest-frame wavelength.} \citep[\eg,][]{Calzetti_1994, Leitherer_1995,Calzetti_1997}, and their infrared excess (IRX$\equiv L_{\rm IR}/L_{\rm UV}$) of galaxies. While $\beta_{\rm UV}$ is a measure of the reddening of UV colour (assuming that the variation in the intrinsic UV spectral slope is negligible), IRX is a proxy for dust attenuation. A higher dust attenuation should increase both IRX and the amount of reddening, and $\beta_{\rm UV}$ and IRX may be correlated. Observationally, it was at first revealed by the seminal work of \citet[hererafter M99]{Meurer_1995, Meurer_1999} that their selected nearby starburst sample (originally complied by \citealt{Calzetti_1994}) exhibited a fairly tight sequence in the IRX-$\beta_{\rm UV}$ plane. Their result suggested that $L_{\rm IR}$ could potentially be reliably constrained when only UV measurements ($L_{\rm UV}$ and $\beta_{\rm UV}$) were available. 

From then on, efforts have been made to extend the study of this empirical relationship using a wider range of diverse galaxy populations. Despite the promising nature of this technique, growing observational evidence has shown that galaxies of a broad range of types can exhibit a non-trivial degree of variations in the IRX-$\beta_{\rm UV}$ relation and show much larger scatter than the original result of M99. For instance, some studies have discovered that the local ultra-luminous infrared galaxies (ULIRGs) tend to have bluer $\beta_{\rm UV}$ in comparison to the canonical M99 relation at their IRX \citep[\eg,][]{Goldader_2002, Buat_2005, Howell_2010, Casey_2014}. On the other hand, observations of nearby normal star-forming and quiescent galaxies have shown that these galaxies appear to be systematically `redder' than the M99 relation and occupy a fairly wide range of positions on the diagram \citep[\eg][]{Bell_2002, Buat_2002, Kong_2004, Buat_2005, Boquien_2012, Grasha_2013}. These findings from the local observations suggest that the general galaxy populations may not follow a tight, universal IRX-$\beta_{\rm UV}$ relation. 

In recent years, a growing number of studies have focused on the IRX-$\beta_{\rm UV}$ relation at higher-$z$ in order to probe the evolution of the dust attenuation properties of galaxies \citep[\eg,][]{Reddy_2010, Heinis_2013, Bouwens_2016, Alvarez_Marquez_2016, Reddy_2018, Koprowski_2018}. {Many of these studies} are based on large samples of UV/optical-selected Lyman-break galaxies (LBGs) due to the efficiency of the source selection techniques \citep{Steidel_1996}. {The IR dust emission of individual high-$z$ LBGs, however, is often undetected} \citep[\eg,][]{Adelberger_2000, Reddy_2006}, and their IRXs are derived instead with a stacking method. Here, galaxies are binned by their measured $\beta_{\rm UV}$ and for each bin, IRX is derived from the stacked UV and IR photometry of the binned galaxies.

The results reported by these studies are not entirely conclusive. While some found results to be in good agreement with the original M99 relation derived using local starbursts \citep[\eg,][]{Heinis_2013, Bourne_2017, Fudamoto_2017, McLure_2018, Koprowski_2018, Alvarez_Marquez_2019, Fudamoto_2020}, others reported a redder and much shallower relation \citep[\eg,][]{Alvarez_Marquez_2016, Bouwens_2016, Reddy_2018}. As a consequence, different conclusions have been reached regarding the evolution of dust attenuation properties with redshift.

What is missing from the analysis of the stacked high-$z$ LBGs, however, is that the results do not truly reflect the level of scatter among the individual systems, but instead, represents only the luminosity-averaged properties of the galaxies at a given $\beta_{\rm UV}$ {and can be easily dominated by several high-luminosity outliers}. Probing this scatter observationally is challenging, as typically only a small subset of the sample {are} detected in the dust continuum, even with the unprecedented sensitivity of ALMA \citep[\eg][]{Bouwens_2016, Dunlop_2017, Fudamoto_2020}. For the rest of the samples, only upper limits on their IRX are known. The few observations able to study the scatter in individual objects suggest that it is significant \citep[$\sigma_{\rm IRX}\simgreat$0.3 dex;][]{Oteo_2013, Fudamoto_2020}. However, our knowledge of this scatter in the general galaxy population at high-$z$ is still fairly limited.

Apart from the approach of using UV-selected LBGs, a few recent studies have analysed samples of high-$z$ dusty star-forming galaxies (DSFGs), which are generally the galaxy population selected at IR/submm bands \citep[\eg,][]{Penner_2012, Casey_2014}. IR-selected samples typically have complete UV detections, enabling a measurement of the IRX-$\beta_{\rm UV}$ relation of individual galaxies. These studies showed that DSFGs have systematically bluer $\beta_{\rm UV}$ compared to the local M99 relation as well as the LBG samples at their given IRX \citep{Penner_2012, Casey_2014P, Casey_2014, Safarzadeh_2017}. Furthermore, deviation of the DSFGs from the M99 relation appears to show a clear correlation with $L_{\rm IR}$ \citep{Casey_2014P, Casey_2014, Narayanan_2018}. These findings suggest that the derived IRX-$\beta_{\rm UV}$ relation of high-$z$ samples may be susceptible to selection effects.

The high sensitivity and resolution of ALMA has allowed the detection of dust emission of very early galaxies (\ie,~$z\simgreat5$) which was previously not possible with single-dish telescopes {\citep[\eg,][and Novak et al. submitted]{Capak_2015, Watson_2015, Walter_2016, Bouwens_2016, Laporte_2017, Venemans_2017, Carniani_2018, Jin_2019, Novak_2019, Matthee_2019, Banados_2019, Neeleman_2020, Bakx_2020, Faisst_2020, Bouwens_2020}}. To date, there have been a handful of objects at this epoch that have reported observational constraints on their IRX-$\beta_{\rm UV}$ relation but with large scatter. One major challenge is that the galaxies at such high redshifts are often observed at fewer photometric bands (both UV and IR) compared to those at low- or intermediate redshifts {\citep{Casey_2012, Casey_2018, Popping_2017, Reddy_2018, Liang_2019}}, and therefore observational constraints on $\beta_{\rm UV}$ and $L_{\rm IR}$ of these galaxies have much larger uncertainties. For example, $L_{\rm IR}$ extrapolated from a single-band ALMA flux density depends strongly on the adopted `dust temperature'\footnote{{The `dust temperature' here does not mean a physical temperature. Observations and simulations have shown that ISM dust has a wide distribution of (physical) temperature \citep[\eg][]{Harvey_2013, Lombardi_2014, Behrens_2018, Liang_2019}. Observationally, often a simplified SED function is adopted for extrapolating $L_{\rm IR}$ from single-band submm flux density, and the `$T$' parameter in the function is referred to as the `dust temperature' of galaxy in the literature \citep{Casey_2012, Casey_2014P}.}}  associated with the assumed functional form of the SED {\citep{Capak_2015, Bouwens_2016, Faisst_2017, Casey_2018, Liang_2019}}. Using 45 K instead of 35 K {with a standard modified blackbody (MBB) function \citep{Hildebrand_1983}} will lead to a factor of $\sim3$ increase in the derived IRX. It is therefore of paramount importance to constrain the uncertainties in the measurements of $\beta_{\rm UV}$ and $L_{\rm IR}$ and to compare them to the intrinsic scatter of the IRX-$\beta_{\rm UV}$ relation at high redshift.

Over the years, there have been a range of theoretical works that explore the physical nature of the IRX-$\beta_{\rm UV}$ relation and the scatter in this relation, including those that adopt analytic and semi-analytic approaches \citep[\eg,][]{Granato_2000, Noll_2009, Ferrara_2017, Faisst_2017, Popping_2017, Reddy_2018, Qiu_2019, Salim_2019} as well as the ones utilising idealised/cosmological hydrodynamic galaxy formation simulations {\citep[\eg,][]{Jonsson_2006, Safarzadeh_2017, Narayanan_2018, Behrens_2018, Ma_2019, Schulz_2020, Shen_2020}}. With different modelling methodologies, these attempts have been successful in accounting for the general power-law trend in the IRX-$\beta_{\rm UV}$ relation as well as reproducing the observed `secondary dependence' of this relation on other variables (\eg~$L_{\rm IR}$ and sSFR). 

The current general consensus amongst the different studies is that while dust optical depth is the key for driving a galaxy's location along the IRX-$\beta_{\rm UV}$ relation, the displacement off the relation arises from variations in the intrinsic UV spectral slope and the shape of the dust attenuation curve \citep[see \eg][and the references therein]{Salim_2020}. Several mechanisms have been suggested that influence the shape of the attenuation curve of galaxies, including changes of the intrinsic dust properties {\citep[\eg~dust composition and dust grain sizes;][]{Pei_1992, Fitzpatrick_1999, Weingartner_2001, Gordon_2003, Safarzadeh_2017, Narayanan_2018}, the spatial configuration of the dust and UV-emitting stars \citep[\eg][]{Natta_1984, Calzetti_1994, Witt_1996, Gordon_1997, Charlot_2000, Witt_2000, Narayanan_2018b, Trayford_2019}}, and the level of ISM turbulence \citep{Fischera_2003, Seon_2016, Popping_2017}. However, the importance of these mechanisms has not been conclusively determined, leaving open the significance of such variations in comparison with those resulting from changes of the intrinsic UV spectral slope. 

High resolution cosmological `zoom-in' galaxy formation simulations are ideal tools to quantify the relative importance of the various sources to the scatter in the IRX-$\beta_{\rm UV}$ relation and to understand how they are related to the observed `secondary dependence' of the relation. These simulations can model the various physical processes, including cosmic gas accretion, gas cooling, metal/dust production, and feedback processes, that are essential for reproducing the realistic star formation histories of galaxies as well as the complex geometry of dust distribution within galaxies {\citep{Somerville_2015, Vogelsberger_2020}}. Synthetic spectral energy distribution (SED) and multi-frequency imaging of the simulated sample can be produced through dust radiative transfer (RT) modelling {\citep[\eg][]{Hayward_2015, Snyder_2015, Torrey_2015, Camps_2018, Narayanan_2020}}. Subsequently, various observational properties of galaxies can be derived and a direct comparison to observations becomes possible.

In this work, we study the IRX-$\beta_{\rm UV}$ relation using a galaxy sample at $z=2-6$ that is extracted from the \textsc{\small MassiveFIRE} simulation suite \citep{Feldmann_2016, Feldmann_2017}. We focus on exploring the origin of the relation, and quantify the relative importance of the several main contributors to the intrinsic scatter. {We also analyse} how they contribute to the observed `secondary dependence' of the relation on other galaxy properties and how the relation depends on the evolutionary stage of a galaxy. Moreover, we also compare the level of the intrinsic scatter driven by the different mechanisms with the observational uncertainties {of} $\beta_{\rm UV}$ and $L_{\rm IR}$ measurements of high-$z$ galaxies. 

This paper is structured as follows. In Section~\ref{Sec:2}, we summarise the simulation methodology and details of the radiative transfer analysis of our sample. In Section~\ref{Sec:3}, we compare the predicted IRX-$\beta_{\rm UV}$ relation of our sample with recent observational data at similar redshifts. In Section~\ref{Sec:4}, we explore in detail the nature of the IRX-$\beta_{\rm UV}$ relation and investigate the various {contribution to} the \textit{intrinsic} scatter {of} this relation. In Section~\ref{Sec:5}, we compare {this} intrinsic scatter with the uncertainties {of} $\beta_{\rm UV}$ and $L_{\rm IR}$ measurements of high-$z$ galaxies. We summarise the findings of this paper and conclude in Section~\ref{Sec:6}. Throughout this paper, we adopt cosmological parameters in agreement with the nine-year data from the Wilkinson Microwave Anisotropy Probe \citep{Hinshaw_2013}, specifically $\Omega_{\rm m}=0.2821$, $\Omega_{\Lambda}=0.7179$, and $H_0=69.7\; \rm km\;s^{-1}\;Mpc^{-1}$.

\vspace{-10pt}
\section{Simulation Methodology}
\label{Sec:2}

In this section, we introduce the simulation methodology. In Section~\ref{Sec:2a}, we briefly introduce the simulation suites from which our galaxy sample is extracted. And in Section~\ref{Sec:2b}, we summarise the methodology of the dust RT analysis on our galaxy sample. 

\subsection{Simulation setup and galaxy catalogue}
\label{Sec:2a}

We adopt the galaxy sample {($M_*\approx 10^9-10^{12} \,M_\odot$)} extracted from the \textsc{\sc MassiveFIRE} cosmological  `zoom-in' suite \citep{Feldmann_2016, Feldmann_2017}, which is part of the Feedback in Realistic Environments (\textsc{fire}) project\footnote{\url{fire.northwestern.edu}} \citep{Hopkins_2014}. The simulation methodology of \textsc{\sc MassiveFIRE} has been described in the above papers, and we refer the interested readers to them for more details. We summarise only the salient points here. 

The \textsc{\small MassiveFIRE} simulations are run with the gravity-hydrodynamics code \textsc{gizmo}\footnote{A public version of \textsc{gizmo} is available at \url{http://www.tapir.caltech.edu/phopkins/Site/GIZMO.html}} (\textsc{\small FIRE-1} version) in the Pressure-energy Smoothed Particle Hydrodynamics (``P-SPH") mode \citep{Hopkins_2013, Hopkins_2015}. The initial conditions of the simulations are generated using the \textsc{MUSIC} (Multi-Scale Initial Conditions) code \citep{Hahn_2011} within the periodic simulation boxes of the low-resolution (LR) dark matter (DM)-only runs with the WMAP cosmology \citep{Hinshaw_2013}. From the outputs of the LR runs, we select a number of model halos to resimulate at much higher resolution and with baryons included. The selected halos have a variety of masses, accretion history, and environmental overdensities.

The catalogue used for this paper includes 18 massive halos selected from a $(100\,{\rm Mpc\,h^{-1}})^3$ comoving simulation box at $z_{\rm final}=2$ (from Series A, B, and C in \citealt{Feldmann_2017}) and 11 additional massive halos selected from two larger boxes (400 and 762 ${\rm Mpc\,h^{-1}}$ on a side) at $z_{\rm final}=6$ \citep{Liang_2019}. Initial conditions for the  `zoom-in' runs are set up with a convex hull surrounding all particles within $3R_{\rm vir}$ at $z_{\rm final}$ of the chosen halo defining the Lagrangian high-resolution (HR) region following the method introduced by \citet{Hahn_2011}. The mass resolution of the HR runs {for dark matter and gas particles} are $m_{\rm DM}=1.7\times10^5\,M_\odot$ and $m_{\rm gas}=3.3\times10^4\,M_\odot$, respectively. The most massive progenitors (MMPs) of the galaxies are identified using the \textsc{amiga} Halo Finder \citep{Gill_2004, Knollmann_2009}. 

The simulations incorporate various gas cooling processes (free-free, photoionization/recombination, Compton, photoelectric, metal-line, molecular and fine-structure processes) {and a uniform UV background using the FG09 model \citep{Faucher_Giguere_2009}, and self-consistently account for 11 separately tracked metal species}. Star formation occurs in self-gravitating, dense and self-shielding molecular gas based on a sink-particle prescription. Specifically, gas that is locally self-gravitating and has density exceeding $n_{\rm crit} = 5\,\rm cm^{-3}$ is assigned an SFR $\dot{\rho}_* = f_{\rm mol}\, \rho_{\rm gas} /t_{\rm ff}$, where $t_{\rm ff}$ is the local free-fall time of gas and $f_{\rm mol}$ is the self-shielding molecular mass fraction calculated following \citet{Krumholz_2011}. Due to the self-gravity criterion, the mean gas density at which star formation occurs is actually significantly higher ($\sim100\,\rm cm^{-3}$ for the resolution of the simulations) than $n_{\rm crit}$. 

The initial mass of a star particle is set to be equal to the mass of the parent gas particle from which it is spawned. Once the star particle is formed, it acts as a single stellar population (SSP) with given metallicity and age. The simulations explicitly incorporate several different stellar feedback channels including (1) local and long-range momentum flux from radiative pressure, (2) energy, momentum, mass, and metal injection from supernovae (Types Ia and II), (3) and stellar mass-loss (both OB and AGB stars), and (4) photoionization and photoelectric heating processes. The relevant stellar feedback quantities are tabulated in the simulations based on the stellar population model \textsc{\small starburst99} (hereafter SB99) with a Kroupa initial mass function (IMF) \citep{Leitherer_1999}, without subsequent adjustment or fine-tuning. We refer the readers to \citet{Hopkins_2014} for details of the feedback prescriptions.

\textsc{fire} simulations have successfully reproduced a variety of observed galaxy properties relevant for this work, including the stellar-to-halo-mass relation \citep{Hopkins_2014, Feldmann_2017}, the sSFRs of galaxies at the cosmic noon ($z\sim2$) \citep{Hopkins_2014, Feldmann_2016}, the {gas-phase and} stellar mass-metallicity relation \citep{Ma_2016a}, the submm flux densities at 850 $\rm \mu m$ \citep{Liang_2018}, the observational effective dust temperatures at $z\sim2$ \citep{Liang_2019} as well as the UV luminosity functions and cosmic star formation rate density at $z>5$ \citep{Ma_2019}.

\vspace{-10pt}
\subsection{Predicting dust SED with \textsc{skirt}}
\label{Sec:2b}

We generate the UV-to-mm continuum SEDs for the galaxy catalogue using \textsc{skirt}\footnote{\textsc{skirt} home page: \url{http://www.skirt.ugent.be}.}, an open-source\footnote{\textsc{skirt} code repository: \url{https://github.com/skirt}} 3D Monte Carlo dust RT code \citep{Baes_2011,Baes_2015, Camps_2015a}. \textsc{skirt} provides full treatment of absorption and multiple anisotropic scattering by dust, and self-consistently computes the dust thermal re-emission and the dust temperature distribution for various astrophysical systems. To prepare our galaxy snapshots as RT input models, we follow the prescription of \citet{Camps_2016} (see also \citealt{Trayford_2017, Camps_2018}).  Here we only summarise the main points of the prescription and refer the readers to the above-mentioned papers for the details.

For the radiative transfer (RT) analysis, each star particle is treated as a SSP, and a spectrum is assigned to each star particle according to the age, {initial} metallicity, and initial mass of the particle. The RT calculations are performed based on an equally spaced logarithmic wavelength grid ranging from $\lambda=0.05-1000\,\rm \mu m$. We launch $10^6$ photon packages for each point in the wavelength grid and for each of the stellar emission and following dust emission phases. To produce the mock images and SEDs of galaxies, we place mock detectors at an arbitrary ``local" distance of 10 Mpc from galaxy along different viewing angles to accumulate both spatially resolved as well as integrated fluxes at each wavelength grid point. 

{Dust mass is assumed to trace metal mass in the ISM.} We discretise the spatial domain using an octree grid and keep subdividing grid cells until the cell contains less than $f=3\times10^{-6}$ of the total dust mass and the $V$-band (0.55 $\rm \mu m$) optical depth in each cell is less than unity. The highest grid level corresponds to a cell width of $\sim20$ pc, \ie~about twice the minimal SPH smoothing length. Gas hotter than $10^6$ K is assumed to be dust-free because of sputtering \citep{Hirashita_2015}. We self-consistently calculate the self-absorption of dust emission and include the transient heating function to calculate non-local thermal equilibrium (NLTE) dust emission by transiently heated small grains and PAH molecules \citep{Baes_2011, Camps_2015b}. To account for the heating of dust by the cosmic microwave background (CMB), we adopt a correction to the dust temperature following Eq. 12 of \citet{Cunha_2013}. 

\begin{figure}
 \begin{center}
 \includegraphics[width=80mm]{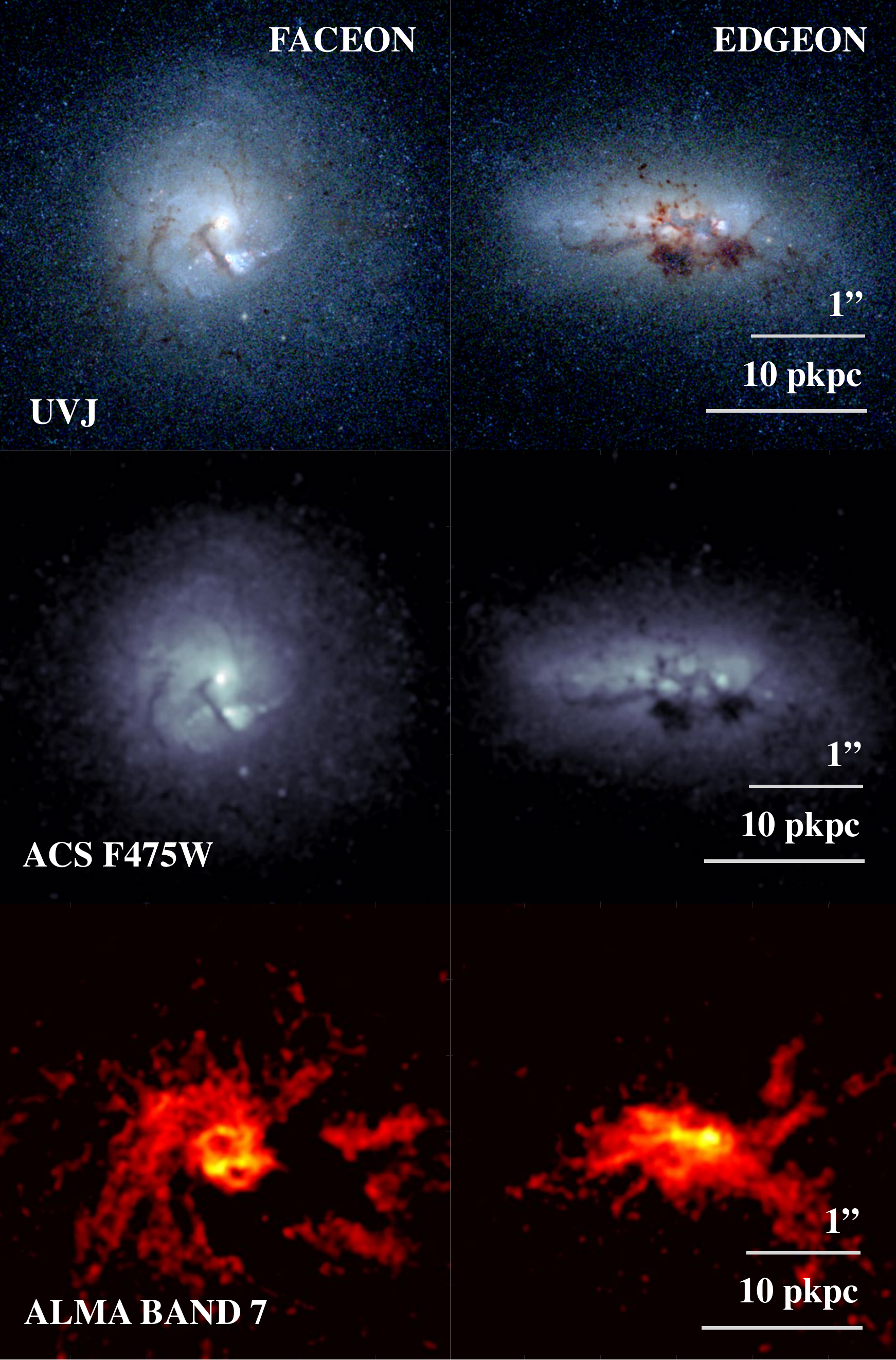}
 \caption{Synthetic images of a disc-like \textsc{\small MassiveFIRE} galaxy at $z=2$ {for} face-on (left panels) and edge-on (right panels) viewing directions. The top, middle and bottom panels {show} the composite UVJ, the \textit{HST} ACS F475W-band, and the ALMA band 7 images, respectively. }
 \label{fig.1}
  \end{center}
   \vspace{-10pt}
\end{figure}

\begin{figure}
 \begin{center}
 \includegraphics[width=75mm]{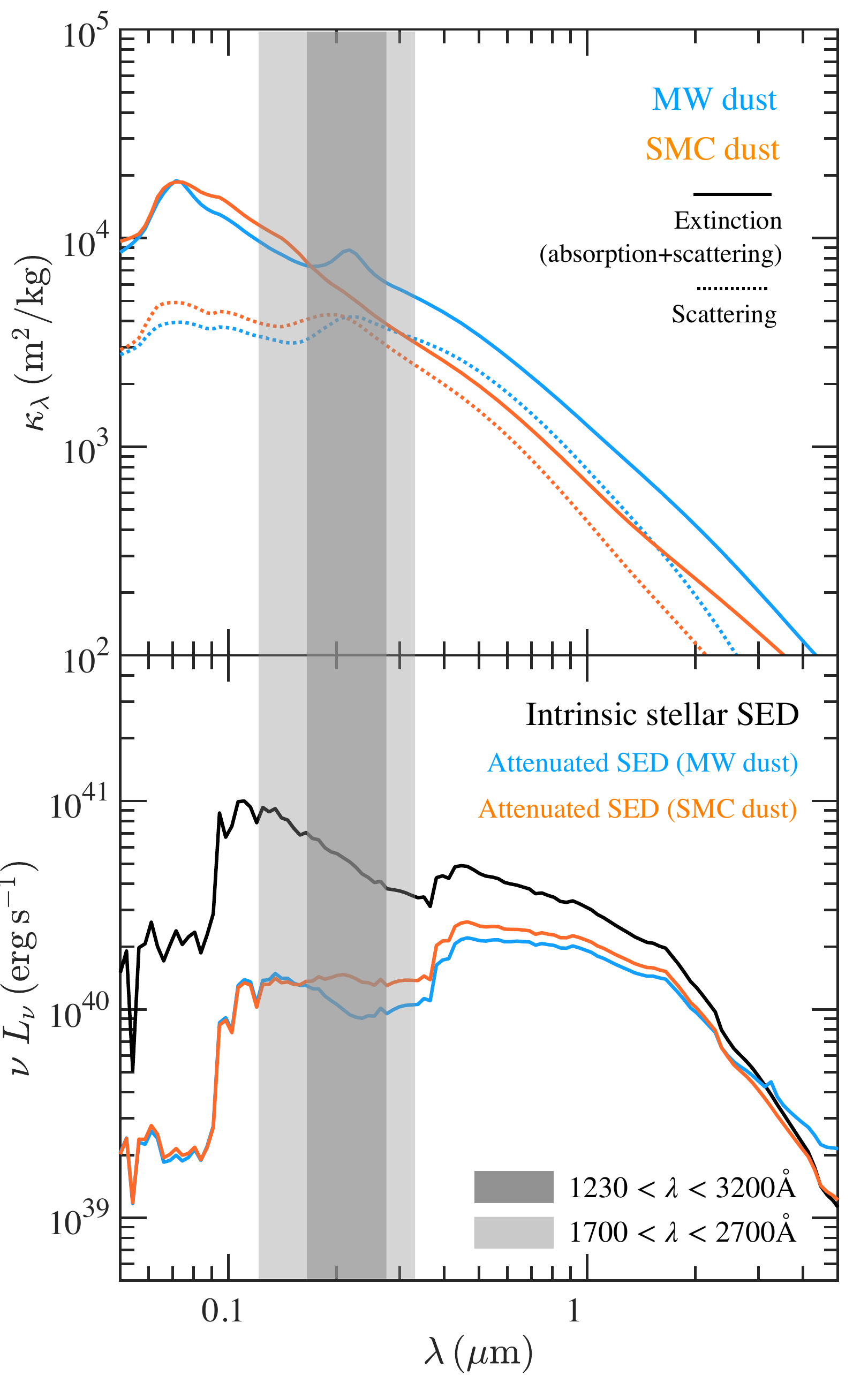}
 \caption{\textit{Upper} panel: Dust extinction (absorption+scattering, solid lines) and dust scattering curves (dotted lines) of the \citet{Weingartner_2001} dust {models}. The blue and orange lines correspond to the Milky Way (MW) and SMC dust {models}, respectively. \textit{Lower} panel: The SEDs of a selected $z=2$ \textsc{\small MassiveFIRE} galaxy. The black line indicates the intrinsic stellar SED of this galaxy. The blue and orange lines show the dust-attenuated SEDs {for a selected viewing angle} that are computed with \textsc{skirt} using the MW and SMC dust {models}, respectively. The light grey area in both panels show the wavelength range $1230<\lambda<3200\,\angstrom$, within which the measured photometry are used for estimating $\beta_{\rm UV}$ by the different studies. The dark grey area marks the regime of the `bump' feature in the MW extinction curve. The derived $\beta_{\rm UV}$ can differ significantly depending on whether the photometry within this regime is included or not if a `bump' feature exists.}
 \label{fig.2}
  \end{center}
   \vspace{-20pt} 
\end{figure}

To understand how several uncertainties in the stellar population and dust properties of high-$z$ galaxies can influence the IRX-$\beta_{\rm UV}$ relation of galaxies, we run several series of RT simulations with different parametrisation of inputs for each galaxy in our sample. This includes a change in 1) {dust extinction law (due to different grain composition and grain size distribution, see Section~\ref{Sec:4e1})},  2) the dust-to-metal mass ratio ($\delta_{\rm dzr}$, {see} Section~\ref{Sec:4e3}) and 3) the stellar population model (single vs. binary stellar evolution, {see} Section~\ref{Sec:4e4}). In the \emph{default} RT model, we adopt the SB99 SED libraries, the Milky Way (MW) dust {model} of \citet[][hereafter WD01]{Weingartner_2001} (for the case of $R_{\rm V}=3.1$, see Fig~\ref{fig.2} for the shape of the extinction curve {of this model}), and assume a constant $\delta_{\rm dzr}=0.4$ \citep{Dwek_1998, Draine_2007, Li_2019}. In addition to these runs, we also perform three additional RT calculations for each galaxy, with alternative choice for the stellar population model (\textsc{bpass}, \citealt{Eldridge_2012, Eldridge_2017}), {dust model (SMC dust} of WD01, see Fig.~\ref{fig.2}), and $\delta_{\rm dzr}$ (0.2 and 0.8), with the other input parameters fixed. We compare the difference in the IRX-$\beta_{\rm UV}$ relation caused by each of the three changes in the RT model in Section~\ref{Sec:4e}. Unless stated otherwise, we refer to the results of the default model throughout the paper.

Finally, we note that while our simulations have better resolution than many previous simulations modelling dust extinction and emission \citep[\eg,][]{Jonsson_2006, Narayanan_2010, Hayward_2011, De_Looze_2014, Camps_2016} and can directly incorporate various important stellar feedback processes, they might still be unable to resolve the emission from H \textsc{\small II} and photodissociation regions (PDR) from some of the more compact birth-clouds surrounding star-forming cores. The time-average spatial scale of these H \textsc{\small II}+PDR regions typically varies from $\sim5$ to $\sim800$ pc depending on the local physical conditions \citep{Jonsson_2010}. Therefore, we also perform additional RT calculations, where star particles are split into two sets according to their age. Star particles that formed less than 10 Myrs ago are identified as `young star-forming' particles, while older star particles are treated as in the default model. To account for the pre-processing of radiation by birth-clouds, we follow \citet{Camps_2016} in assigning a source SED from the \textsc{mappingsiii} \citep{Groves_2008} family to young star-forming particles. Dust associated with the birth-clouds is removed from the neighbouring gas particles to avoid double-counting \citep[see][]{Camps_2016}. We will discuss the effect of the variations in the conditions of the birth-clouds on the IRX-$\beta_{\rm UV}$ relation in Section~\ref{Sec:4e5}.   

We show in Fig.~\ref{fig.1} the synthetic images produced {by} \textsc{skirt} on one of our \textsc{\small MassiveFIRE} galaxies \citep[galaxy ID: MF A2:0,][]{Feldmann_2016, Feldmann_2017} at $z=2$ for both face-on (left panels) and edge-on viewing directions (right panels). In particular, we show compositive U, V, J false-colour images (top panels), images of the flux densities at \textit{Hubble Space Telescope (HST)} ACS-F475W band (middle panels) and ALMA band 7 (bottom panels). The broadband flux densities are calculated by convolving the simulated SED output from \textsc{skirt} with the transmission function of each band filter. The \textit{HST} ACS-F475W band corresponds to rest-frame {$\rm \lambda\approx1600\,\angstrom$} at $z=2$, \ie~in the far-ultraviolet (FUV) regime. The regions of higher dust extinction in the UV/optical corresponds to the most luminous regions at submm wavelength due to dust thermal emission. 

We compute $L_{\rm IR}$ of the \textsc{\small MassiveFIRE} galaxies by integrating the simulated SEDs over the wavelength range $\rm \lambda=8-1000\,\mu m$. Without explicit notification, $\beta_{\rm UV}$ is calculated using the flux densities measured at rest-frame $\lambda=1230$ and  $3200\,\angstrom$ to avoid the contamination by the 2175 $\angstrom$ `bump' feature {(indicated by dark grey area in Fig.~\ref{fig.2}, see also \citealt{Behrens_2018})} in the MW extinction curve, \ie~

\begin{equation}
\beta_{\rm UV} = \frac{{\rm log}\,(f_{\lambda,\,0.12})-{\rm log}\,(f_{\lambda,\,0.32})}{
{\rm log}\,(\lambda_{0.12})-{\rm log}\,(\lambda_{0.32})}
\label{eq.1}
\end{equation}

\noindent where $f_{\lambda,\,0.12}$ and $f_{\lambda,\,0.32}$ are the specific flux (in units of $\rm erg\,s^{-1}\,m^{-3}$) at $\lambda=1230$ and 3200 $\angstrom$, respectively. Throughout the paper, we adopt {the MW and SMC dust models of WD01}; However, we note that {the dust properties (\ie~composition and grain size distribution) and the resulting shape of the dust extinction curve} of high-$z$ galaxies are uncertain \citep{Stratta_2007, Zafar_2011, Salim_2020}, {in particular} the strength of the `bump' feature at around $\lambda=2175\,\angstrom$ \citep{Kriek_2013, Ma_2015, Ma_2017, Narayanan_2018b}. 

\vspace{-10pt}

\section{Comparing simulations with observations}
\label{Sec:3}

In this section, we compare the predicted IRX vs. $\beta_{\rm UV}$ relation of the \textsc{\small MassiveFIRE} sample with recent observational data. We first introduce the `canonical' relations derived using local starburst galaxies in Section~\ref{Sec:3a}. In Section~\ref{Sec:3b}, we compare the simulation data with the stacked data derived using the high-$z$ Lyman-break galaxy (LBG) samples. And in Section~\ref{Sec:3c}, we compare it with the data of IR-selected samples. We also discuss secondary {dependences} of the IRX vs. $\beta_{\rm UV}$ relation and the impact of selection effect in observation of galaxies in Section~\ref{Sec:3c}.

\begin{figure*}
 \begin{center}
 \includegraphics[width=88mm, height=81 mm]{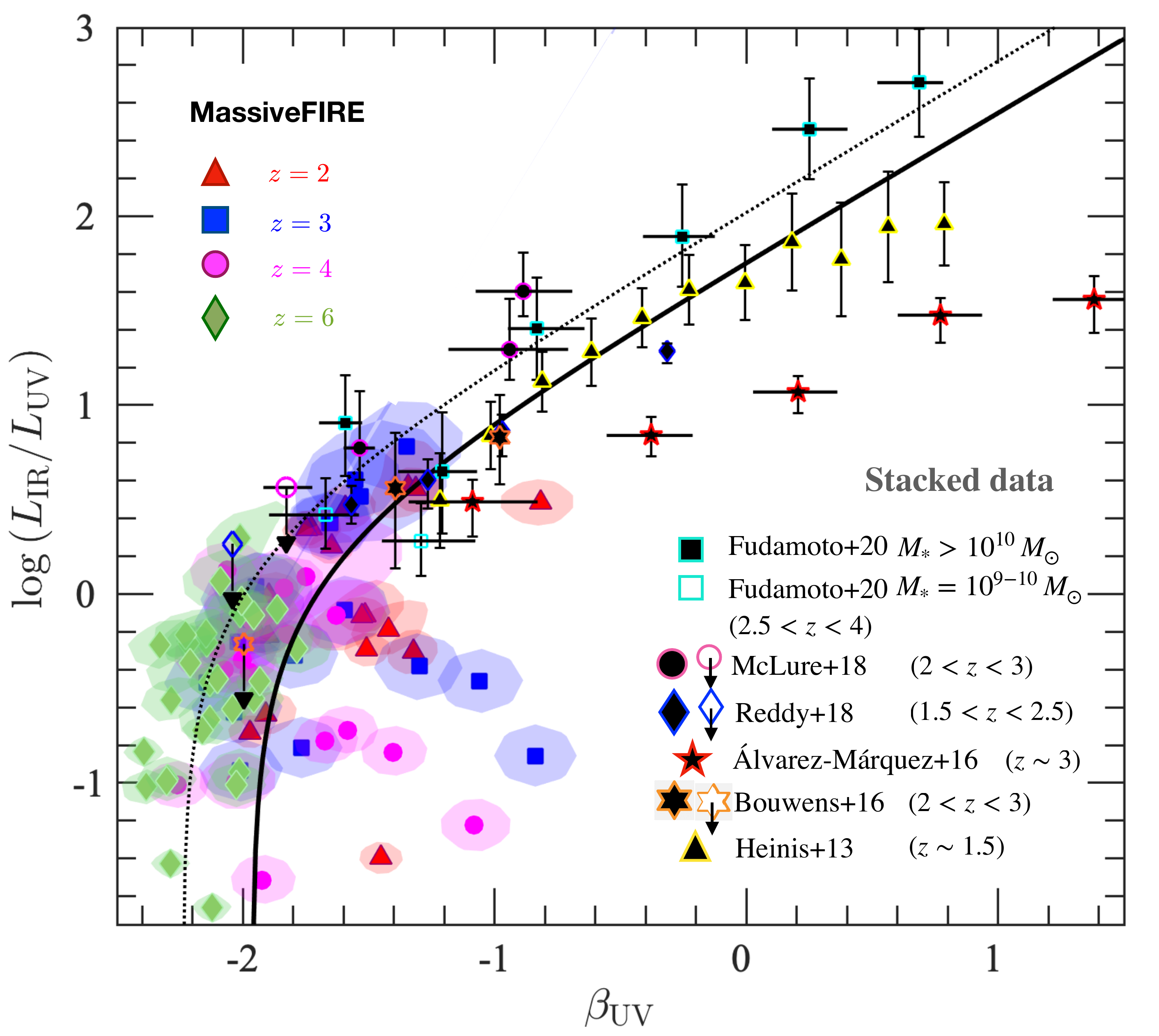}
 \includegraphics[width=88mm, height=81 mm]{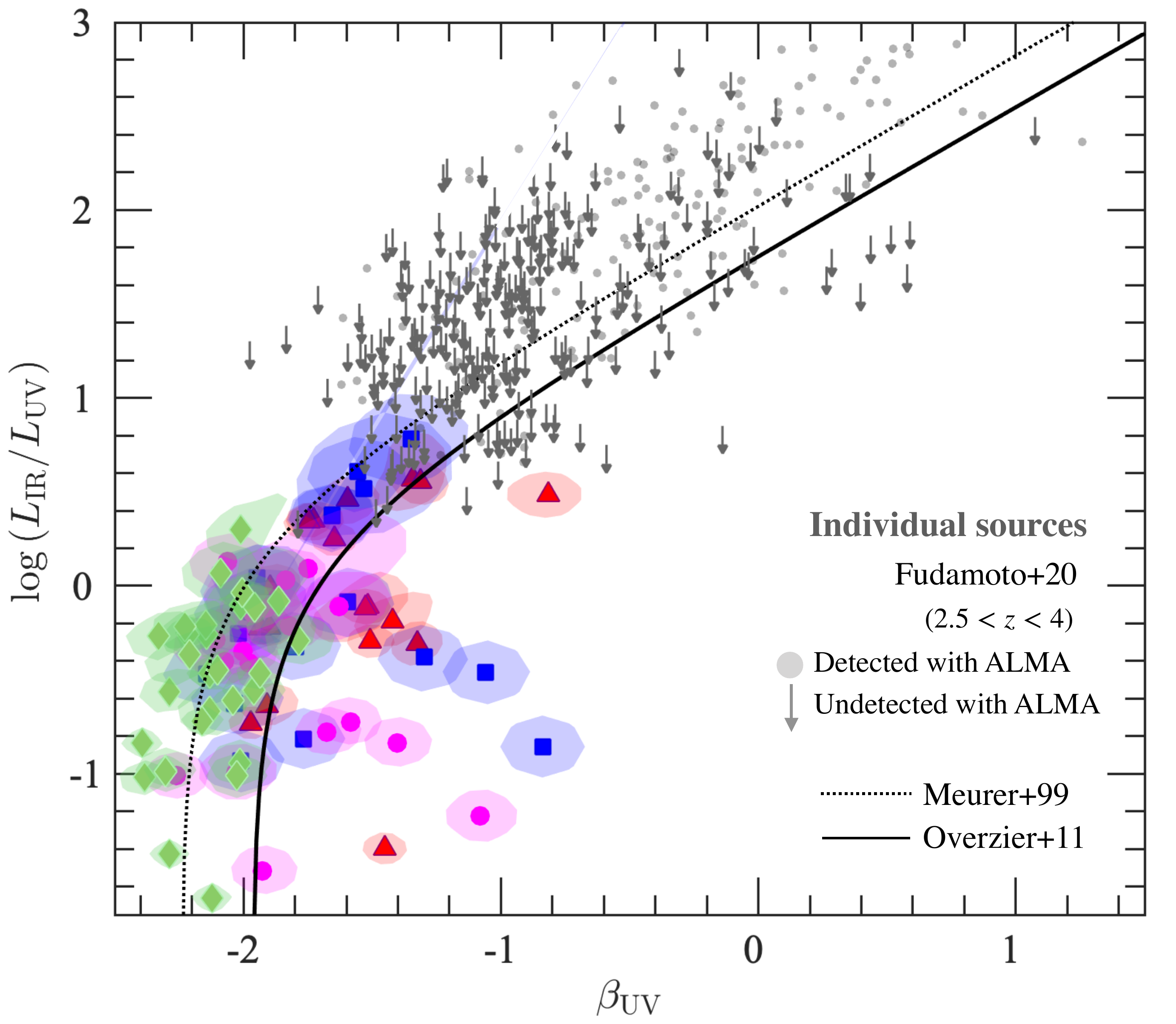}
 \vspace{-5pt}
 \caption{Comparison between the IRX-$\beta_{\rm UV}$ relation of the \textsc{\small MassiveFIRE} galaxies with the recent observational data derived using UV-selected samples. In both panels, red triangles, blue squares, magenta circles and green diamonds correspond to the \textsc{\small MassiveFIRE} sample at $z=2$, $z=3$, $z=4$ and $z=6$, respectively. The contour around each data point indicates the $3\sigma$ (\ie~$99.8\%$ confidence level) distribution of the result of 24 randomly selected viewing directions. In the \textit{left} panel, the observational data are stacked results, including \citet{Heinis_2013} (yellow-edged triangles), \citet{Alvarez_Marquez_2016} (red-edged stars), \citet{Bouwens_2016} (orange hexagram), \citet{McLure_2018} (magenta-edged circles), \citet{Reddy_2018} (blue-edged diamonds) and \citet{Fudamoto_2020} (cyan-edged squares). The results of the individual sources in the \citet{Fudamoto_2020} sample are explicitly shown in the \textit{right} panel. The ALMA-detected objects are indicated by grey dots whereas the $3\sigma$ upper confidence limit of the non-detected objects in the sample are marked with grey downward arrows. In both panels, the dotted and solid black lines indicate the original and the aperture-corrected M99 relations derived using the local starburst samples, respectively (Section~\ref{Sec:3a}). {While the majority of \textsc{\small MassiveFIRE} galaxies are in broad agreement with the local M99 relation, a number of them appear to be noticeably below (or rightwards) due to the recent quenching of star formation in those galaxies (Section~\ref{Sec:4d}). The stacked results derived from the various LBG samples exhibit non-trivial discrepancies among different studies. The majority ($\sim90\%$) of the LBGs in these samples are not individually resolved by ALMA. }} 
  \label{fig.3}
  \end{center}
     \vspace{-15pt}
\end{figure*}

\subsection{The canonical relation of local starbursts}
\label{Sec:3a}

The relation between IRX and $\beta_{\rm UV}$ of galaxies was {first derived based on a sample of $\sim60$ nearby compact starbursts (M99)}. In this work, $\beta_{\rm UV}$ {was} measured using the \textit{International Ultraviolet Explorer (IUE)} satellite \citep{Kinney_1993}, while far-IR luminosity (over $40\le\lambda\le120\,\rm \mu m$) were extrapolated from the two-band (60 and 100 $\rm \mu m$) photometry measured by the \textit{Infrared Astronomical Satellite (IRAS)}. Since then, IRX is defined more often using the bolometric IR luminosity in the literature, which {includes} emission over a larger wavelength range of $8\le\lambda\le1000\,\rm \mu m$. The {revision of} the definition of IRX results in an increase of IRX by 0.15 dex with respect to the original equation {\citep[\eg][]{Calzetti_2000}}, Eq. 10 of M99. With the new definition, the equation is revised to be

\begin{equation}
{\rm log}\,{\rm IRX} = {\rm log}\,(10^{0.4\,(4.43+1.99\,\beta_{\rm UV})}-1)+0.076.
\label{eq.2}
\end{equation} 

\noindent Hereafter, we refer to this relation as the `M99 relation' in this paper. Given that this result has widely been adopted for correcting dust-attenuated SFRs of galaxies over a range of redshifts by many different studies, we also refer to it as the `canonical relation', as {in} many other previous works.

One major problem with the M99 relation is that the UV fluxes measured with the \textit{IUE} satellite were incomplete due to its limited aperture size ($10" \times 20"$), which is typically much smaller than the full spatial extent of a local galaxy ($\sim$arcmins). Hence, $L_{\rm UV}$ were underestimated and because \textit{IUE} had focused only on the galaxies' core regions, a colour bias of $\beta_{\rm UV}$  was introduced because of the colour gradient. The same sample was later remeasured using the \textit{Galaxy Evolution Explorer} \citep[\textit{GALEX};][]{Morrissey_2007} by other groups \citep[][]{Overzier_2011, Takeuchi_2012, Casey_2014}, which has a much larger aperture size. These studies find lower IRX and redder $\beta_{\rm UV}$ of the exact same galaxies generally. We quote here the relation derived by  \citet{Overzier_2011} 
  
\begin{equation}
{\rm log}\,{\rm IRX} = {\rm log}\,(10^{0.4\,(4.54+2.07\,\beta_{\rm UV})}-1)+0.225.
\label{eq.3}
\end{equation} 

\noindent Hereafter, we will refer to Eq.~\ref{eq.3} as the `$\rm M99_{\rm corr}$ relation' in the paper, which stands for `aperture-corrected M99 relation', to distinguish it from the original result derived by M99 (Eq.~\ref{eq.2}). 

\subsection{The high-$z$ IRX-$\beta$ relation}
\label{Sec:3b}

\subsubsection{The results derived by UV-selected galaxies}

The majority of the current observational constraints on the IRX-$\beta_{\rm UV}$ relation at high-$z$ are derived using LBG samples due to the efficiency of the selection methods \citep{Steidel_1996}. Most studies have reported stacked results due to the difficulty in getting reliable detection of the dust continuum of many individual sources.

In the \textit{left} panel of Fig.~\ref{fig.3}, we show the stacked results of the LBG samples at $1.5\simless z\simless4$ obtained by \citet{Heinis_2013} (yellow-edged triangles), \citet{Bouwens_2016} (orange-edged hexagram), \citet{Alvarez_Marquez_2016} (red-edged astericks),  \citet{Reddy_2018} (blue-edged diamonds), \citet{McLure_2018} (pink-edged circles) and \citet{Fudamoto_2020} (cyan-edged squares). In these studies, galaxies are binned by their measured $\beta_{\rm UV}$, which are determined by fitting the power-law, $f_\lambda\propto\lambda^{\beta_{\rm UV}}$, to the available UV photometry over different wavelength ranges between rest-frame $1200\,\angstrom$ and $3200\,\angstrom$. For each bin, an IRX is extrapolated using the stacked UV and IR photometry of the objects in that bin. Specifically, $L_{\rm UV}$ {($\equiv\lambda_{0.16}L_{\lambda,\,0.16}$) is} extrapolated from the best-fit power-law function at $\rm \lambda=1600\,\angstrom$, while  $L_{\rm IR}$ is derived by fitting the assumed dust SED templates {\citep[\eg][]{Dale_2002}} or {MBB functions \citep[see \eg][]{Casey_2012, Casey_2014P}} to the available stacked \textit{Herschel} or ALMA broadband fluxes. We summarise the detailed methodology for deriving $\beta_{\rm UV}$, $L_{\rm UV}$ and $L_{\rm IR}$ used by each study in Table~\ref{T1}.

\begin{table*}
\caption{The selection criteria and the methods for deriving $L_{\rm IR}$, $L_{\rm UV}$ and $\beta_{\rm UV}$  adopted by each observation that is referenced in Fig.~\ref{fig.3} (Section~\ref{Sec:3b}).}
\begin{tabular}{ p{2.6 cm} p{6 cm}  p{8 cm} }
 \hline
\multicolumn{1}{}{} \; Paper \;\;\; & Selection criteria of the sample &  Methods for deriving $L_{\rm IR}$,  $\beta_{\rm UV}$ and $L_{\rm UV}$  \\
 \hline
 \citet{Heinis_2013} &  The sample contains 42,184 galaxies selected from the optical imaging of the COSMOS field \citep{Capak_2007} in the $u^*$ band ($1.2<z_{\rm phot}<1.7$ and $u^*<26$ mag at $5\sigma$). The mean redshift of the sample is $<z_{\rm phot}>=1.43$. & $L_{\rm IR}$ {is} derived from fitting the \citet{Dale_2002} SED templates to the stacked fluxes at 250, 350 and 500 $\rm \mu m$ extracted from \textit{Herschel} Space Observatory Spectral and Photometric Imaging Receiver \citep[SPIRE,][]{Griffin_2010, Swinyard_2010} imaging of the COSMOS field. $\beta_{\rm UV}$ {is} computed by fitting the photometry to a single power-law SED, $f_\lambda\propto\lambda^{\beta_{\rm UV}}$. The rest-UV photometry is obtained from the Subaru $u^*$ and $V$ broad-band, 12 intermediate- and 2 narrow-band filters that cover the wavelength range $1200<\lambda<3000\angstrom$. $L_{\rm UV}$ {is} calculated at rest-frame 1600 $\angstrom$ using the best-fit SEDs.\\
  \hline 
  \citet{Alvarez_Marquez_2016} & The sample contains $\sim$22,000 LBGs at $2.5<z_{\rm phot}<3.5$ that are selected within the COSMOS field using broad-band filters $u^*$, $V_{\rm J}$ , and $i^+$ ($V_{\rm J}<26.5$ and $i^+<26.1$ mag at $5\sigma$) and are included in the \citet[][version 2.0]{Ilbert_2009} photometric redshift catalogue. The mean redshift of the sample is $<z_{\rm phot}>=3.02$.  & $L_{\rm IR}$ {is} estimated by fitting the \citet{Dale_2014} SED templates to the stacked fluxes at \textit{Herschel} Photodetector Array Camera and Spectrometer \citep[PACS,][]{Poglitsch_2010} (100 and 160 $\rm \mu m$) and SPIRE (250, 350 and 500 $\rm \mu m$), and AzTEC (1.1 mm) bands. $\beta_{\rm UV}$ {is} computed by fitting the power-law SED to the rest-UV photometry within the wavelength range $1250<\lambda<2000\angstrom$ from the \citet{Capak_2007} catalogue. The photometry is obtained using the Subaru $B_{\rm J}$, $V_{\rm J}$, $g^+$, $r^+$, $i^+$, $z^+$ broad-band, 12 intermediate- and 2 narrow-band filters. $L_{\rm UV}$ {is} calculated at rest-frame $1600\,\angstrom$ from the best-fit SEDs.\\
    \hline 
     \citet{Bouwens_2016} & The sample includes 330 LBGs spanning the redshift range $2<z_{\rm phot}<10$ selected from the \textit{Hubble Ultra-Deep Field} (HUDF) via dropout technique. & $L_{\rm IR}$ {is} inferred from converting the stacked ALMA 1.2 mm fluxes by a standard modified blackbody function with a dust temperature of 35 K and a power-law spectral index for dust emissivity of $\beta_{\rm IR}=1.6$. $\beta_{\rm UV}$ of the galaxies {is} estimated by fitting the \textit{HST} photometry in various bands (from ACS-F606W to WFC3-F160W) to the power-law SEDs. $L_{\rm UV}$ is calculated from the best-fit SEDs at $1600\,\angstrom$. 
     \\
    \hline 
 \citet{McLure_2018}   &  The sample consists of the star-forming galaxies at $2<z_{\rm phot}<3$ within the deep 1.2-mm ALMA mosaic (35 $\rm \mu Jy\;beam^{-1}$ at $1\sigma$) of the HUDF presented by \citet{Dunlop_2017}. The UV-to-MIR photometry were assembled from the Great Observatories Origins Deep Survey-South (GOODS-S) \citep{Guo_2013} and the Ultra-Deep Survey (UDS) \citep{Galametz_2013} catalogues provided by the Cosmic Assembly Near-infrared Deep Extragalactic Legacy Survey (CANDELS) team, incorporated with the catalogue for the Unltra-Visible and Infrared Survey Telescope for Astronomy (UVISTA) survey Data Release 3 (DR3) \citep{Mortlock_2017}.   & $L_{\rm IR}$ {is} estimated from converting the 1.2-mm flux, assuming an optically thin modified blackbody spectrum with a dust temperature of 35 K and a dust emissivity index of $\beta_{\rm IR}=1.6$.  $\beta_{\rm UV}$ is determined by fitting the photometry from the GOOD-S, UDS and UVISTA catalogues that cover the wavelength range $1268<\lambda<2580\angstrom$ with the power-law SED. $L_{\rm UV}$ is calculated from the best-fit SEDs at $1600\,\angstrom$. \\
  \hline
 \citet{Reddy_2018} & The sample contains $\sim3,500$ galaxies at $1.5\le z_{\rm phot} \le 2.5$ extracted from the ground- and space-based photometry compiled by the 3D-\textit{HST} survey \citep{Skelton_2014} with newly obtained imaging from the HDUV Legacy Survey \citep{Oesch_2018} in the GOODS-N and GOODS-S fields. The UV and optical depth of the sample are $H\approx27$ and $m_{\rm UV}\approx27$, respectively. Objects identified as X-ray active galactic nucleus (AGN) \citep{Shao_2010, Xue_2011} or classified as quiescent galaxies by the UVJ method \citep{Williams_2009} are excluded. & $L_{\rm IR}$ {is} derived by fitting the \citet{Elbaz_2011} main-sequence dust SED template to the stacked fluxes at 100 and 160 $\mu m$ obtained from the \textit{Herschel}/PACS imaging. $\beta_{\rm UV}$ {is} computed by fitting the power-law function to the broadband photometry covering the wavelength range of $1268<\lambda<2580\,\angstrom$. $L_{\rm UV}$ is calculated from the best-fit SEDs at $1600\,\angstrom$. \\
 \hline
  \citet{Fudamoto_2020} & The sample includes 1, 512 galaxies at $2.5<z_{\rm phot}<4$ selected from the COSMOS2015 catalogue \citep{Laigle_2016} that are part of the ALMA archival band 6 and 7 observations that were publicly available as of January 2018. The COSMOS2015 catalogue are extracted from a combined near-IR image from the UltraVISTA survey \citep[$J$, $H$, $K$ bands,][]{McCracken_2012} and the $z^+$ image from the Subaru telescope.  & $L_{\rm IR}$ {is} estimated by scaling a dust SED template, which is previously derived for $z\sim3$ galaxies by \citet{Alvarez_Marquez_2016}, to the stacked fluxes in ALMA band 6 or 7 of the sample.  $\beta_{\rm UV}$ {is} estimated by employing SED fitting to photometric data over the wavelength range of $1500<\lambda<2500\angstrom$. $L_{\rm UV}$ is calculated from the best fit SEDs at $1600\,\angstrom$. \\
 \hline
\end{tabular}
\label{T1}
\end{table*}

Looking at the \textit{left} panel, we can see that these studies have reported fairly diverse results of IRX-$\beta_{\rm UV}$ relation. Specifically, \citet{McLure_2018} and \citet{Fudamoto_2020} show fairly blue $\beta_{\rm UV}$ and the derived relations agree with the canonical M99 relation (dotted black line). The relations by \citet{Heinis_2013}, \citet{Bouwens_2016} and \citet{Reddy_2018}, however, are more compatable with the corrected M99 relation (solid black line). And yet the relation by \citet{Alvarez_Marquez_2016} appears to be significantly `redder' than the other observations and {shows} shallower slope compared to the M99 or $\rm M99_{\rm corr}$ relations. The shape of this curve resembles the expected relation of a SMC-type dust extinction curve (see Section~\ref{Sec:4a}). 

The evolution of the IRX-$\beta_{\rm UV}$ relation can be interpreted as a sign of change in stellar population age or the shape of attenuation curve \citep[\eg][and see references therein]{Salim_2020}, and the difference in the derived relation can obviously lead to different estimate of such changes. However, it should also be noted that the uncertainties in the measurements of both $\beta_{\rm UV}$ and IRX can be non-trivial, which makes the interpretation of the observed evolution of the IRX-$\beta_{\rm UV}$ relation challenging. Specifically, it can be seen from Table~\ref{T1} that different studies have adopted different photometry spanning over different wavelength range for estimating $\beta_{\rm UV}$ and $L_{\rm UV}$ of their sample. \citet{Alvarez_Marquez_2016}, for example, have only adopted the photometry blueward to $\lambda=2000\,\angstrom$, which is significantly shorter compared to the upper limit of the other studies. This can lead to non-negligible difference in the estimated $\beta_{\rm UV}$, because the true SED shape can deviate from a simple power-law (\ie,~$f_\lambda\propto\lambda^{\beta_{\rm UV}}$, and see the \textit{lower} panel of Fig.~\ref{fig.2} for an example of the SEDs produced by using the dust extinction curves of the WD01 model), which most studies have assumed. We will explore this issue in more details in Section~\ref{Sec:5a}. 

Apart from that, we note that different studies have adopted different \textit{Herschel}/ALMA photometry as well as different fitting techniques for extrapolating $L_{\rm IR}$, as summarised in Table~\ref{T1}. While some studies have fit different dust SED templates to the stacked fluxes at multiple IR-to-mm bands \citep[\eg,][]{Heinis_2013, Alvarez_Marquez_2016, Reddy_2018}, others also have derived $L_{\rm IR}$ by fitting single-band stacked flux densities with an assumed template SED \citep[\eg,][]{Fudamoto_2020} or {MBB} function with an assumed `dust temperature' \citep[\eg,][]{Bouwens_2016, McLure_2018}. For those depending on single-band flux densities, the derived IRX therefore strongly depends on the assumed template or `dust temperature'. We therefore point out the implicit uncertainties in the results from the different studies due to the inconsistencies in the methodology for deriving $L_{\rm IR}$ and the limited constraint on the dust SED shape at high-$z$.

One important issue about the stacked data is that they do not reflect the dispersion of individual sources, but instead, only represents the \textit{luminosity-weighted} results. In the \textit{right} panel of Fig.~\ref{fig.3}, we explicitly show the result of the individual source in the \citet{Fudamoto_2020} sample as an example. The sample of \citet{Fudamoto_2020} contains 1512 galaxies selected from the COSMOS2015 catalogue \citep{Laigle_2016} that are part of the ALMA archival band 6 and 7 observations. 172 out of 1512 galaxies ($11.4\%$) are detected {with} more than $3\sigma$ with ALMA. In the figure, the data of the detected sources and the 3$\sigma$ (\ie~$99.8\%$) upper limits of the undetected objects are marked with grey filled circles and grey downward arrows, respectively. 

The scatter among individual sources is non-trivial. The IRX of the ALMA-detected objects has a dispersion as large as $\sim$0.3 dex at given $\beta_{\rm UV}$. Since the IRX and $\beta_{\rm UV}$ of all the objects in the same sample are measured using the same methodology, the scatter present in the figure is largely intrinsic. We also emphasise that the exact location of the undetected sources on the diagram is unknown and can in principle be offset from the $\rm M99_{corr}$ relation. Given their large population, the dispersion of the complete LBG sample of \citet{Fudamoto_2020} is in fact uncertain. The stacked data, which is biased by the IR-luminous objects, appears to well agree with the canonical M99 relation (\textit{left} panel), while the individual objects {may} deviate from it. This again highlights the issue that stacked results of the high-$z$ LBG samples may not reflect the distribution of the location of the individual sources in the {IRX-$\beta_{\rm UV}$} plane.

\subsubsection{Comparing the simulation results with the observations}

We now {compare the prediction of \textsc{\small MassiveFIRE} with} the observational data. In Fig.~\ref{fig.3}, we show the IRX-$\beta_{\rm UV}$ relation of {our \textsc{\small MassiveFIRE} sample} at $z=2-6$. The coloured symbols represent the data that are averaged over 24 random viewing angles of each galaxy, and the semi-transparent coloured contours around those filled symbols indicate the $3\sigma$ (\ie,~$99.8\%$ confidence level) probability distribution of the results of the different viewing angles. The redshifts of the galaxies are indicated by the colour and shape of the symbols as labeled. We show in this figure only the results of our fiducial RT model (SB99 stellar evolution model, $\delta_{\rm dzr}=0.4$, and the {MW dust model} of WD01, see Section~\ref{Sec:2b}). 

From the figure, we can see that the simulated data exhibit fairly large scatter on the diagram. While a large fraction of the galaxies are in broad agreement with the canonical M99 relation, there are also a number of galaxies that appear to have significantly redder $\beta_{\rm UV}$ for their IRX (or significantly lower IRX at their $\beta_{\rm UV}$). To better quantify the location of galaxies on the diagram, we define the variable, $\delta\beta_{\rm UV}$, as the horizontal offset of the galaxy's UV spectral slope from the $\rm M99$\footnote{Although the original M99 relation has the problem of missing UV fluxes and colour bias in $\beta_{\rm UV}$ (see Section~\ref{Sec:3a}) and has been corrected by several subsequent works, we still use the uncorrected relation as the benchmark because it has been very widely adopted for correcting dust-obscured SFRs by many studies in the past.} relation on the diagram, \ie~

\begin{equation}
	\delta\beta_{\rm UV} \equiv \beta_{\rm UV} - \beta_{\rm M99}\,({\rm IRX}),
\label{eq.4}
\end{equation}

\noindent where $\beta_{\rm M99}({\rm IRX})$ is the inverse function of Eq.~\ref{eq.2}.

We find that 54 out of 83 galaxies ($66\%$, \ie~$1\sigma$) in our sample at $z=2-6$ lie within $\delta\beta_{\rm UV}=\pm0.2$, while 21 (8) galaxies lie redwards (bluewards) to that region. The standard deviation of $\delta\beta_{\rm UV}$ of the entire sample is 0.32. We notice a mild redshift evolution of $\delta\beta_{\rm UV}$, from a median value of $-0.08$ at $z=6$ (green diamonds) to $0.14$ at $z=2$ (red triangles). This is mainly driven by the increase of the intrinsic UV spectral slope ($\beta_{\rm UV,\,0}$) on average with decreasing redshift \citep[see discussion in Section~\ref{Sec:4d}, and also \eg][]{Reddy_2018}. Note that this evolution from $z=6$ and $z=2$ (by 0.22) does not appear to be significant compared to the dispersion of the entire sample ($1\sigma$ is 0.32) or that of any subsample of a given redshift. This is because $\delta\beta_{\rm UV}$ is strongly correlated to the very recent star formation history of galaxy on the timescale of $\rm \simless100\,Myrs$ and galaxies at a given snapshot have large variations in the star formation history in the past on this timescale {\citep[see Section~\ref{Sec:4d} for a more detailed discussion, and also][]{Faucher_Giguere_2017, Feldmann_2017, Sparre_2017, Flores_2020}}.

The objects that are in broad agreement with the canonical M99 relation in our sample are therefore {also in good agreement} with the observational data at high-$z$ in the overlapping parameter space (\ie,~$1\simless \rm IRX\simless10$ and $-2\simless\beta_{\rm UV}\simless-1$, see the \textit{left} panel of Fig.~\ref{fig.3}). On the other hand, a subset of our simulated galaxies lies well below the canonical M99 relation. They are quiescent galaxies with relatively low {recent SFR} as well as low dust optical depth (see Section~\ref{Sec:4}). {These objects are very faint at the ALMA bands and may constitute a considerable fraction of the submm undetected galaxies in the observations (see the \textit{right} panel of Fig.~\ref{fig.3}) {and may not be selected as LBGs}.}

We also note that variations due to different viewing angles do not appear to be as significant as {galaxy-to-galaxy} variations. The mean $3\sigma$ dispersion of $\beta_{\rm UV}$ {due to viewing angle variations in} the \textsc{\small MassiveFIRE} sample is only 0.10, which is much smaller than the {galaxy-to-galaxy} dispersion. The viewing direction is thus not a major source of scatter in the observed IRX-$\beta_{\rm UV}$ relation. In Section~\ref{Sec:4}, we will assess the relative contribution of the different sources of the scatter in more detail, including stellar population age, viewing direction, as well as other parameter changes of the fiducial RT model.

Finally, we note that our sample does not {completely cover} the parameter space occupied by the observational data, specifically, at high IRX (\ie,~IRX$\simgreat10$) and $\beta_{\rm UV}$ (\ie,~$\beta_{\rm UV}\simgreat-1$). This arises because observational samples contain larger number of galaxies (typically, hundreds to thousands of sources, see Table~\ref{T1}) spanning a much wider dynamic range of properties. In particular, they include galaxies with higher dust optical depth, which explains the galaxies in the upper right corner of the diagram {(see next section). It is noteworthy that, in practice, this is also the regime where IRX-$\beta_{\rm UV}$ relation becomes useful for estimating dust-obscured SFR with UV data alone, because a small error in $\beta_{\rm UV}$ does not lead to a significant difference in the derived IRX (and hence $L_{\rm IR}$) for a given IRX-$\beta_{\rm UV}$ relation.} 

\begin{figure}
 \begin{center}
 \includegraphics[width=86mm, height=75mm]{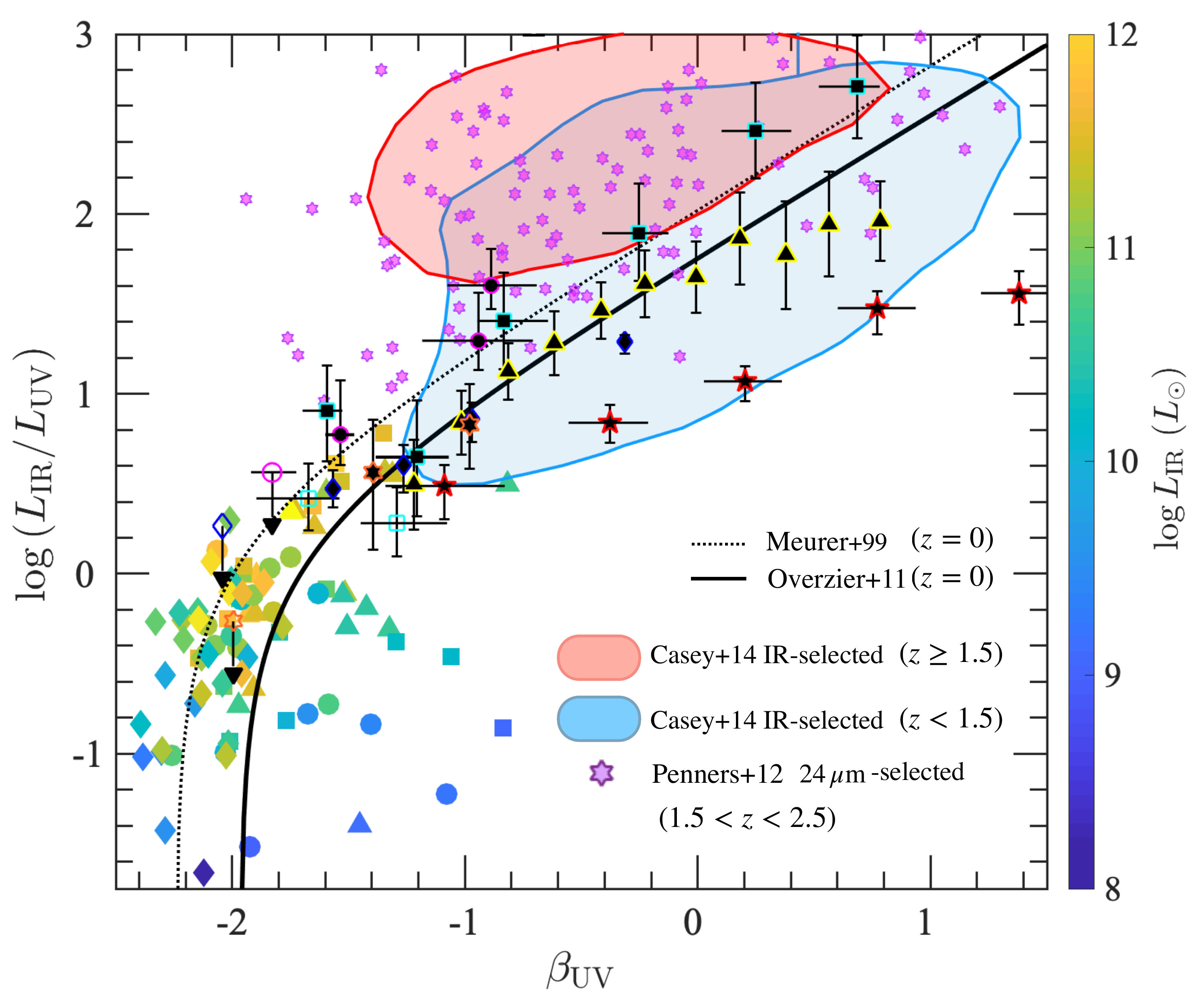}
     \vspace{-10pt}
 \caption{The `secondary dependence' of the IRX-$\beta_{\rm UV}$ relation on $L_{\rm IR}$. The simulation data are angle-averaged results, and are colour-coded by $L_{\rm IR}$ of the galaxy. The redshifts of the galaxies {(over the range of $z=2-6$)} are indicated by the shape of the symbols as in Fig.~\ref{fig.3}. The magenta hexagrams represent the data derived by \citet{Penner_2012} using a 24 $\rm \mu m$-selected galaxy sample {at $1.5<z<2.5$, which spans $10^{12}\simless \,L_{\rm IR} \,\simless 10^{13}\,L_\odot$}. The red and blue shaded areas represent the $1\sigma$ distribution of the \textit{Herschel}-selected galaxies in the \citet{Casey_2014} sample within the redshift range $1.5\le z<3.5$ {(<$L_{\rm IR}$>$=10^{12.5}\,L_\odot$)} and $0<z<1.5$ {(<$L_{\rm IR}$>$=10^{11.5}\,L_\odot$)}, respectively. The local M99 and $\rm M99_{\rm corr}$ relations, as well as the stacked data of the UV-selected samples are the same as shown in the \textit{left} panel of Fig.~\ref{fig.3}. {The IR-selected samples at high-$z$ show bluer $\beta_{\rm UV}$ at given IRX compared to the UV-selected samples.}}
 \label{fig.4}
  \end{center}
    \vspace{-15pt}
\end{figure}

\subsection{The dependence of IRX-$\beta$ relation on $L_{\rm IR}$}
\label{Sec:3c}

\begin{table*}
\caption{The selection criteria and the methods for deriving $L_{\rm IR}$, $L_{\rm UV}$ and $\beta_{\rm UV}$ used by the two studies that are referenced in Fig.~\ref{fig.4} ({see also} Section~\ref{Sec:3c}). Both studies analysed IR-selected samples. }
\begin{tabular}{ p{2.6 cm} p{5.8 cm}  p{8.2 cm} }
 \hline
\multicolumn{1}{}{} \; Paper \;\;\; & Selection criteria of the sample &  Methods for deriving $L_{\rm IR}$,  $\beta_{\rm UV}$ and $L_{\rm UV}$  \\
 \hline
    \citet{Penner_2012} & The sample contains $\sim60$ dust-obscured galaxies \citep[DOGs, ][]{Dey_2008} at $1.5<z_{\rm phot}<2.5$ that are selected from a catalogue of 24 $\rm \mu m$ sources produced for the \textit{Spitzer}/MIPS survey of the GOODS-N region \citep[M. Dickinson: PI;][]{Magnelli_2011}.  & $L_{\rm IR}$ {is} estimated by integrating the redshifted SED template from the \citet{Chary_2001} SED libraries that most closely matches the observed 100 $\rm \mu m$ flux density (measured by \textit{Herschel}/PACS) over the wavelength range $\lambda=8-1000\,\rm \mu m$. $\beta_{\rm UV}$ of the galaxies {is} derived by fitting the single power-law SED, $f_\lambda\propto \lambda^{\beta_{\rm UV}}$, to the $>3\sigma$ flux densities measured in the $B$, $V$, $R$, $I$, and $z$ bands that are extracted from Subaru images \citep{Capak_2004}. $L_{\rm UV}$ {is} computed using the best-fit SED at $1600\,\angstrom$.   \\
 \hline
    \citet{Casey_2014} &  The sample includes $>4000$ \textit{Herschel}-selected dusty star-forming galaxies (DSFGs) at $0<z_{\rm phot}<3.5$ in the COSMOS field that have $>3\sigma$ detection at two or more of the five PACS+SPIRE bands. & For each galaxy, a best-fit SED of the \citet{Casey_2012} functional form is found by fitting all the available IR-to-mm data at $\rm \lambda=8-2000\,\mu m$, and $L_{\rm IR}$ is derived from integrating the best-fit SED over the range $\lambda=8-1000\,\rm \mu m$. $\beta_{\rm UV}$ is calculated by fitting the power-law SED to the multiple photometric measurements available in the COSMOS that are in the $1230-3200\,\angstrom$ range. $L_{\rm UV}$ is calculated at $1600\,\angstrom$ as in the other works.  \\    
 \hline
 \end{tabular}
\label{T2}
\end{table*}

{Several} works in the recent years {have studied} the IRX-$\beta_{\rm UV}$ relation at high-$z$ using galaxy samples selected at IR or submm bands. In contrast to the UV/optical-selected samples, which often do {not} have reliable detections of the IR emission from most of the sources, IR/submm-selected samples have nearly complete detections at rest-UV bands so that constraints on individual sources are possible. The general finding from these studies is that IR-detected samples, on average, show bluer $\beta_{\rm UV}$ {at given IRX} than UV-selected samples. 

Specifically, we show in Fig.~\ref{fig.4} the data from two studies, \citet[][hererafter P12]{Penner_2012} and \citet[][hereafter C14]{Casey_2014}. P12 have adopted a sample of $\sim60$ galaxies at $1.5<z<2.5$ selected at the \textit{Spitzer}/MIPS band at 24 $\rm \mu m$, whereas C14 have used a much larger sample that contains $>4000$ \textit{Herschel}-selected objects that span over a larger redshift range of $0<z<3.5$ (see Table~\ref{T2} for the details). We show the data of the individual sources of P12 by magenta asterisks in the figure. For the C14 sample, we separately show the result of the galaxies within two redshift ranges, $0<z<1.5$ and $1.5\le z<3.5$. Because of the large sample size of C14, we only show the $1\sigma$ probability contours of the two redshift categories. The lower- (blue semi-transparent area) and higher-$z$ (red semi-transparent area) categories contain 3246 and 919 galaxies, respectively. 

The IR-selected samples show {a} fairly large scatter in the relation similar to the UV-selected samples (\textit{right} panel of Fig.~\ref{fig.3}). Comparing the data of the P12 sample and the high-$z$ data of C14 with the stacked results of the various UV-selected samples at similar redshifts, however, we see that the IR-selected samples have on average bluer $\beta_{\rm UV}$ at a given IRX (or high IRX at a given $\beta_{\rm UV}$). This discrepancy {follows from} the `secondary dependence'  of the IRX-$\beta_{\rm UV}$ relation on $L_{\rm IR}$ --- the horizontal deviation of a galaxy's location from the canonical relation anti-correlates to $L_{\rm IR}$ \citep{Casey_2014} --- combined with the higher IR luminosities in the IR-selected sample. {This `secondary dependence' is reproduced by our simulations, as is shown in Fig.~\ref{fig.4}.}


The trend that the high-$z$ population of C14 (red semi-transparent area) appears to be systematically above the low-$z$ population (blue semi-transparent area) is now easy to understand. Galaxies selected at higher redshifts {are biased to higher $L_{\rm IR}$} \citep{Casey_2014P}, and hence they appear to be offset from the low-$z$ counterparts in the upper left direction.

Finally, we note that our simulations do not reproduce some of the objects with extremely high IRX in the observational samples (P12 and C14), as can be seen from Fig.~\ref{fig.4}. Despite the uncertainties in the measurements of $\beta_{\rm UV}$ and $L_{\rm IR}$, which we will discuss in Section~\ref{Sec:5} in detail, a straightforward interpretation is that our simulations do not produce {as IR-luminous systems as} the ones included in the observational samples. The galaxies in the P12 and C14 samples with $\rm IRX\simgreat2$ have on average  $L_{\rm IR}\approx10^{13}\,L_\odot$, which exceeds the most luminous object in our sample by an order of magnitude. It should thus be emphasised that the apparent IRX-$\beta_{\rm UV}$ relation depends on the selection of the galaxy population. 

\vspace{-15pt}
\section{Dissecting the IRX-$\beta$ relation}
\label{Sec:4}

In the last section, we have shown that both the simulation and observational data exhibit non-trivial scatter in the IRX-$\beta_{\rm UV}$ relation of high-$z$ galaxies. Therefore, in this section, we focus on examining the physical origins of the intrinsic scatter in this relation. 

We start {by} analysing {a simple} dust slab model in Section~\ref{Sec:4a}, which provides useful insights into the nature of the IRX-$\beta_{\rm UV}$ relation. In Section~\ref{Sec:4b}, we reveal the role of {the} dust optical depth in driving this relation and show {the} tight correlation between dust optical depth and IRX of galaxies. In Section~\ref{Sec:4c}, we investigate the physical mechanisms that determine both properties. In Section~\ref{Sec:4d}, we analyse the evolutionary trajectories of galaxies in the IRX-$\beta_{\rm UV}$ plane. Finally, we discuss uncertainties in the stellar population and dust models and estimate their impact on the resulting IRX-$\beta_{\rm UV}$ relation in Section~\ref{Sec:4e}. 

\subsection{Physical insights from the dust slab model}
\label{Sec:4a}

\begin{figure}
 \begin{center}
 \includegraphics[width=70mm]{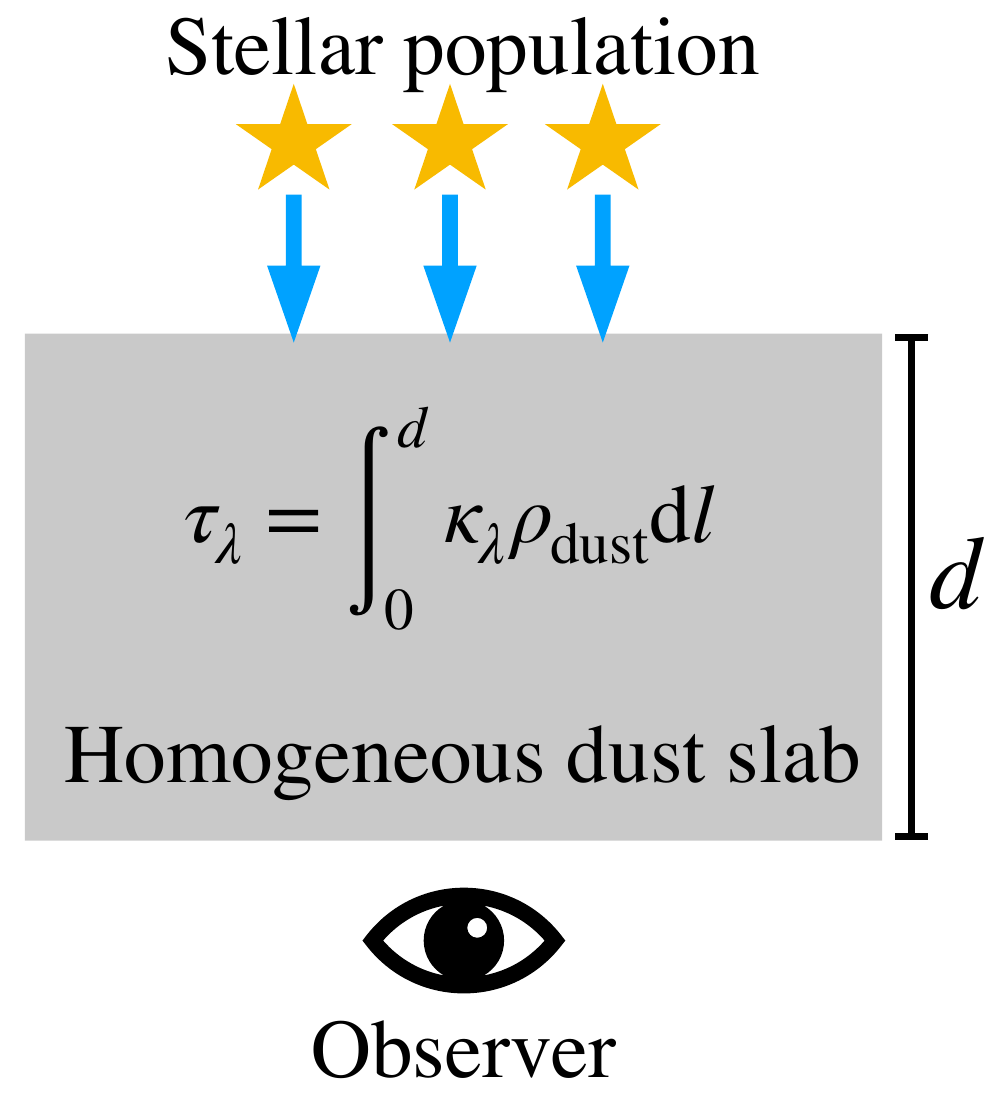}
 \caption{Schematic plot of the toy model of a homogeneous dust slab intervening between a fixed stellar population and the observer. For this toy model, the attenuation curve, $\tau_\lambda$, is simply a function of dust mass column density along the sightline.}
    \label{fig.5}
  \end{center}
    \vspace{-10pt}
\end{figure}

A simple toy model can often offer important physical insights into more complex physics problems. To understand the nature of the IRX-$\beta_{\rm UV}$ relation of high-$z$ galaxies, we at first derive this relation for a simplified model of a homogeneous dust slab intervening between a fixed stellar population and the observer (see Fig.~\ref{fig.5} for a schematic plot of the model). The toy model does not capture two major complexities in real galaxies --- 1) stellar population and thus the intrinsic stellar SEDs evolve with time, and 2) the dust column density for each star particle in a galaxy is not a constant. This toy model, however, is sufficient to show the role that the optical depth plays in shaping the IRX-$\beta_{\rm UV}$ relation \citep[\eg][]{Calzetti_2000, Popping_2017, Imara_2018}. 

Energy balance implies that the $L_{\rm IR}$ emitted by dust equals to the amount of energy of the stellar light that is absorbed by dust per unit time. Therefore, IRX($\equiv L_{\rm IR}/L_{\rm UV}$) is equivalent to the ratio of the absorbed to the unabsorbed stellar radiation. For the dust slab model, $L_{\rm IR}$ can be expressed as

\begin{equation}
L_{\rm IR}=L_{\rm abs} = \int^{\lambda_{\rm max}}_0 (1-{\rm e}^{-\tau_\lambda})\;L^*_{\lambda} \;{\rm d}\lambda,
 \label{eq.5}
\end{equation}

\noindent where $L_{\rm abs}$ is the stellar luminosity absorbed by the dust slab (in units of $\rm erg\,s^{-1}$), $\tau_\lambda$ is the dust optical depth at $\lambda$, $L^*_{\lambda}$ is the specific intrinsic luminosity of the fixed stellar population in the background (in units of $\rm erg\,s^{-1}\,m^{-1}$), and $\lambda_{\rm max}$ is the wavelength beyond which direct light from stars constitutes to the total emission at a negligible level ($\approx1\mu m$). IRX can subsequently be expressed as 

\begin{equation}
    {\rm IRX}=\frac{L_{\rm IR}}{L_{\rm UV}}=\frac{L_{\rm abs}}{\lambda_{0.16}\,L_{\lambda,\,0.16}}=\frac{\int^{\lambda_{\rm max}}_{0}(1-{\rm e}^{-\tau_\lambda})\,L^*_{\lambda}\,{\rm d} \lambda}{\lambda_{0.16}\,({\rm e}^{-\tau_{0.16}}  L^*_{\lambda,\,0.16})},
 \label{eq.6}
\end{equation}

\noindent where $\tau_{0.16}$ represents the dust optical depth at $\rm \lambda_{0.16}=1600\,\angstrom$, and $L_{\lambda,\,0.16}$ ($={\rm e}^{-\tau_{0.16}}L^*_{\lambda,\,0.16})$ and $L^*_{\lambda,\,0.16}$ correspond to the attenuated and unattenuated (intrinsic) specific luminosity of the stellar population at $\rm \lambda_{0.16}=1600\,\angstrom$, respectively. 

By re-arrangement, Eq.~\ref{eq.6} can be written as

\begin{align}
    {\rm IRX}&=\left(\frac{1-{\rm e}^{-\tau_{0.16}}}{ {\rm e}^{-\tau_{0.16} }} \right) \frac{\int^{\lambda_{\rm max}}_{0}(1-{\rm e}^{-\tau_\lambda})/(1-{\rm e}^{-\tau_{0.16}})\,L^*_{\lambda}\,{\rm d} \lambda}{L^*_{\lambda,\,0.16}\,\lambda_{0.16}} \nonumber \\
    &=({\rm e}^{\tau_{0.16}}-1)\;\mathcal{Y},
     \label{eq.7}
\end{align}

\noindent where we define a new dimensionless parameter, \ie~

\begin{equation}
    \mathcal{Y} \equiv \frac{\int^{\lambda_{\rm max}}_{0}(1-{\rm e}^{-\tau_\lambda})/(1-{\rm e}^{-\tau_{0.16}})\,L^*_{\lambda}\,{\rm d} \lambda}{L^*_{\lambda,\,0.16}\,\lambda_{0.16}}. 
  \label{eq.8}
\end{equation}

\noindent It can immediately be seen from the above equation that $\mathcal{Y}$ is dependent on both intrinsic stellar SED ($L^*_{\lambda}$) as well as dust attenuation law ($\tau_\lambda$). For a homogeneous dust slab, $\tau_\lambda$ can be expressed as\footnote{{We neglect the light scattered back into the sightline from dust.}} 

\begin{equation}
	\tau_\lambda =\int^{d}_0 \kappa_\lambda \rho_{\rm dust}  \;{\rm d}l = \kappa_\lambda \rho_{\rm dust} d = \kappa_\lambda N_{\rm dust}.
\label{eq.9}
\end{equation}

\noindent $\tau_\lambda$ is simply a function of dust column mass density (\ie,~$N_{\rm dust}=\rho_{\rm dust}d$), assuming the dust extinction curve\footnote{In this paper, we use the term  `dust extinction law (curve)' to refer to the dust opacity function, $\kappa_\lambda$, which is in unit of $\rm m^2\,kg^{-1}$. We also frequently use the term  `dust attenuation law (curve)' to refer to the optical depth function, $\tau_\lambda$, which is dimensionless. {For the dust slab model}, $\tau_\lambda$ is proportional to $\kappa_\lambda$ and thus they have the same functional shape {(Eq.~\ref{eq.9})}. $\tau_\lambda$ of a galaxy, however, does not necessarily have the same functional shape as $\kappa_\lambda$, depending on the spatial configuration of dust and star distribution \citep[see \eg,][and references therein]{Salim_2020}. The terminology `law' and  `curve' are interchangeably utilised in this paper.}, $\kappa_\lambda$, is consistent everywhere in the slab.

\begin{figure}
 \begin{center}
 \includegraphics[width=80mm]{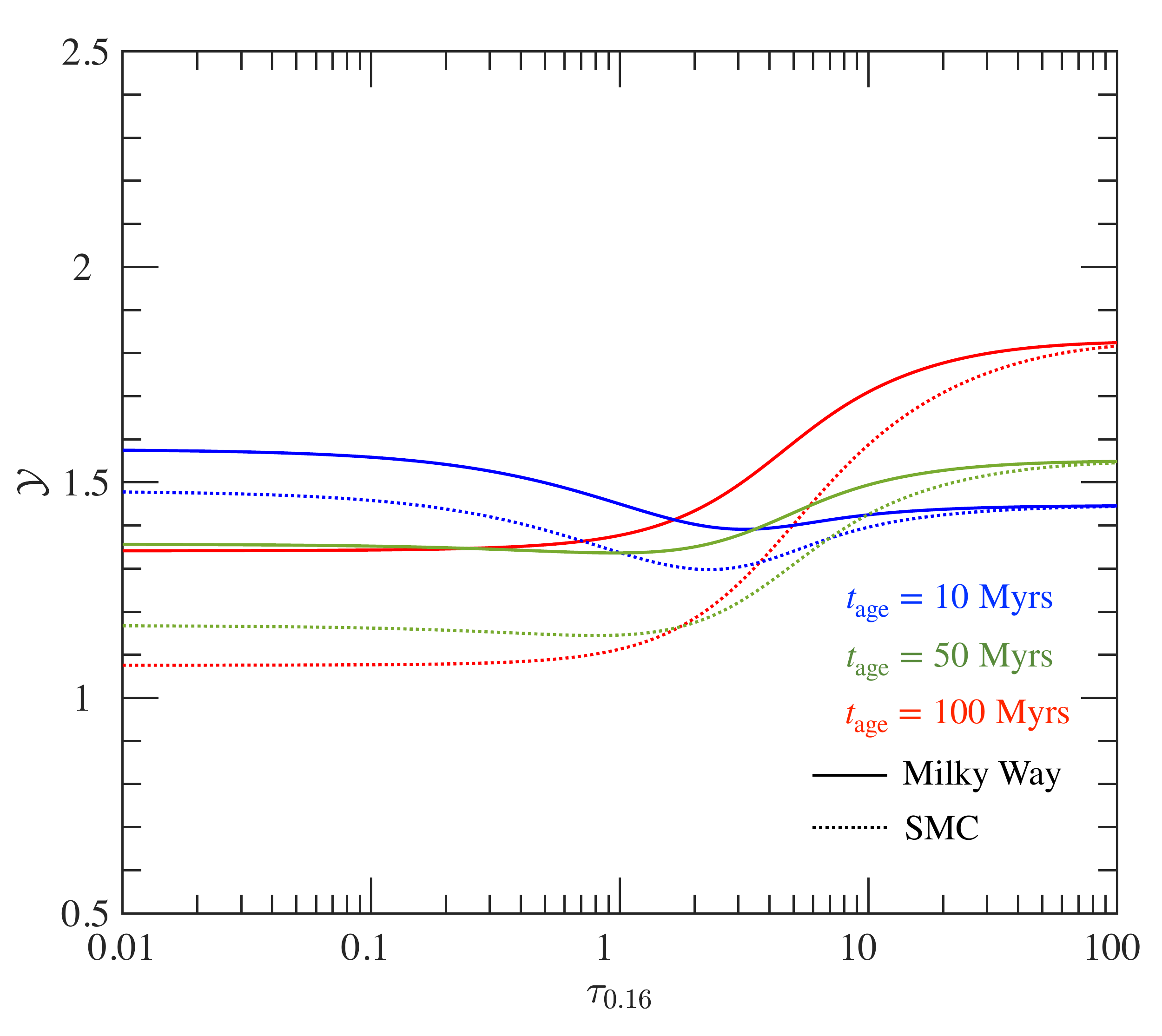}
 \caption{ $\mathcal{Y}$ parameter as a function of $\tau_{0.16}$ (see Eq.~\ref{eq.7} for the definition of $\mathcal{Y}$) for various stellar population ages and dust extinction curves. The blue, green and red lines correspond to the cases of $t_{\rm age}=10$, 50, and 100 Myrs, respectively, with stellar metallicities $Z_*=0.1Z_\odot$. The stellar SEDs are extracted from the \textsc{\small starburst99} template libraries based on $t_{\rm age}$ and $Z_*$. The results of the MW and SMC {dust models of WD01} are shown by solid and dotted lines, respectively. $\mathcal{Y}$ is a weak function of $\tau_{0.16}$. } 
    \label{fig.6}
  \end{center}
  \vspace{-15pt}
\end{figure}

\begin{figure}
 \begin{center}
 \includegraphics[width=80mm]{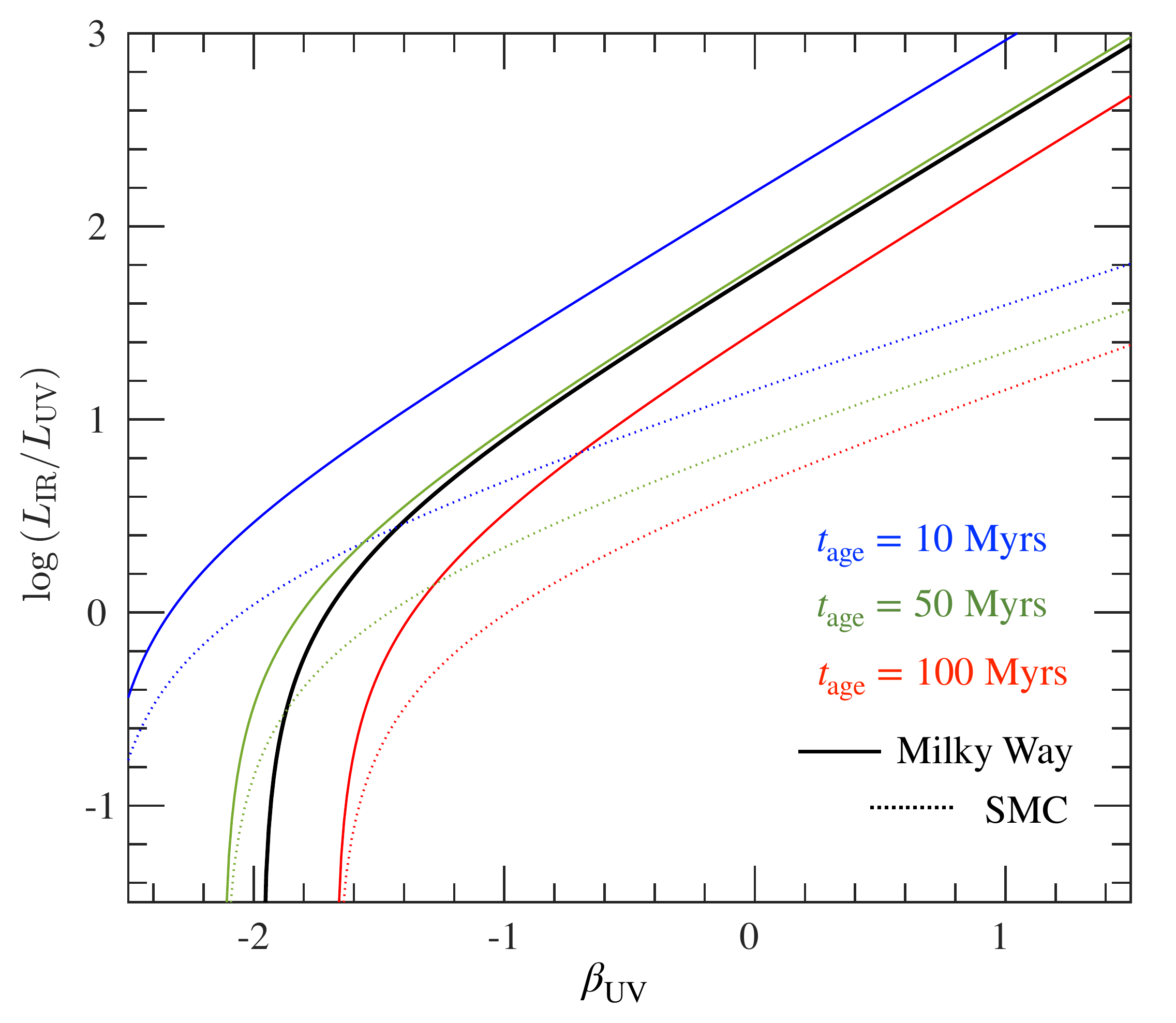}
 \caption{The analytic solution of IRX-$\beta_{\rm UV}$ relation for the dust slab model (by using Eq.~\ref{eq.17}). The coloured lines correspond to the same models as in Fig.~\ref{fig.6}.  The solid black line indicates the local $\rm M99_{corr}$ relation. }
    \label{fig.7}
  \end{center}
  \vspace{-15pt}
\end{figure}

Eq.~\ref{eq.7} implies that if $\mathcal{Y}$ shows a relatively small variation with different dust and stellar properties (\eg~$\tau_\lambda$ and $L_{\rm *,\,\lambda}$), then IRX and $\tau_{0.16}$ simply follow an exponential {relation}. Let us now consider two extreme conditions. First of all, when the dust optical depth is sufficiently small ($\ie~\tau_\lambda\ll1$), Eq.~\ref{eq.8} can simplify to

\begin{equation}
	\mathcal{Y}\approx \frac{\int^{\lambda_{\rm max}}_0 (\tau_\lambda/\tau_{0.16})L^*_{\lambda}\;{\rm d}\lambda}{L^*_{\lambda,\,0.16}\,\lambda_{0.16}} = \frac{\int^{\lambda_{\rm max}}_0 \hat{\tau}_\lambda L^*_{\lambda}\;{\rm d}\lambda}{L^*_{\lambda,\,0.16}\,\lambda_{0.16}}.
 \label{eq.10}
\end{equation}

\noindent It can be seen that $\mathcal{Y}$ depends on both $L_{\rm *,\,\lambda}$ and the \textit{normalised} attenuation curve, \ie~$\hat{\tau}_\lambda=\tau_\lambda/\tau_{0.16}$, but is independent of $\tau_{0.16}$. On the other hand, when the dust slab is very optically thick ($\ie~\tau_\lambda\gg1$), Eq.~\ref{eq.8} approximates {to}

\begin{equation}
\mathcal{Y}\approx \frac{\int^{\lambda_{\rm max}}_0 L^*_{\lambda}\;{\rm d}\lambda}{ L^*_{\lambda,\,0.16}\,\lambda_{0.16}}.
    \label{eq.11}
\end{equation}

\noindent In this case, $\mathcal{Y}$ becomes the ratio of the total intrinsic stellar luminosity to $L^*_{\lambda,\,0.16}\,\lambda_{0.16}$, the proxy for stellar UV luminosity. It is interesting to note that $\mathcal{Y}$ is independent of the dust-related properties (neither $\hat{\tau}_\lambda$ or $\tau_{0.16}$ {enters the equation}). 

Overall, $\mathcal{Y}$ is a weak function of $\tau_{0.16}$. To illustrate this, we show the $\mathcal{Y}$ vs. $\tau_{0.16}$ relation for a number of different stellar SEDs ($t_{\rm age}=10$, 50 and 100 Myrs with $Z_*=0.1Z_\odot$, extracted from the SB99 libraries) and dust {attenuation} curves {(yielded from the MW and SMC dust models of WD01)} in Fig.~\ref{fig.6}. It is clear that for all the cases, $\mathcal{Y}$ does not vary by more than a factor of 2 over four orders of magnitude in $\tau_{0.16}$. Therefore, for a dust slab model, IRX and $\tau_{0.16}$ roughly follow the relation \citep[\cf][]{Meurer_1999, Calzetti_2000, Safarzadeh_2017, Narayanan_2018}

\begin{equation}
	\rm IRX \propto {\rm e}^{\tau_{0.16}}-1.
\label{eq.12}
\end{equation}

\noindent The relation ensures that when $\tau_{0.16}\rightarrow0$, $\rm IRX\rightarrow0$, meaning that no light is reemitted by dust when {the} optical depth {approaches} zero.

Now we examine how $\beta_{\rm UV}$ relates to $\tau_{0.16}$. By definition, $\beta_{\rm UV}$ is set by the slope between FUV and NUV luminosity, both of which scale exponentially with optical depth as 

\begin{equation}
    L_{\rm \lambda,\,FUV | NUV} = {\rm e}^{-\tau_{\rm FUV | NUV }} L^*_{\rm \lambda,\,FUV | NUV},
    \label{eq.13}
\end{equation}

\noindent where $L_{\rm \lambda,\,FUV}$ ($L_{\rm \lambda,\,NUV}$) and $L^*_{\rm \lambda,\,FUV}$ ($L^*_{\rm \lambda,\, NUV}$) represent the attenuated and intrinsic specific stellar luminosity at a given FUV (NUV) band, respectively. $\beta_{\rm UV}$ can therefore be expressed as 

\begin{align}
    \beta_{\rm UV} &= \frac{{\rm log}\,( {\rm e}^{-\tau_{\rm FUV}} L^*_{\rm \lambda,\,FUV}) - {\rm log}\,({\rm e}^{-\tau_{\rm NUV}} L^*_{\rm \lambda,\,NUV} )}{{\rm log}\,\lambda_{\rm FUV}-{\rm log}\,\lambda_{\rm NUV}} \nonumber \\
    &= \frac{{\rm log}\,[ {\rm e}^{-\tau_{0.16}\hat{\tau}_{\rm FUV}} L^*_{\rm \lambda,\,FUV}] - {\rm log}\,[{\rm e}^{-\tau_{0.16}\hat{\tau}_{\rm NUV}}L^*_{\rm \lambda,\,NUV} ] }{{\rm log}\,\lambda_{\rm FUV}-{\rm log}\,\lambda_{\rm NUV}} 
    \label{eq.14}
\end{align}

\noindent where we have expressed $\tau_\lambda$ in terms of $ \tau_{\rm 0.16}$, \ie~$\tau_\lambda= \hat{\tau}_\lambda \times \tau_{\rm 0.16}$, for both FUV and NUV bands. By re-arrangement of this equation, we get

\begin{align}
    \beta_{\rm UV} &= \frac{{\rm log}\,(L^*_{\rm \lambda,\,FUV}/L^*_{\rm \lambda,\,NUV})+[\hat{\tau}_{\rm NUV}-\hat{\tau}_{\rm FUV}] \tau_{0.16}}{{\rm log}\,(\lambda_{\rm FUV}/\lambda_{\rm NUV})}\nonumber \\
    &= \beta_{\rm UV,\,0} + \mathcal{Z} \, \tau_{0.16},
    \label{eq.15}
\end{align}

\noindent where 

\begin{equation}
    \beta_{\rm UV,\,0} =  \frac{{\rm log}\,(L^*_{\rm \lambda,\,FUV}/L^*_{\rm \lambda,\,NUV})}{{\rm log}\,(\lambda_{\rm FUV}/\lambda_{\rm NUV})} \;\;\;{\rm and}\;\;\;  \mathcal{Z} =  \frac{\hat{\tau}_{\rm NUV}-\hat{\tau}_{\rm FUV}}{{\rm log}\,(\lambda_{\rm FUV}/\lambda_{\rm NUV})}
    \label{eq.16}
\end{equation}

\noindent are pure functions of stellar SED and dust attenuation law, respectively. $\beta_{\rm UV,\,0}$ is in fact the UV spectra slope of the unattenuated stellar SED. 

Eq.~\ref{eq.15} and Eq.~\ref{eq.16} imply that for the dust slab model, the increment (reddening) of the UV spectral slope due to the dust extinction scales \textit{linearly} {with} $\tau_{0.16}$ {\citep[see \eg][for the empirical relation derived using the local starburst sample from \citealt{Kinney_1993}]{Calzetti_2000}}. The slope of the inclination, $\mathcal{Z}$, depends on the steepness of the dust attenuation curve in the FUV-to-NUV wavelength range. 

By combining Eq.~\ref{eq.7} and~\ref{eq.15}, we can relate IRX and $\beta_{\rm UV}$ by a simple formula {\citep[\cf][]{Meurer_1999, Hao_2011}},  

\begin{equation}
    {\rm IRX} = [{\rm e}^{(\beta_{\rm UV}-\beta_{\rm UV,\,0})/\mathcal{Z}}-1]\mathcal{Y},
    \label{eq.17}
\end{equation}

\noindent where the three parameters, $\beta_{\rm UV,\,0}$, $\mathcal{Z}$, and $\mathcal{Y}$, are functions of $L^*_{\lambda}$ and $\tau_\lambda$. For the dust slab model, the IRX-$\beta_{\rm UV}$ relation therefore has this well-defined analytic solution. 

We show in Fig.~\ref{fig.7} the analytic solution given by Eq.~\ref{eq.17} for several different stellar SEDs and {dust models} as in Fig.~\ref{fig.6}. We can see that by increasing $t_{\rm age}$, the predicted IRX-$\beta_{\rm UV}$ relation shifts horizontally to higher $\beta_{\rm UV}$ without having its shape much affected. In this case, $\beta_{\rm UV, 0}$ in Eq.~\ref{eq.17} noticeably increases with $t_{\rm age}$. $\mathcal{Z}$ is unaffected since it is independent of stellar properties (Eq.~\ref{eq.16}).  In contrast, when the MW {dust} is replaced with the SMC {dust}, the expected IRX-$\beta_{\rm UV}$ relation becomes much shallower in addition to having a horizontal shift to higher $\beta_{\rm UV}$. In this case, $\mathcal{Z}$ significantly increases (from 0.57 of the MW {dust model} to 1.01 of the {SMC model}, {calculated} using $\lambda_{\rm FUV}=1230\,\angstrom$ and $\lambda_{\rm NUV}=3200\,\angstrom$ in Eq.~\ref{eq.16}) whereas $\beta_{\rm UV,\,0}$ is unchanged. For both cases, $\mathcal{Y}$ only mildly changes {(with either $t_{\rm age}$ or dust model) and thus has} only a minor effect on the location and shape of the IRX-$\beta_{\rm UV}$ relation.

\subsection{UV Optical depth as driver of the IRX-$\beta$ relation}
\label{Sec:4b}

\begin{figure}
 \begin{center}
 \includegraphics[width=83mm]{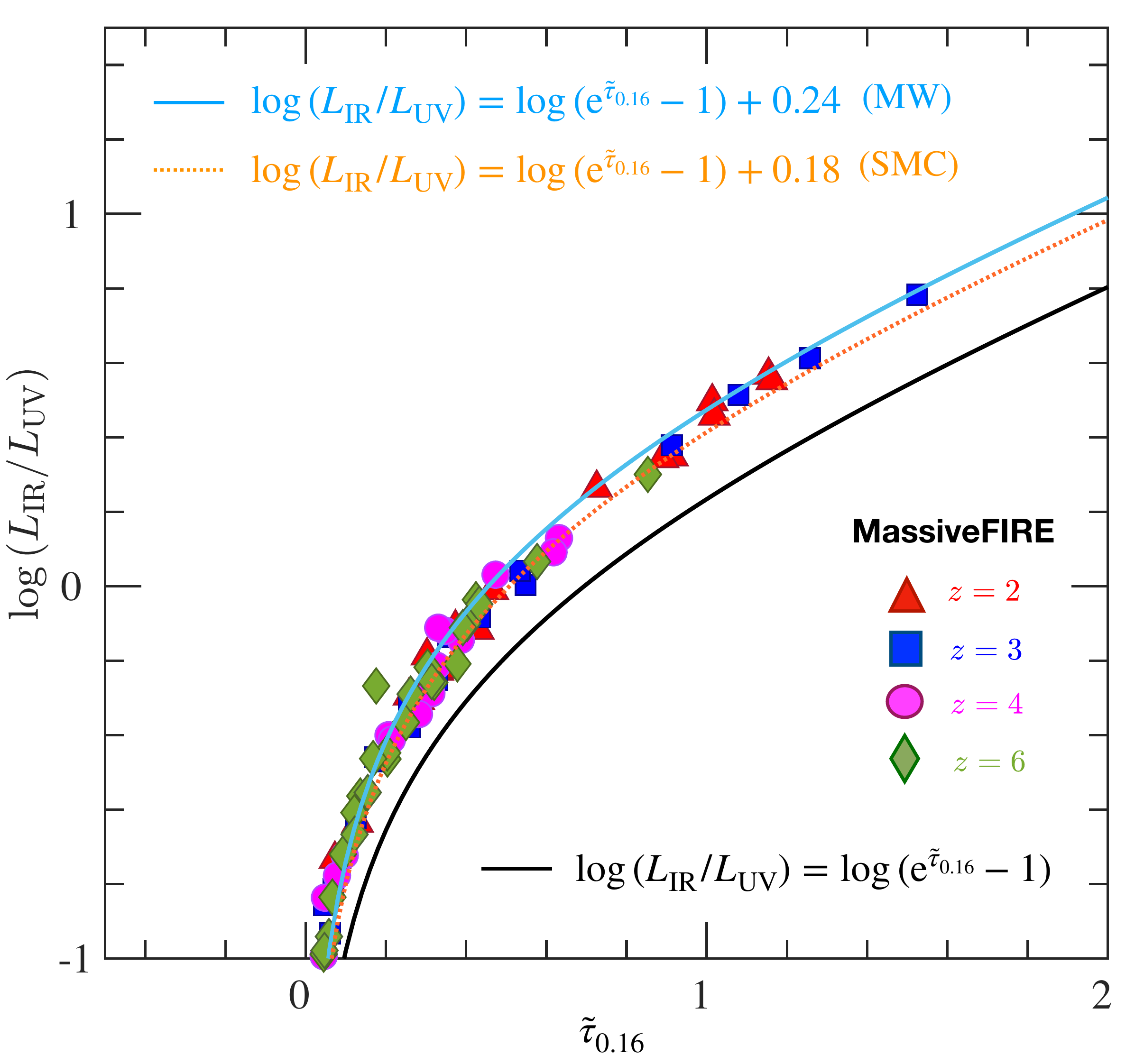}
 \caption{Relation between \textit{angle-averaged} $\tilde{\tau}_{0.16}$ and IRX of the \textsc{\small MassiveFIRE} galaxies. The redshifts of the galaxies are marked with the shape of the symbols as labelled. The data shown are obtained from the fiducial RT model with MW dust. The solid cyan line shows the best-fit exponential-law relation to the data of the fiducial model, while the dotted orange line represents the best-fit relation to the data yielded by using an alternative model where {SMC dust is implemented}. The solid black line represents the benchmark relation of Eq.~\ref{eq.7} with $\mathcal{Y}=1$. The angle-averaged IRX and $\tau_{0.16}$ of the \textsc{\small MassiveFIRE} galaxies can be well fit by an exponential {curve} (\ie~Eq.~\ref{eq.7}), {with $\mathcal{Y}=1.73$ (1.51) for the MW (SMC) dust model. The discrepancy between the best-fit relations (blue and orange lines) and the benchmark relation (black line) is mainly due to the bolometric correction of the UV-to-NIR stellar light relative to the UV emission at 1600 $\angstrom$ (see Eq.~\ref{eq.11}, \cf~\citealt{Meurer_1999, Calzetti_2000, Hao_2011}).}}
    \label{fig.8}
  \end{center}
  \vspace{-15pt}
\end{figure}

We have shown that for the dust slab model, the IRX-$\beta_{\rm UV}$ relation has the simple analytic form of Eq.~\ref{eq.17}. {For a given dust extinction law}, $\tau_{0.16}$ can be viewed as the underlying driver of this relation \citep[see also][]{Popping_2017, Narayanan_2018}. While IRX and $\tau_{0.16}$ follow an exponential-law relation (Eq.~\ref{eq.7}), $\beta_{\rm UV}$ and $\tau_{0.16}$ are simply linearly correlated to each other (Eq.~\ref{eq.15}). Both IRX and $\beta_{\rm UV}$ increase with $\tau_{0.16}$ monotonically. 

It is {not obvious that these results also apply in real galaxies}. As we noted at the beginning of Section~\ref{Sec:4a}, one key aspect of galaxy that is not captured by the dust slab model is that the dust column density differs for different star-forming regions of a galaxy. Therefore, $\tau_\lambda$ should more generally represent the \textit{effective} amount of light lost in aggregate for a number of sightlines between each star and the observer, {with a compensation of light scattered back into the sightlines}\footnote{ {We find that on average $\sim7\%$ of $L_{\rm UV}$ of the \textsc{\small MassiveFIRE} galaxies is from the light scattered into the camera from dust from our RT calculations.}}, which can be defined as

\begin{equation}
    \tilde{\tau}_\lambda \equiv -\ln \left(\frac{L_{\rm \lambda}}{L^*_{\lambda}}\right) = -\ln \left(\frac{\Sigma^{i=N_{\rm star}}_{i=0}L_{\rm i,\,\lambda}{\rm e}^{-\tau_{\rm i,\,\lambda}}+L_{\rm scattered,\,\lambda}}{\Sigma^{i=N_{\rm star}}_{i=0} L^*_{\rm i,\,\lambda}}\right),
     \label{eq.18}
\end{equation}

\noindent where {$L_\lambda$} and $L^*_{\lambda}$ correspond to the attenuated and intrinsic specific luminosity of the galaxy at $\lambda$. {$L^*_{\rm i,\,\lambda}$} and $\tau_{\rm i,\,\lambda}$ in the above equation represent the intrinsic specific luminosity and the optical depth of each individual star or star-forming region. $\tau_{\rm i,\,\lambda}$ is simply the product of $\kappa_\lambda$ and the dust column mass density along the sightline, which has the same form as the dust slab model (Eq.~\ref{eq.9}). {Finally, $L_{\rm scattered, \lambda}$ represents the specific luminosity of light scattered back into the sightlines from dust. Hereafter we use the notation `$\tilde{\tau}_\lambda$' to refer to the \textit{effective} dust optical depth of galaxy, and to distinguish it from the definition using Eq.~\ref{eq.9}. }

We now examine whether $\tilde{\tau}_{0.16}$, using the more generalised definition (Eq.~\ref{eq.18}), follows the same relation with IRX and $\beta_{\rm UV}$ of galaxy as those expected from the simple dust slab model (\ie~Eq.~\ref{eq.7} and Eq.~\ref{eq.15}).

In Fig.~\ref{eq.8}, we at first show the relation between $\tilde{\tau}_{0.16}$ and IRX of the \textsc{\small MassiveFIRE} galaxies at different redshifts ($z=2-6$). The coloured symbols represent the result of our fiducial RT model, where we adopt {MW dust} and average over 24 random viewing angles. It is clear from the figure that the angle-averaged IRX and $\tilde{\tau}_{0.16}$ are well correlated, and their relation can be well fit by an exponential {curve in the form of} Eq.~\ref{eq.7}, with $\mathcal{Y}$ being a free parameter. Using least-$\chi^2$ method, we obtain the best-fit relation

\begin{equation}
    {\rm log}\, {\rm IRX} = {\rm log} ({\rm e}^{\tilde{\tau}_{0.16}}-1)+(0.24\pm0.04).\;\;\;(\rm MW)
     \label{eq.19}
\end{equation}

\noindent We show this relation with solid cyan line in Fig.~\ref{fig.8}. The relation indicates $\mathcal{Y}=1.73\pm0.15$.  

The $\tilde{\tau}_{0.16}$ vs. IRX relation depends {(weakly)} on the dust attenuation law, as indicated by Eq.~\ref{eq.8} and Eq.~\ref{eq.10}. In Fig.~\ref{fig.8}, we also show the best-fit exponential-law curve (dotted orange line) to the data yielded by an alternative RT model where we adopt SMC dust instead of MW dust as adopted in our fiducial RT model. We find that the best-fit relation of SMC dust is 0.06 dex below that of MW dust, indicating that a slightly higher $\tilde{\tau}_{0.16}$ is needed for the same IRX with SMC dust. This is consistent with what is indicated by the analytic solutions for the dust slab model shown in Fig.~\ref{fig.6}. SMC dust always yields a lower $\mathcal{Y}$ value than MW dust for the different stellar SEDs. We also note that for clarity of presentation, we do not explicitly show the individual data of SMC dust in Fig.~\ref{fig.8}, for they overlap much with the data of MW dust.

\begin{figure}
 \begin{center}
 \includegraphics[width=78mm]{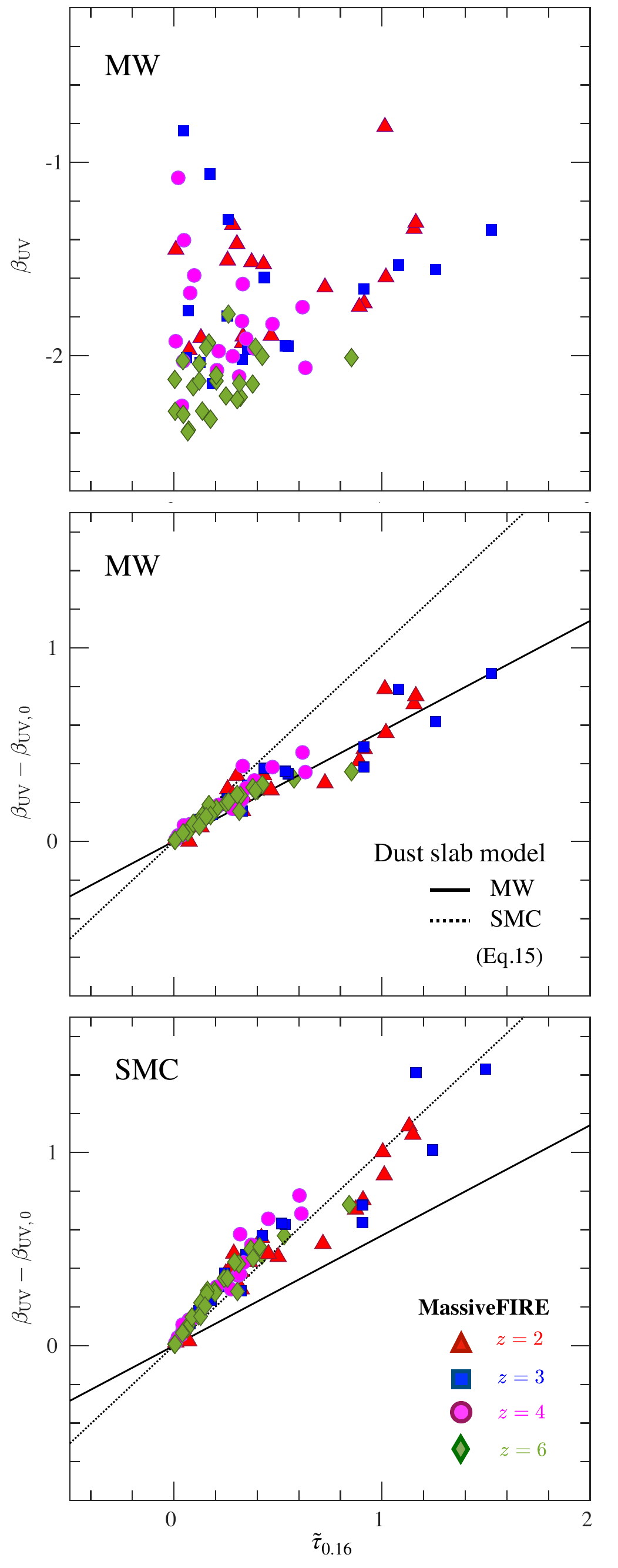}
 \caption{The $\tilde{\tau}_{0.16}$ vs. $\beta_{\rm UV}$ relation (\textit{top} panel) and the $\tilde{\tau}_{0.16}$ vs. $\beta_{\rm UV}-\beta_{\rm UV,\,0}$ relation (\textit{middle} and \textit{bottom} panels) of the \textsc{\small MassiveFIRE} galaxies at different redshifts. The redshifts of the galaxies are indicated by the shape of the symbols as labelled. The \textit{top} panel shows the result for MW dust. The \textit{middle} and \textit{bottom} panels show the results for MW and SMC dust {models}, respectively. In the \textit{middle} and \textit{bottom} panels, the solid and {dotted} lines indicate the analytic solution (Eq.~\ref{eq.15}) of the dust slab model for MW and SMC dust, respectively. {The $\tilde{\tau}_{0.16}$ vs. $\beta_{\rm UV}-\beta_{\rm UV,\,0}$ relations of the \textsc{\small MassiveFIRE} galaxies are in broad agreement with the analytic solutions of the dust slab model, whereas the dispersion of the data in these relations indicate the variation in the shape of the attenuation curve due to the complex dust-to-star geometry of the galaxies.}}
     \label{fig.9}
  \end{center}
  \vspace{-10pt}
\end{figure}

\begin{figure}
 \begin{center}
 \includegraphics[width=83mm]{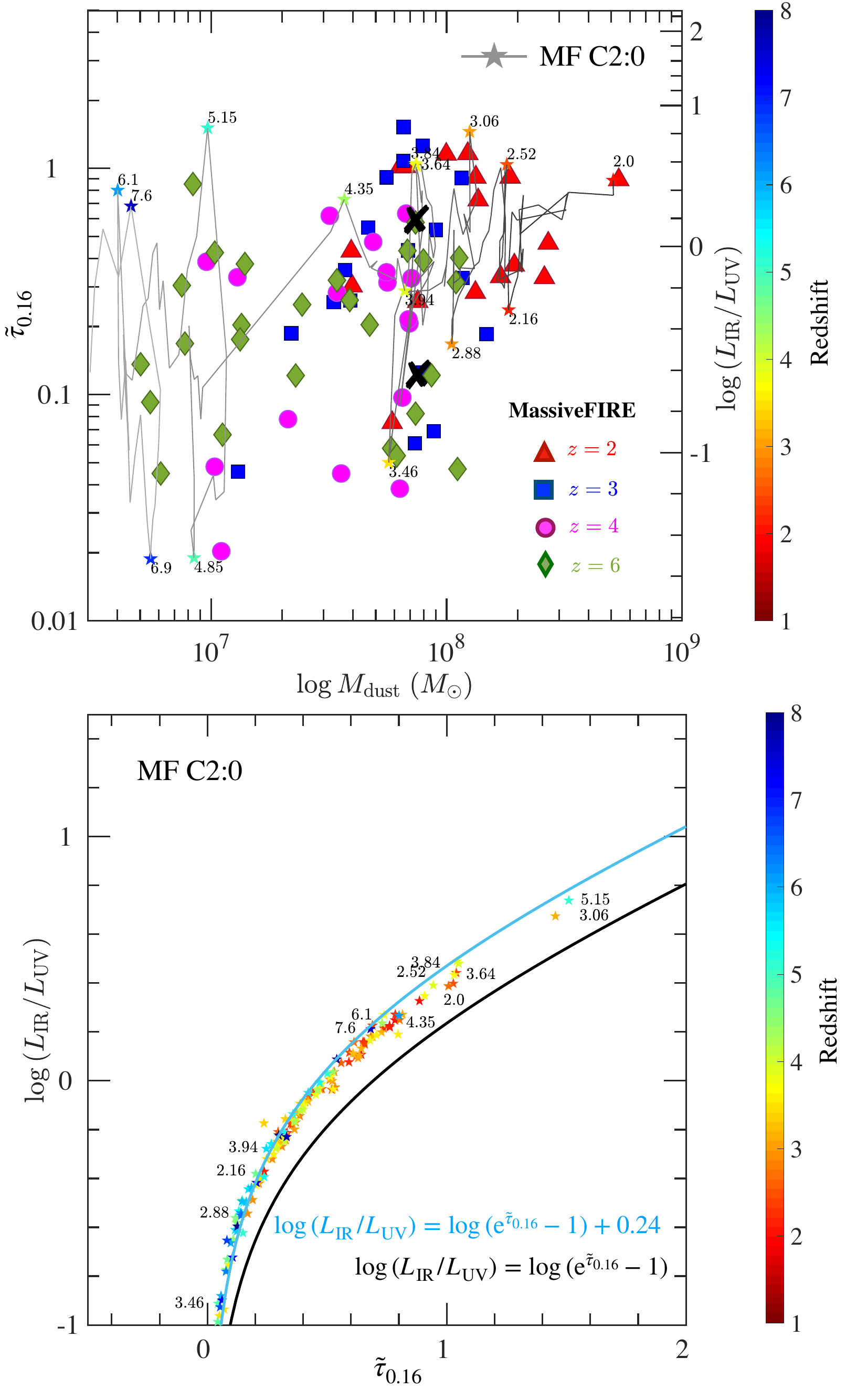}
 \caption{\textit{Upper} panel: Relation between $M_{\rm dust}$ and angle-averaged $\tilde{\tau}_{0.16}$ (IRX) of the \textsc{\small MassiveFIRE} sample at different redshifts. The redshifts of the galaxies are indicated by the shape of the symbols as labelled. The dark grey line presents the evolutionary trajectory of a selected individual \textsc{\small MassiveFIRE} galaxy (galaxy ID: MF C2:0) on the diagram. A number of instantaneous peaks and troughs in $\tilde{\tau}_{0.16}$ of this galaxy are explicitly marked by asterisks colour-coded by their corresponding redshifts. The two black crosses mark the location of a $z=6$ (galaxy ID: MF D3:0) and a $z=3$ galaxies (galaxy ID: MF B3:0). The 2D maps of the various properties of the two galaxies are shown in Fig.~\ref{fig.12}. The y-axis on the right shows the IRX, which maps to $\tilde{\tau}_{0.16}$ by Eq.~\ref{eq.19}. {$\tilde{\tau}_{0.16}$, which is well correlated with IRX, is not a good proxy of $M_{\rm dust}$ of galaxy.} \textit{Lower} panel: The angle-averaged $\tilde{\tau}_{0.16}$ vs. IRX relation of the selected galaxy, MF C2:0, at different snapshots. Each data point represents a simulation snapshot and is colour-coded by the corresponding redshift. The cyan and black lines are identical as the ones in Fig.~\ref{fig.8}. }
    \label{fig.10}
  \end{center}
  \vspace{-15pt}
\end{figure}

Let us now examine the $\tilde{\tau}_{0.16}$ vs. $\beta_{\rm UV}$ relation. We show the result of the \textsc{\small MassiveFIRE} galaxies at $z=2-6$ {for MW dust in the \textit{top}} panel of Fig.~\ref{fig.9}. We can see from the \textit{top} panel that $\tilde{\tau}_{0.16}$ and $\beta_{\rm UV}$ are poorly correlated, among either the entire sample or each individual redshift {\citep[see also \eg][]{Boquien_2012, Narayanan_2018}}. This is in stark contrast with the simple linear correlation expected from {Eq.~\ref{eq.15}. However, the equation includes the term $\beta_{\rm UV,\,0}$}.  The intrinsic $\beta_{\rm UV,\,0}$ may vary between galaxies, due to different star formation histories and hence different age distributions of their stars. 

The remaining two panels of Fig.~\ref{fig.9} prove that this explanation is the correct one. In particular, there is a strong correlation between $\tilde{\tau}_{0.16}$ and $\Delta\beta_{\rm UV}=\beta_{\rm UV}-\beta_{\rm UV,\,0}$. As expected, the $\tilde{\tau}_{0.16}$ vs. $\Delta\beta_{\rm UV}$ relation of the \textsc{\small MassiveFIRE} sample is in broad agreement with the analytic solution derived using the dust slab model, Eq.~\ref{eq.15}. This is true for both MW (\textit{middle} panel) and SMC dust (\textit{bottom} panel). The slope of inclination of the analytic solution ($\mathcal{Z}$) for MW and SMC dust are 0.57 and 1.01 (calculated by Eq.~\ref{eq.16}), respectively. The data of \textsc{\small MassiveFIRE} galaxies roughly follow the predicted \textit{linear} relation for both dust models.

We notice, however, that the simulation data {shows} scatter, even though they are produced using a constant extinction curve. The deviation of the galaxies from the analytic solution indicates the variations in the shape of the dust attenuation curve ($\tilde{\tau}_\lambda$) of the galaxies from the underlying extinction curve ($\kappa_\lambda$). The shape of $\tilde{\tau}_\lambda$ depends not only on $\kappa_\lambda$, but also on the geometry of the dust and star particle distribution in the galaxies {\citep[see \eg][]{Witt_1996, Witt_2000, Charlot_2000, Narayanan_2018b}}.

To summarise this section, $\tilde{\tau}_{0.16}$ is well correlated with IRX and the relation between the two resembles an exponential {relation} predicted by the dust slab model. In contrast, $\tilde{\tau}_{0.16}$ and $\beta_{\rm UV}$ are not well correlated due to the variations in the intrinsic UV spectral slope of the galaxies, $\beta_{\rm UV,\,0}$. The important implication is that $\beta_{\rm UV}$ should not be viewed as a reliable proxy for the dust optical depth (or level of dust attenuation) of high-$z$ galaxies \citep{Narayanan_2018}. Instead, $\Delta\beta_{\rm UV}=\beta_{\rm UV}-\beta_{\rm UV,\,0}$ is linearly correlated with $\tilde{\tau}_{0.16}$.

\begin{figure}
 \begin{center}
 \includegraphics[width=86mm]{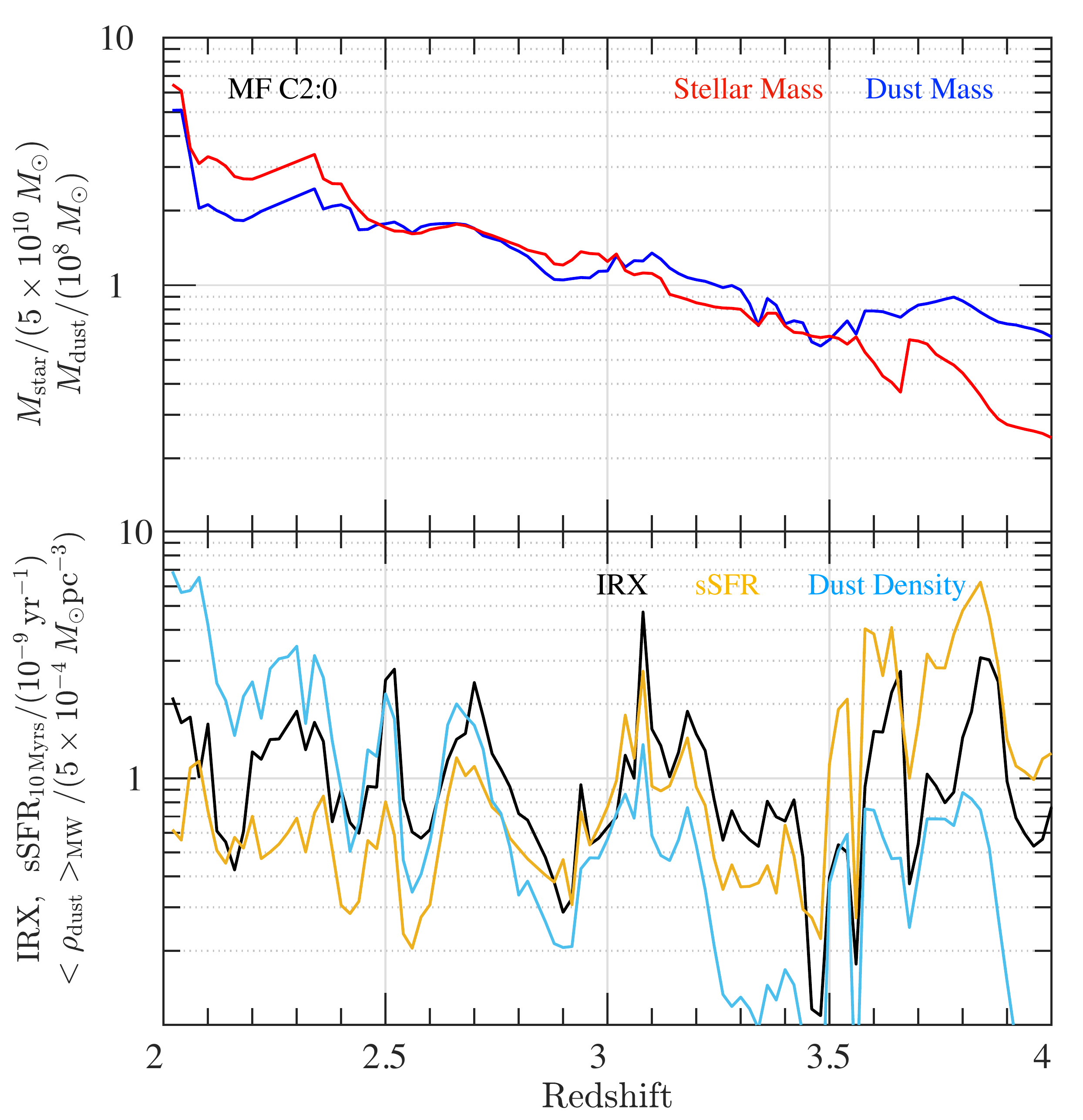}
 \caption{\textit{Upper} panel: Evolution of the normalised $M_*$ (red) and $M_{\rm dust}$ (blue) of a {selected} \textsc{\small MassiveFIRE} galaxy between $z=2$ and $z=4$. \textit{Lower} panel: Evolution of IRX (black), as well as the normalised sSFR (brown) and mass-weighted dust density (cyan) over the same time period. IRX is well correlated with sSFR and the dust density while $M_{\rm dust}$ slowly grows. {This result indicates that IRX (or $\tilde{\tau}_{0.16}$) of galaxy depends on the spatial configuration of dust {with respect to star-forming regions} in galaxy rather than $M_{\rm dust}$.} }
    \label{fig.11}
  \end{center}
  \vspace{-15pt}
\end{figure}

\begin{figure*}
 \begin{center}
 \includegraphics[width=177mm]{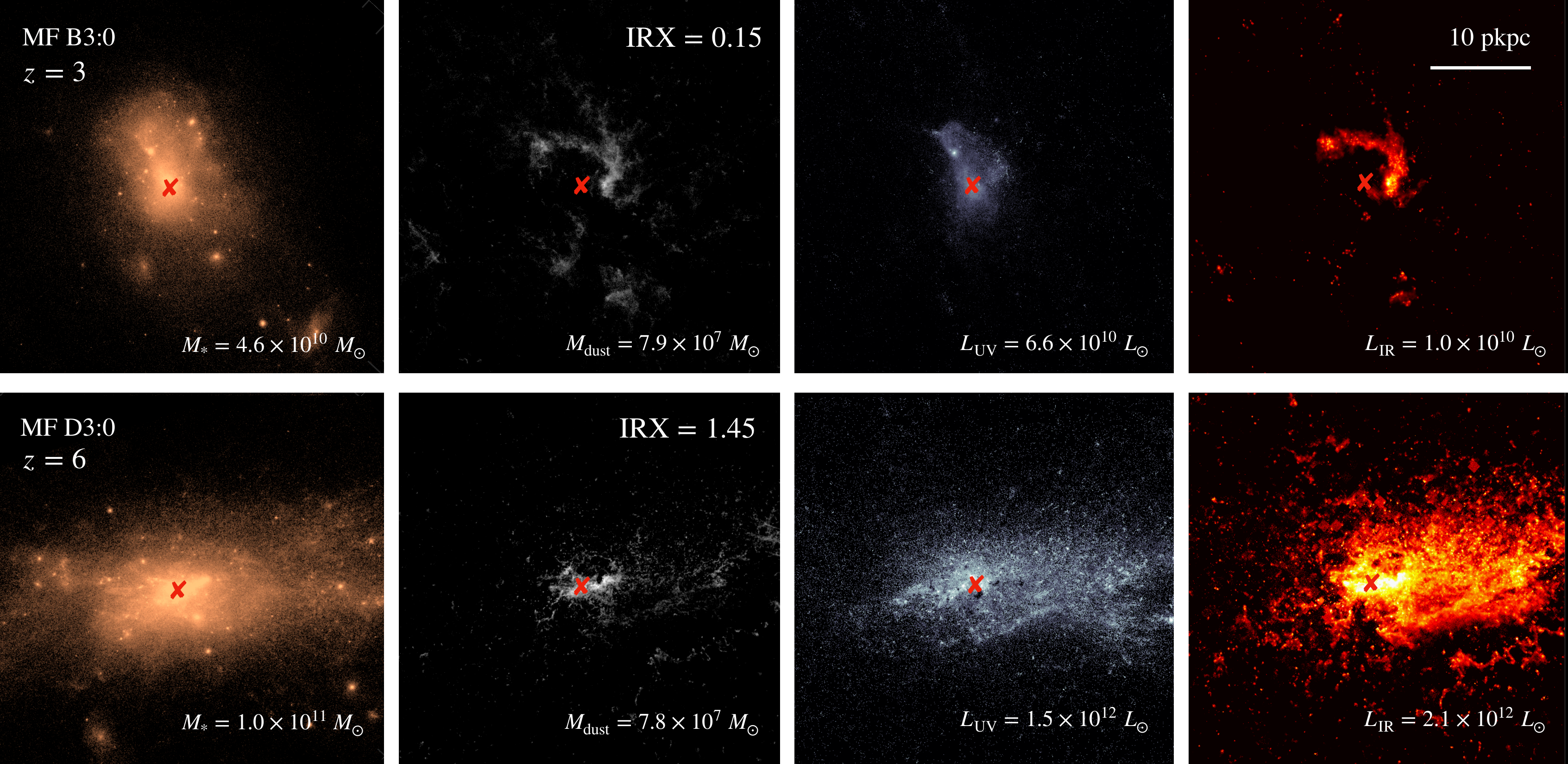}
   \vspace{-10pt}
 \caption{2D maps of stellar mass (\textit{left}), dust mass (\textit{middle left}), UV (\textit{middle right}) and IR surface brightness (\textit{right}) of two \textsc{\small MassiveFIRE} galaxies at $z=3$ (\textit{upper} panels) and $z=6$ (\textit{lower} panels). The two galaxies have similar $M_{\rm dust}$ ($\approx10^8\,M_\odot$) but very different IRX (0.15 vs. 1.45). {The red crosses in the \textit{upper} and \textit{lower} panels mark the location of the maximum surface density of stellar mass of the two galaxies. These also correspond to the local maximum points of the UV surface brightness (\textit{right middle} panels)}. IRX strongly depends on the {dust-to-star geometry} in galaxies.}
    \label{fig.12}
  \end{center}
  \vspace{-15pt}
\end{figure*}

\begin{figure*}
 \begin{center}
 \includegraphics[width=150mm]{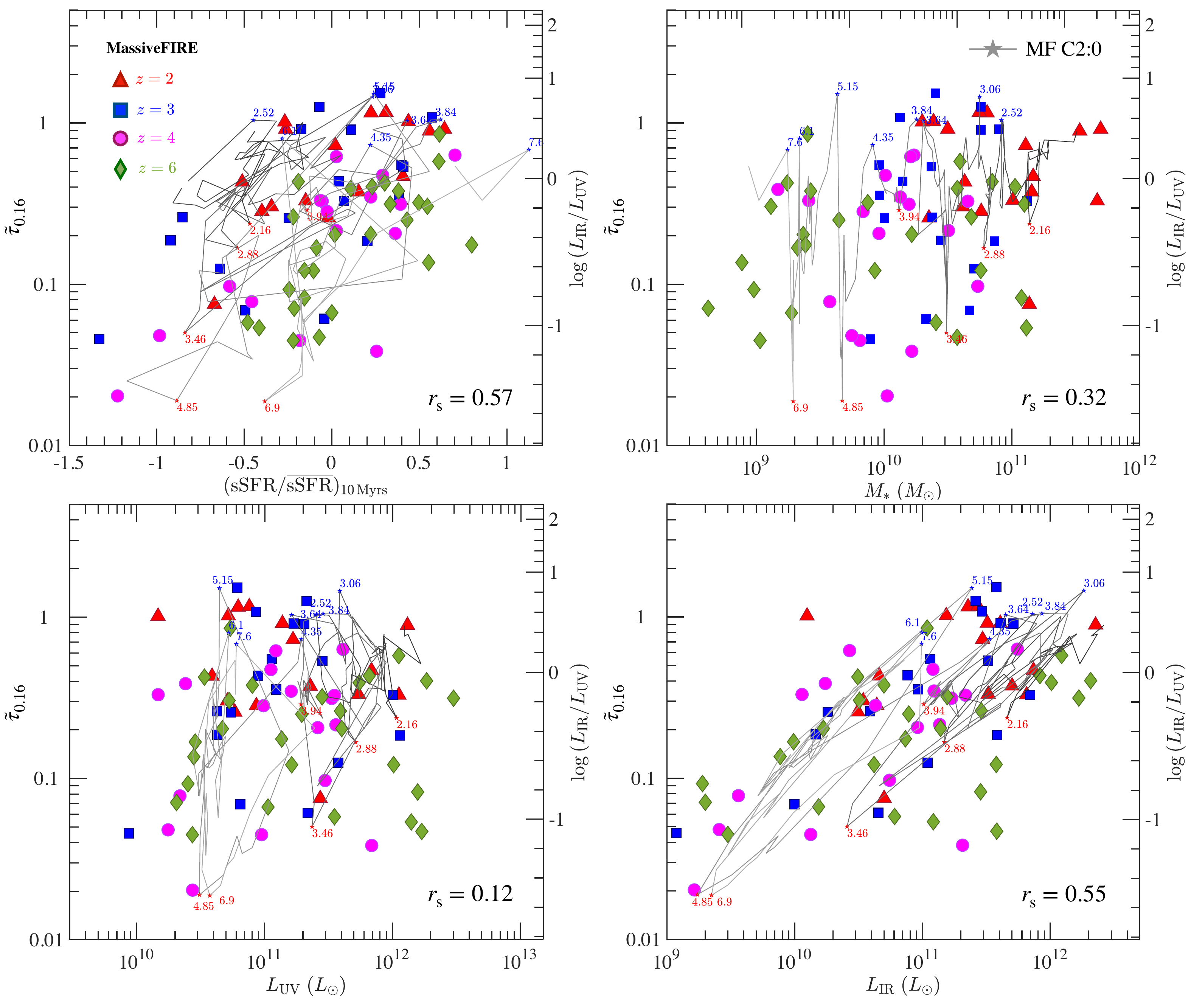}
\vspace{-5pt}
 \caption{The $\tilde{\tau}_{0.16}$ vs. starburstiness (\textit{upper left}), $\tilde{\tau}_{0.16}$ vs. $M_*$ (\textit{upper right}), $\tilde{\tau}_{0.16}$ vs. $L_{\rm UV}$ (\textit{lower left}) and $\tilde{\tau}_{0.16}$ vs. $L_{\rm IR}$ (\textit{lower right}) relations of the \textsc{\small MassiveFIRE} galaxies at $z=2-6$. The redshifts of the galaxies are indicated by the shape of the symbols as labelled. The dark grey line in each panel represents the evolutionary trajectory of a selected \textsc{\small MassiveFIRE} galaxy from $z=8$ to $z=2$. A number peaks and troughs in $\tilde{\tau}_{0.16}$ (IRX) are marked with blue and red asterisk in each panel. $\tilde{\tau}_{0.16}$ shows a {moderate} correlation with starburstiness (Spearman's correlation coefficient $r_{\rm s}$=0.57) and $L_{\rm IR}$ ($r_{\rm s}$=0.55) of galaxies, and a weak correlation with $M_*$ ($r_{\rm s}$=0.32) and $L_{\rm UV}$ ($r_{\rm s}$=0.12).}
    \label{fig.13}
  \end{center}
  \vspace{-20pt}
\end{figure*}

\subsection{The nature of infrared excess of galaxies}
\label{Sec:4c}

We have shown in the last subsection that the IRX of galaxies are well correlated with $\tilde{\tau}_{0.16}$. It is therefore important to understand what determines $\tilde{\tau}_{0.16}$ of galaxies at different stages of their evolution.

The total \textit{effective} optical depth of galaxies may depend on the total dust mass and/or the spatial configuration of dust distribution. We first examine the relation between $\tilde{\tau}_{0.16}$ and $M_{\rm dust}$ \footnote{{Physical properties of galaxies (\ie~$M_{\rm dust}$, $M_*$, SFR, $L_{\rm UV}$, $L_{\rm IR}$ and etc.) reported in this paper are estimated using a radial kernel of 30 physical kpc around the dark matter halo center.}} in \textsc{\small MassiveFIRE}. The \textit{upper} panel of Fig.~\ref{fig.10} shows that $\tilde{\tau}_{0.16}$ and $M_{\rm dust}$ are weakly correlated. We also overplot in the panel the trajectory of a selected galaxy (galaxy ID: MF C2:0) between $z=8$ and $z=2$ with a dark grey line. This galaxy goes through several periods of significant rise and decline in $\tilde{\tau}_{0.16}$, while its dust mass slowly increases. On the other hand, $\tilde{\tau}_{0.16}$ is always well correlated with IRX (both are angle-averaged) and their relation does not deviate much from the best-fit exponential-law curve that we have derived using the entire sample, as shown in the \textit{lower} panel of the figure. This is because the $\mathcal{Y}$ parameter in Eq.~\ref{eq.7} has limited variation as the galaxy evolves with time.  

The evolution of the galaxy's $\tilde{\tau}_{0.16}$ (or equivalently, IRX) is well correlated to dust mass density, as is shown in Fig.~\ref{fig.11} (\textit{lower} panel). This signifies that $\tilde{\tau}_{0.16}$ of galaxies is driven by the compactness of the spatial configuration of dust rather than the total amount of dust. Interestingly, the evolution of both IRX (black line) and dust mass density (cyan line) also coincide with that of sSFR (golden line), since more star formation is triggered when the ISM gas/dust becomes more compact and thus the local free-fall timescale of star-forming clouds decreases ($t_{\rm ff}\propto\rho^{-1/2}$). IRX (or $\tilde{\tau}_{0.16}$), sSFR, and dust mass density can significantly vary on relatively short timescales ($\simless100\,\rm Myrs$), while {the dust and stellar masses of galaxy} gradually grow. 

\begin{figure}
 \begin{center}
 \includegraphics[width=87mm]{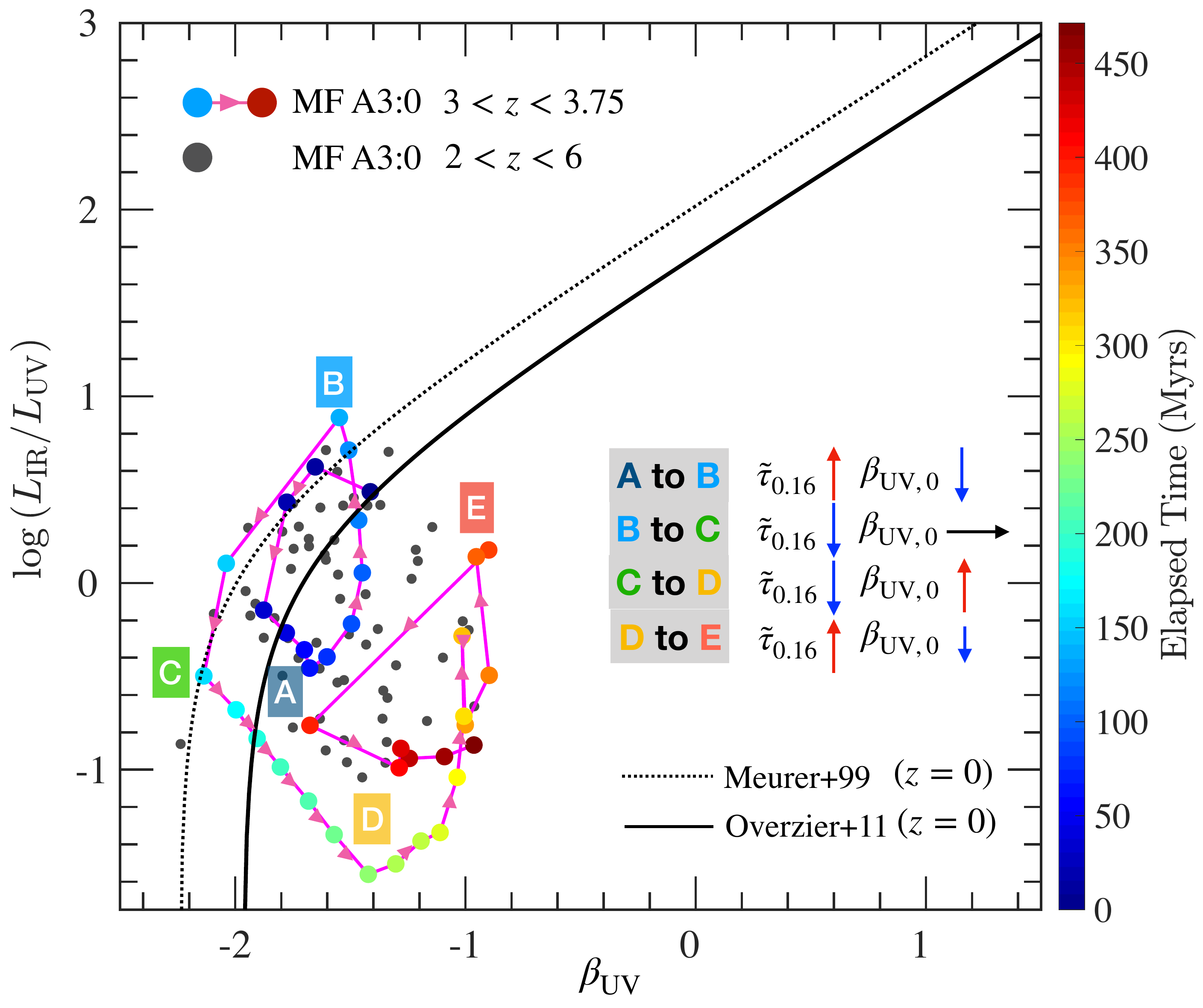}
 \caption{Evolution of selected \textsc{\small MassiveFIRE} galaxy in the IRX-$\beta_{\rm UV}$ plane. The magenta curve shows the evolutionary trajectory of this galaxy between $z=3.75$ and $z=3$. The filled dots linked by the curve represent the output snapshots of the simulation, which are colour-coded by the time that has elapsed since $z=3.75$ (in unit of Myrs). The grey dots indicate the location of the galaxy at the other snapshots between $z=2-6$. We specifically mark the five characteristic snapshots (A, B, C, D and E) on the galaxy's trajectory to distinguish the different evolutionary stages of the galaxy (Section~\ref{Sec:4d}). The dotted and solid black lines indicate the local M99 and $\rm M99_{\rm corr}$ relations, respectively. The location of galaxy in the IRX-$\beta_{\rm UV}$ plane exhibits a large spread over cosmic time and its evolutionary trajectory shows counter-clockwise rotation.}
    \label{fig.14}
  \end{center}
  \vspace{-20pt}
\end{figure}

\begin{figure}
 \begin{center}
 \includegraphics[width=88mm]{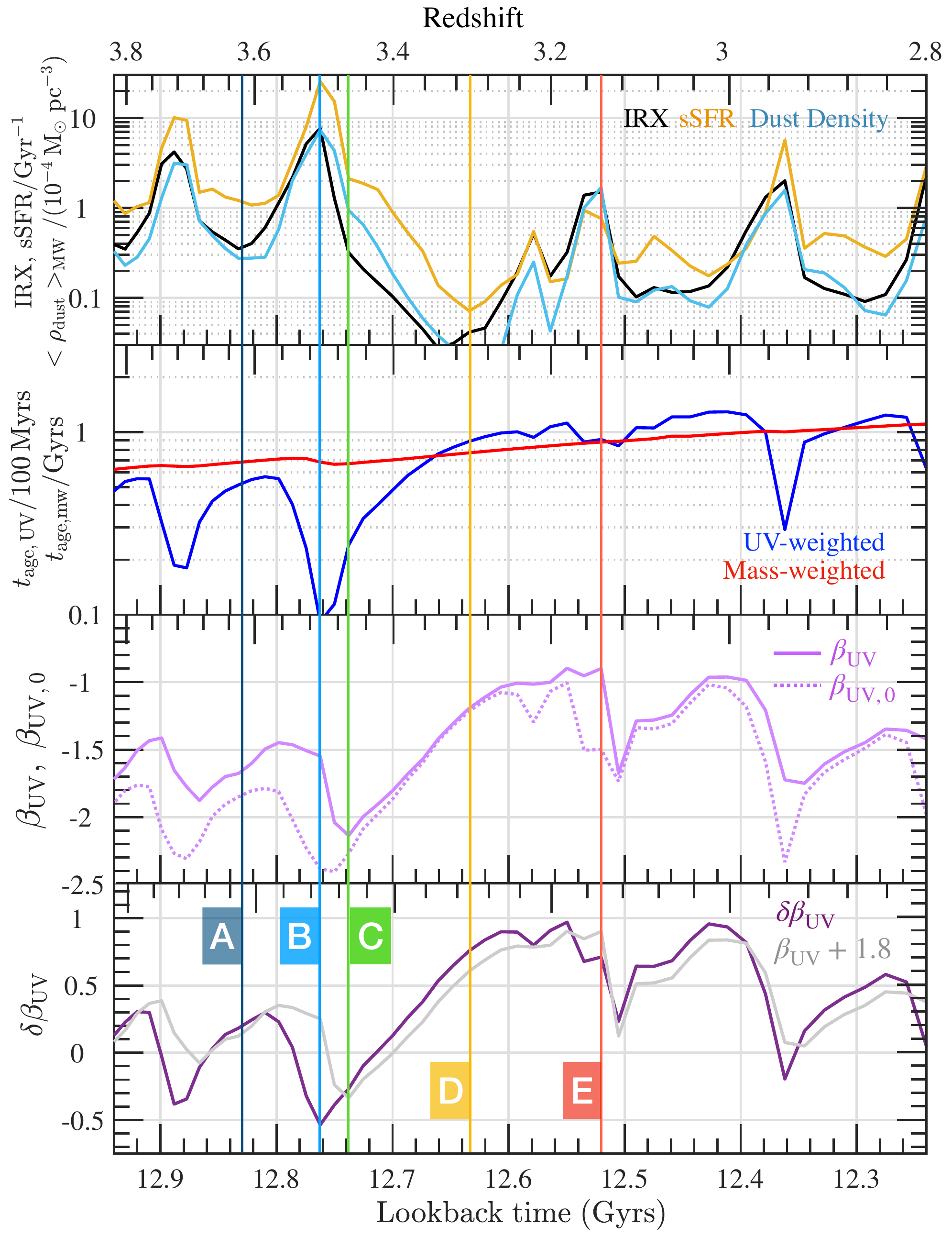}
 \caption{Evolution of different properties of a {typical} \textsc{\small MassiveFIRE} galaxy ({same as shown in Fig.~\ref{fig.14}}) from $z=3.8$ to $z=2.8$. \textit{Top} panel: Evolution of the IRX (black line), the normalised sSFR (yellow line), and mass-weighted dust density (cyan line). \textit{Upper middle} panel: Evolution of the UV-luminosity-weighted (blue line) and the mass-weighted (red line) stellar age of the galaxy. \textit{Lower middle} panel: Evolution of $\beta_{\rm UV}$ (solid violet line) and $\beta_{\rm UV,\,0}$ (dotted violet line). \textit{Bottom} panel: Evolution of $\delta\beta_{\rm UV}$ (purple line), the horizontal offset of the galaxy's location from the canonical M99 relation. The grey line in this panel represents the rescaled UV spectral slope, $\beta_{\rm UV}$+1.8. The vertical dark blue, blue, green, yellow and red lines mark the five characteristic snapshots that distinguish the different evolutionary stages of the galaxy, as in Fig.~\ref{fig.14}. }
    \label{fig.15}
  \end{center}
    \vspace{-10pt}
\end{figure}

Therefore, the observationally derived IRX of a galaxy mainly reflects the spatial configuration of dust with respect to star-forming regions, and provides limited constraint on its dust mass. To better illustrate this, we present the 2D maps of the dust mass column density of two \textsc{\small MassiveFIRE} galaxies at two different redshifts ($z=3$ and $z=6$) in Fig.~\ref{fig.12}. These two galaxies have about the same $M_{\rm dust}$ ($\approx10^8\,M_\odot$) but significantly different IRX (0.15 vs. 1.45). The location of the two galaxies on the $M_{\rm dust}$ vs. $\tilde{\tau}_{0.16}$ diagram are marked with black crosses in Fig.~\ref{fig.10} (\textit{upper} panel). The $z=6$ galaxy (galaxy ID: MF D3:0) is at an instantaneous peak of star formation, having a $\rm SFR_{10\,Myrs}=250\,M_\odot\,{\rm yr^{-1}}$. The $z=3$ galaxy (galaxy ID: MF B3:0) is relatively quiescent and has a much lower $\rm SFR_{10\,Myrs}$ of only $7\,M_\odot\,{\rm yr^{-1}}$.  Comparing the \textit{upper} and \textit{lower} panels (\textit{middle left}), it is clear that the $z=6$ object shows a more compact spatial configuration of dust distribution near the star-forming regions at the galaxy center (marked by red cross in each panel), where most of its UV and IR emission originates. The dust in the $z=3$ galaxy, on the contrary, {is dispersed away from the UV-emitting regions}. The relatively high IRX of the $z=6$ galaxy is due to the higher obscuration of its star-forming regions. 

We now investigate the relationship between $\tilde{\tau}_{0.16}$ (equivalently, IRX) and several important observable properties of galaxies, including starburstiness\footnote{We defined `starburstiness' as the ratio of a galaxy's sSFR to the median sSFR of the sample at that redshift, \ie~$\rm sSFR/\overline{sSFR}(z)$.}, $M_*$, $L_{\rm UV}$ and $L_{\rm IR}$, see Fig.~\ref{fig.13}. {In each panel, we show the results for both the $z=2-6$ galaxy sample and the evolutionary trajectory of a selected \textsc{\small MassiveFIRE} galaxy.}

The figure shows that $\tilde{\tau}_{0.16}$ (equivalently, IRX) has a moderate correlation with starburstiness (Spearman's correlation coefficient $r_{\rm s}=0.57$), which is not surprising because the increase in ISM density results in a higher SFR. $\tilde{\tau}_{0.16}$ also correlates with $L_{\rm IR}$ ($r_{\rm s}=0.55$) because galaxies of higher $\tilde{\tau}_{0.16}$ tend to be more bursty and thus dust is exposed to a harder radiation field resulting from the enhanced fraction of young stars \citep[see also \eg,][]{Wang_1996, Adelberger_2000, Bell_2003, Buat_2005, Jonsson_2006, Buat_2007, Buat_2009, Reddy_2010, Hayward_2012}. In addition, a higher \textit{fraction} of stellar light is absorbed and reemitted due to higher $\tilde{\tau}_{0.16}$. $L_{\rm UV}$, on the contrary, shows almost no correlation with $\tilde{\tau}_{0.16}$ ($r_{\rm s}=0.12$) due to increasing absorption with $\tilde{\tau}_{0.16}$ \citep[see also \eg,][]{Buat_2009, Reddy_2010, Casey_2014, Sklias_2014, Reddy_2018}. And finally, $\tilde{\tau}_{0.16}$ and $M_*$ are only weakly correlated ($r_{\rm s}=0.32$). This is consistent with the scenario that the spatial redistribution of dust and the interstellar radiation field can significantly change on the timescale of a few 10 Myrs, while galaxies assemble their dust and stellar masses on much longer timescales \citep[see Fig.~\ref{fig.11}, \cf,][]{Sklias_2014, Bouwens_2016, Fudamoto_2017, Reddy_2018, Koprowski_2018, McLure_2018, Fudamoto_2020, Bouwens_2020}.

\begin{figure}
 \begin{center}
 \includegraphics[width=83mm, height=205mm]{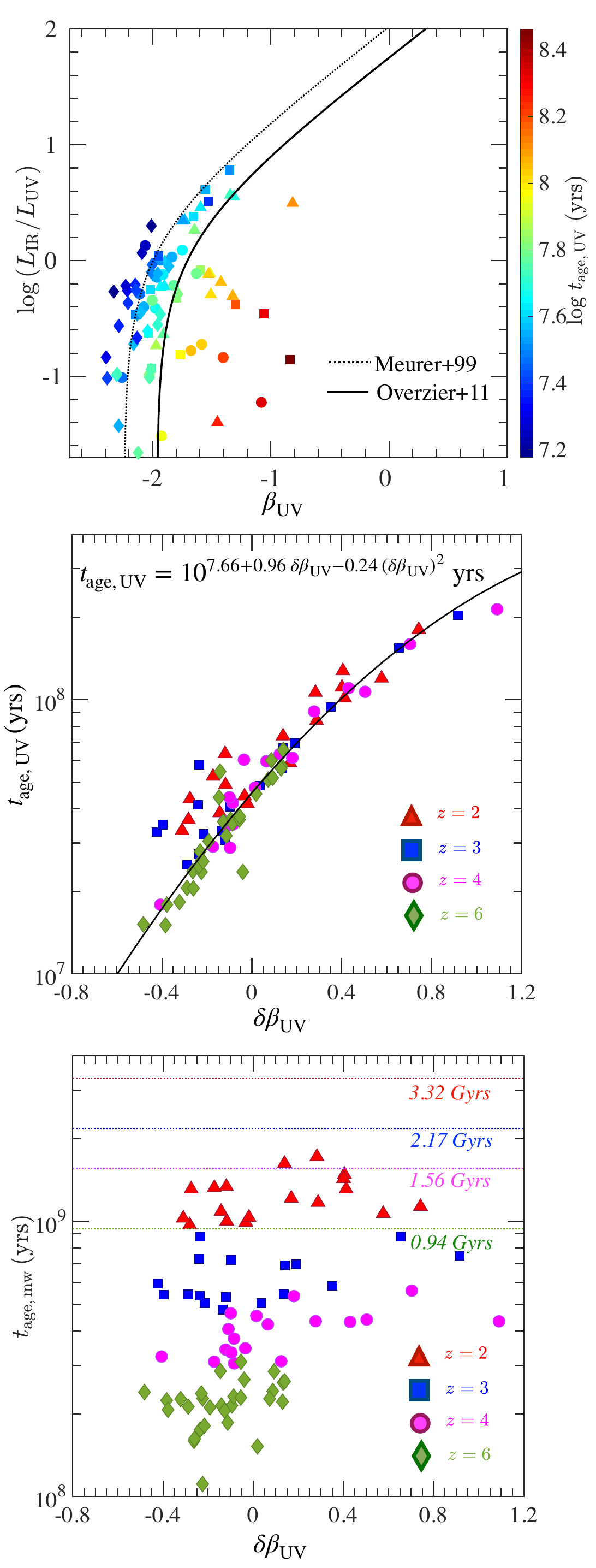}
 \caption{\textit{Top} panel: The `secondary dependence' of the IRX-$\beta_{\rm UV}$ relation on the UV-luminosity-weighted stellar age, $t_{\rm age,\,UV}$. The dotted and solid black lines indicate the local M99 and $\rm M99_{corr}$ relations, respectively. \textit{Middle} panel: The relation between $t_{\rm age,\,UV}$ and $\delta\beta_{\rm UV}$ in \textsc{\small MassiveFIRE}. \textit{Bottom} panel: The relation between the mass-weighted stellar age, $t_{\rm age,\,MW}$, and $\delta\beta_{\rm UV}$. The dotted red, blue, magenta and green horizontal lines mark the age of the Universe at $z=2$, $z=3$, $z=4$ and $z=6$, respectively. While $\delta\beta_{\rm UV}$ is strongly correlated with $t_{\rm age,\,UV}$, $\delta\beta_{\rm MW}$ and $t_{\rm age,\,mw}$ {have} no clear correlation.}
    \label{fig.16}
  \end{center}
  \vspace{-20pt}
\end{figure}

\vspace{-10pt}
\subsection{{The evolution of galaxies in the IRX-$\beta$ plane}}
\label{Sec:4d}

\begin{figure*}
 \begin{center}
 \includegraphics[width=165mm]{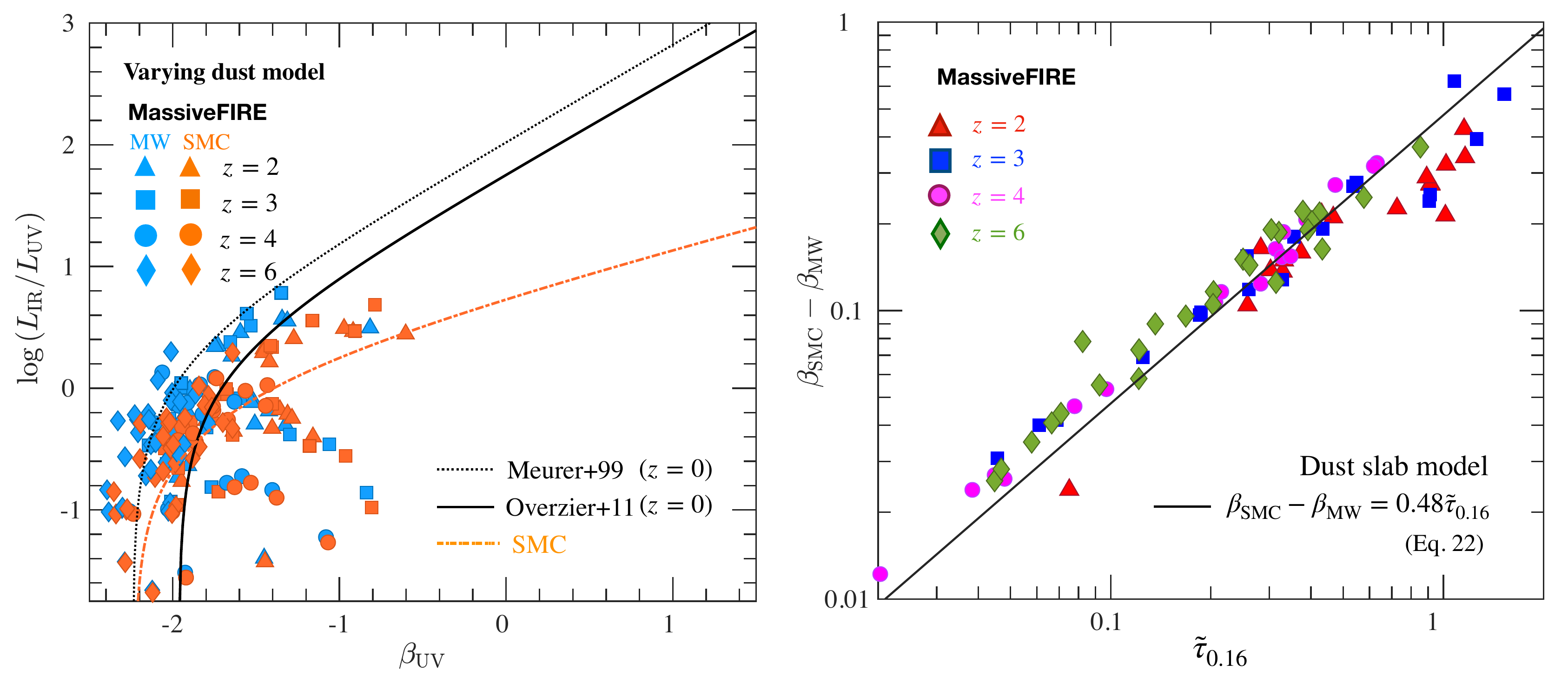}
 \caption{\textit{\textit{Left} panel:} IRX vs. $\beta_{\rm UV}$ relation of the \textsc{\small MassiveFIRE} galaxies for the two dust {models}, MW (blue symbols) and SMC (orange symbols) of WD01. The dotted and solid black lines show the M99 and $\rm M99_{corr}$ relations. The dot-dashed orange line shows the result of the dust slab model for {SMC dust}, with $t_{\rm age}=50$ Myrs and $Z_*=0.1Z_\odot$. \textit{Right} panel: The difference between the $\beta_{\rm UV}$ produced by MW and SMC dust as a function of $\tilde{\tau}_{0.16}$. The solid black line represents the analytic result derived from the dust slab model (Eq.~\ref{eq.22}).}
    \label{fig.17}
  \end{center}
  \vspace{-10pt}
\end{figure*}

Using the dust slab model, we have shown that while dust {UV} optical {depth} drives galaxies along the direction of the IRX-$\beta_{\rm UV}$ relation, {variation of stellar age contributes} to the offsets from the relation. Optical depth and stellar age thus act as two independent factors that affect the location of galaxies in the IRX-$\beta_{\rm UV}$ plane in the case of the toy model. For real galaxies, however, the two quantities may be correlated. For instance, Fig.~\ref{fig.11} shows that during a starburst, {$\tilde{\tau}_{0.16}$ (equivalent to IRX) increases while more young stars are born}. The {anti-correlation} between $\tilde{\tau}_{0.16}$ and stellar {population ages} has consequence for the evolutionary trajectory of galaxies in the IRX-$\beta_{\rm UV}$ plane. 

In Fig.~\ref{fig.14}, we show the location of a {selected} \textsc{\small MassiveFIRE} galaxy (galaxy ID: MF A3:0) in the IRX-$\beta_{\rm UV}$ plane at different snapshots at $z=2-6$. While we show only one example, the overall behaviour of this galaxy is typical for the galaxies in our sample.

The evolutionary trajectory of this galaxy over the period of $z=3-3.75$ is highlighted by the magenta line. We mark five characteristic times that distinguish the different evolutionary stages using capital letters (from `A' to `E'). Fig.~\ref{fig.15} shows the evolution of IRX and $\beta_{\rm UV}$ of the galaxy as a function of lookback time, together with other galaxy properties that are relevant for the physical explanation of the galaxy's trajectory in the IRX-$\beta_{\rm UV}$ plane. 

According to Fig.~\ref{fig.14}, the trajectory of the galaxy in the IRX-$\beta_{\rm UV}$ plane is counter-clockwise. Time `A' corresponds to $z=3.62$, when a minor merger occurs. Over the next $70$ Myrs, the gas/dust distribution of the galaxy becomes more concentrated due to the instabilities induced by the merger {and the star-forming regions become more obscured}, resulting in a significant boost of $\tilde{\tau}_{0.16}$ and hence IRX of the galaxy, {until time} `B' ($z=3.52$). Meanwhile, the sSFR of galaxy is also enhanced due to the growing compactness of the gas distribution (\textit{top} panel of Fig.~\ref{fig.15}). Furthermore, the galaxy's UV-luminosity-weighted age, $t_{\rm age,\,UV}$, computed by weighting the age of each star particle by its luminosity over $\lambda=0-3200\,\angstrom$, decreases from $\sim0.5$ Gyr at time `A'  to $\sim0.1$ Gyr {at time `B'}. This is because more young, massive OB stars are formed over this period, which dominates the rest-UV emission of the galaxy. Consequently, $\beta_{\rm UV,\,0}$ declines from -1.8 to -2.4. $\beta_{\rm UV}$, on the contrary, is nearly constant because the increased reddening due to the larger $\tilde{\tau}_{0.16}$ cancels out the decrease in $\beta_{\rm UV,\,0}$ ({see} Eq.~\ref{eq.15}).

From `B' to `C', the galaxy moves roughly along the M99 relation downwards in the IRX-$\beta_{\rm UV}$ plane (Fig.~\ref{fig.14}). The IRX of the galaxy decreases to nearly the same value as at time `A'. This happens on a much shorter timescale, $\sim20$ Myrs (see Fig.~\ref{fig.15}). Over this period of time, feedback from the newly born stars efficiently {ejects} the material in the star-forming region, which leads to a decrease in $\tilde{\tau}_{0.16}$ of the galaxy. $\beta_{\rm UV,\,0}$, {however}, does not {change significantly} on this short timescale (\textit{lower middle} panel, Fig.~\ref{fig.15}). {Therefore,} $\beta_{\rm UV}$ of the galaxy becomes bluer because of the reduced reddening due to the decrease of $\tilde{\tau}_{0.16}$. 

From `C' to `D', the galaxy undergoes a quenching phase that lasts about 100 Myrs (see \citealt{Feldmann_2017} for an in-depth discussion of quenching in \textsc{\small MassiveFIRE} galaxies). Over this period, stellar feedback continues to {eject} the dust near star-forming regions and reduce the optical depth of the galaxy. As the OB stars die out, $t_{\rm age,\,UV}$ increases to $\sim1$ Gyr, which is similar to the mass-weighted stellar age (\textit{upper middle} panel, Fig.~\ref{fig.15}).  The galaxy's sSFR continuously declines, reaching a minimum value of $7\times10^{-11}\,M_\odot\,{\rm yr^{-1}}$ at Snapshot `D' (\textit{top} panel). Due to the aging of the stellar population, $\beta_{\rm UV,\,0}$ significantly increases from -2.3 to -1.2, driving the increase of $\beta_{\rm UV}$ (\textit{lower middle} panel). Over this period, the galaxy's $\beta_{\rm UV}$ becomes redder mainly because of the aging of the stellar population instead of the optical depth effect.

At Snapshot `D', the galaxy restarts gas accretion which triggers active star formation within the galaxy again. Over a period of $\sim120$ Myrs, its gas and dust density, UV optical depth and IRX increases {until time} `E' (\textit{top} panel, Fig.~\ref{fig.15}). However, the gas accretion over this period is not as violent as the period of `A' to `B', and so the galaxy remains below the M99 relation since the young OB stars do not outshine the more evolved stars at rest-frame UV wavelength.

The galaxy undergoes such starburst-dispersal-quenching-reaccretion cycles throughout its lifetime {\citep{Muratov_2015, Angles-Alcazar_2017, Feldmann_2017, Sparre_2017}}. As a result, the trajectory of galaxy in the IRX-$\beta_{\rm UV}$ plane shows repeated counter-clockwise rotation. The location and the size of these cycles certainly depend on the strength of the starburst and the quenching of each cycle. Overall, the location of an individual galaxy on the diagram shows significant dispersion over cosmic time, as indicated by the coloured and grey dots in Fig.~\ref{fig.14}. The IRX of MF A3:0 varies by more than two orders of magnitude and its $\beta_{\rm UV}$ by {order} unity at $z=2-6$.

From Fig.~\ref{fig.15}, we can see that $\delta\beta_{\rm UV}$ (\textit{bottom} panel), the horizontal offset of the galaxy's $\beta_{\rm UV}$ from the M99 relation, well correlates to $t_{\rm age, \,UV}$ (\textit{upper middle} panel). We therefore expect a good correlation between the two quantities of the general galaxy sample. This is indeed the case. We can see from Fig.~\ref{fig.16} that the `secondary dependence' of the IRX-$\beta_{\rm UV}$ relation on $t_{\rm age,\,UV}$ appears to be remarkable. And by fitting $\delta\beta_{\rm UV}$ vs. $\rm log\,(t_{\rm age,\,UV})$ of the sample by a second-order polynomial law, we obtain

\begin{align}
    {\rm log}\,\left(\frac{t_{\rm age,\,UV}}{10^8\;{\rm yrs}} \right) = &(7.66\pm0.02)+(0.96\pm0.15)\,\delta\beta_{\rm UV}\\ \nonumber
    &-(0.24\pm0.48)\,(\delta\beta_{\rm UV})^2.
\label{eq.20}
\end{align}
 
\noindent We note that {$t_{\rm age,\,mw}$} shows no clear correlation with $\delta\beta_{\rm UV}$ ({\textit{lower}} panel, Fig.~\ref{fig.16}). This reflects that the shape of galaxies' SEDs at rest-frame UV is determined by the galaxies' recent star formation activities, and is not well correlated with the formation history of the more evolved bulk of the stellar population.

\subsection{Additional sources of scatter in the IRX-$\beta$ relation}
\label{Sec:4e}

We will examine in this section a few additional sources of the scatter in the IRX-$\beta_{\rm UV}$ relation, including the changes of {dust extinction law}, dust-to-metal mass ratio, direction of viewing and also the stellar population model (singular vs. binary evolution of stars).  

\vspace{-10pt}
\subsubsection{The variation of the dust extinction law}
\label{Sec:4e1}

The dust extinction law of galaxies at high-$z$ is not well constrained and is one important source of uncertainty in the IRX-$\beta_{\rm UV}$ relation. In this work, we adopt the {MW and SMC dust models} of WD01. The {extinction curve of the SMC dust model} has a steeper slope at UV than {that of the MW model} and has no clear `bump' feature at $\lambda=2175\,\angstrom$ (Fig.~\ref{fig.2}), which is present in the MW curve. Since high-$z$ galaxies {are more metal-poor} than nearby galaxies of the same mass \citep[\eg,][]{Tremonti_2004}, SMC dust is often invoked by the studies of high-$z$ galaxies. 

We produce the IRX-$\beta_{\rm UV}$ relation using the {SMC dust} in addition to the {MW dust} (which is employed in the fiducial model), and show the results of both cases in Fig.~\ref{fig.17} (\textit{left} panel). As discussed before (Section~\ref{Sec:4b}), the two {dust models} lead to very similar IRX. The SMC {dust} generates an IRX {that is} lower, on the average, by only about 0.06 dex (see Fig.~\ref{fig.8}). In contrast, it can lead to a significant offset of $\beta_{\rm UV}$, and the offset becomes more prominent with increasing $\tilde{\tau}_{0.16}$ and IRX. This is clearly shown in \textit{right} panel of Fig.~\ref{fig.17}, where we plot the difference between the $\beta_{\rm UV}$ obtained from the two {dust models} as a function of $\tilde{\tau}_{0.16}$. The relation is in good agreement with the prediction from the dust slab model, which we show by the solid black line in the \textit{right} panel. 

To derive this analytic relation of the toy model, we first re-write the Eq.~\ref{eq.15} for the two {dust models} as

\begin{equation}
\beta_{\rm MW} = \beta_0 + \mathcal{Z}_{\rm MW} \,\tilde{\tau}_{0.16}\;\; {\rm and} \;\;\beta_{\rm SMC} = \beta_0 + \mathcal{Z}_{\rm SMC}\, \tilde{\tau}_{0.16}.
\label{eq.21}
\end{equation}

\noindent Then by subtracting one equation by the other, we obtain

\begin{equation}
\beta_{\rm SMC} - \beta_{\rm MW} = (\mathcal{Z}_{\rm SMC}- \mathcal{Z}_{\rm MW})\,\tilde{\tau}_{0.16} = 0.48\,\tilde{\tau}_{0.16},
\label{eq.22}
\end{equation}

\noindent where we have input $\mathcal{Z}_{\rm SMC}=1.01$ and $\mathcal{Z}_{\rm MW}=0.57$, {calculated using $\lambda_{\rm FUV}=1230\,\angstrom$ and $\lambda_{\rm NUV}=3200\,\angstrom$ in Eq.~\ref{eq.16}}. The equation shows that the difference in the $\beta_{\rm UV}$ of galaxies produced by the two extinction curves is expected to be \emph{linearly} scaled to $\tilde{\tau}_{0.16}$. 

Note that this result is not limited to the two particular {dust models} that we use. Any variation in the steepness of the dust extinction curve at UV will result in a linear relation between the offset of $\beta_{\rm UV}$ and $\tilde{\tau}_{0.16}$. The slope of the linear relation is determined by the difference in the steepness of the two extinction curves being considered.

\subsubsection{The dependence on viewing angle}
\label{Sec:4e2}

\begin{figure*}
 \begin{center}
 \includegraphics[width=163mm]{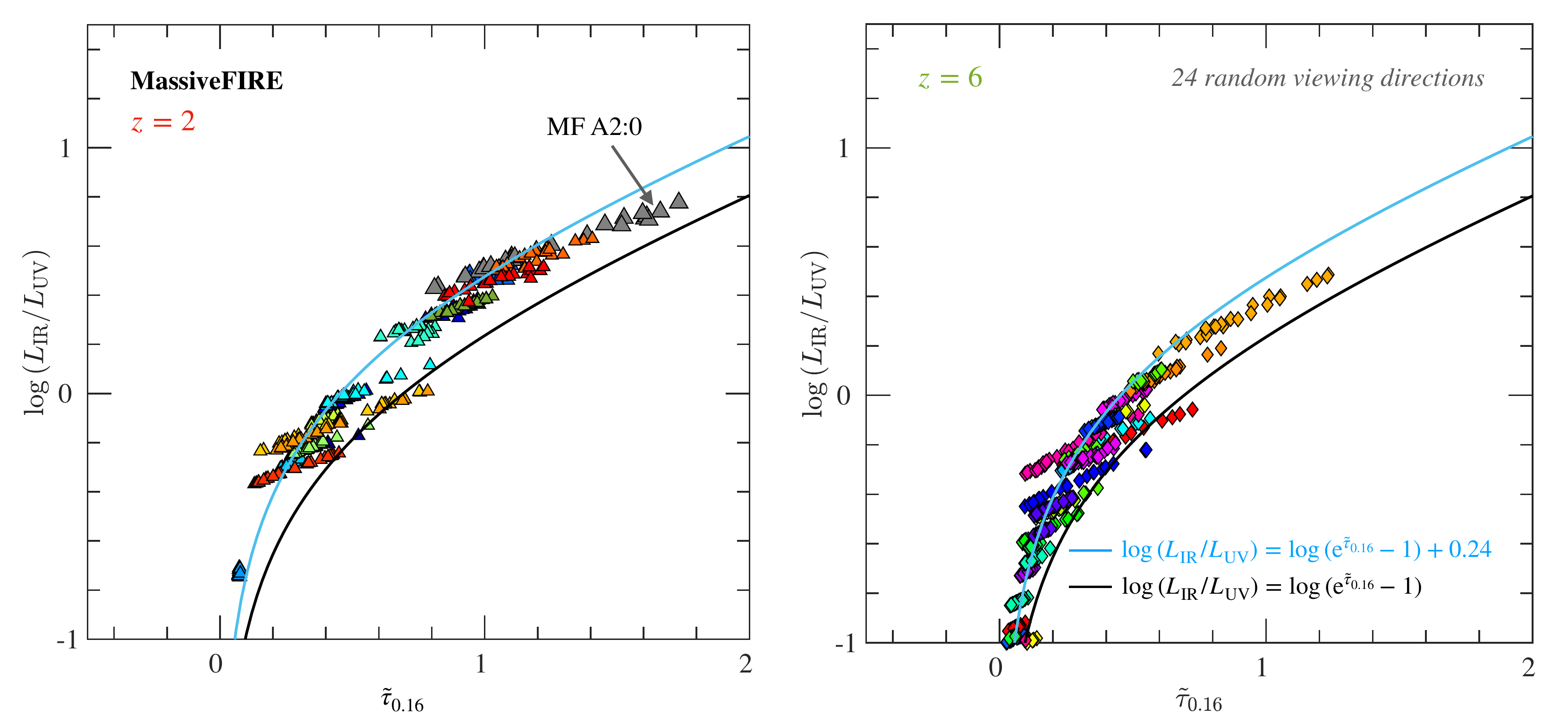}
 \caption{$\tilde{\tau}_{0.16}$ vs. IRX relation of \textsc{\small MassiveFIRE} galaxies at 24 different viewing angles. We show the results for the $z=2$ and $z=6$ galaxies in the \textit{left} and \textit{right} panels, respectively. All viewing directions of the same galaxy are marked by the same colour. The cyan and black solid lines are the same reference relations as in Fig.~\ref{fig.8}. The $\tilde{\tau}_{0.16}$ vs. IRX relation of each individual galaxy at different viewing angles is shallower than the exponential relation that best fit the angle-averaged result of the whole sample (Eq.~\ref{eq.19}). This indicates that the dust attenuation law of galaxies depends on viewing direction.}
    \label{fig.18}
  \end{center}
   \vspace{-10 pt}
\end{figure*}

The dust attenuation curve of galaxies may depend on various directions due to a non-isotropic distribution of dust and stars. Consequently, viewing direction can be one source of scatter {for the} IRX-$\beta_{\rm UV}$ relation.

We examine how the $\tilde{\tau}_{0.16}$ vs. IRX relation depends on viewing direction {in} Fig.~\ref{fig.18}. {There} we show {IRX and $\tilde{\tau}_{0.16}$ at $z=2$ for the} galaxies of our sample at 24 random viewing angles. {The} data points {corresponding to} the different viewing angles exhibit more significant scatter compared with the angle-averaged result (Fig.~\ref{fig.8}). The angle-averaged relation is well fit {by} Eq.~\ref{eq.7}. {Interestingly, viewing the same galaxy from different viewing angles leads to} a shallower {IRX-$\tilde{\tau}_{0.16}$} relation than {Eq.~\ref{eq.7}}, {indicating a smaller $\mathcal{Y}$ at higher $\tilde{\tau}_{0.16}$}. This result signifies the variation of {the} dust attenuation curve with viewing direction {(Eq.~\ref{eq.10})}. 

\begin{figure*}
 \begin{center}
 \includegraphics[width=182mm]{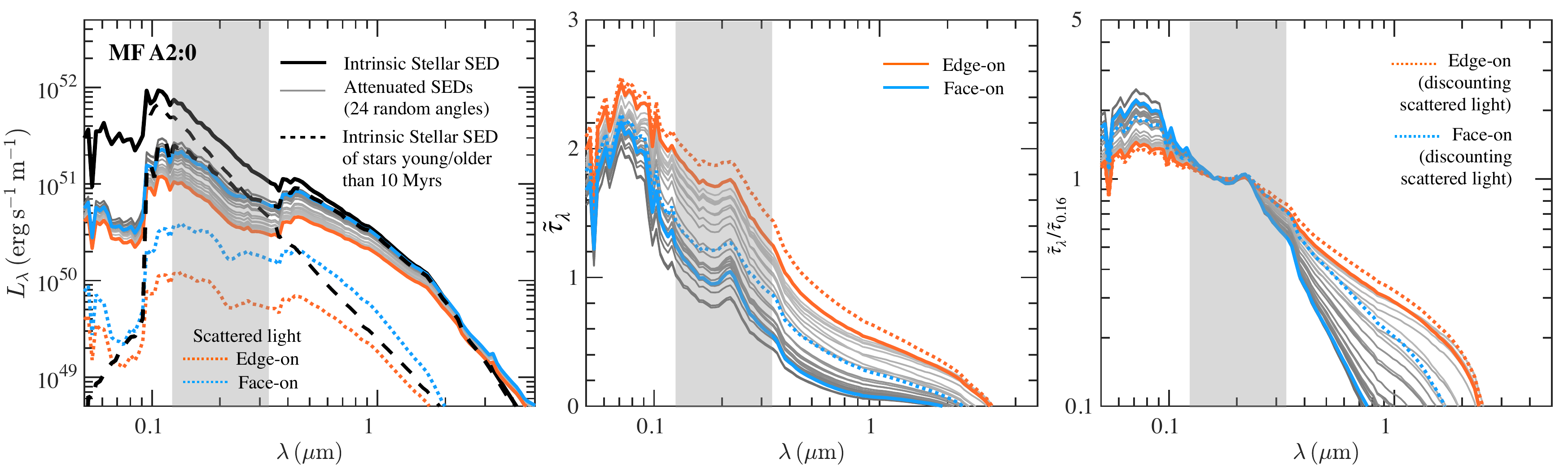}
 \caption{ \textit{Left} panel: The intrinsic (solid black line) and attenuated SEDs (grey+coloured lines) of a selected \textsc{\small MassiveFIRE} galaxy at $z=2$ ({same as in Fig.~\ref{fig.1}}). The two dashed black lines indicate the stellar emission from the stars younger/older than 10 Myrs. The grey lines show the attenuated SEDs of the 24 random viewing angles. The solid orange and blue lines correspond to the attenuated SEDs at the edge-on and face-on directions, respectively. The dotted orange and blue lines show the light that is scattered into the cameras in those two directions. \textit{Middle} panel: The attenuation {curves for} the different viewing angles. The dotted orange and blue lines show the attenuation curves when the light scattered into the cameras is discounted. \textit{Right} panel: The \textit{normalised} attenuation curves (normalised by $\tilde{\tau}_{0.16}$) of the different viewing angles. The grey shaded area in each panel marks the wavelength range of $\lambda=1230-3200\,\angstrom$, where the photometries are used for determining the $\beta_{\rm UV}$ of galaxies by the observations. {$\tilde{\tau}_\lambda/\tilde{\tau}_{0.16}$ appears to be shallower in the edge-on direction, where $\tilde{\tau}_{0.16}$ is larger.}}
    \label{fig.19}
  \end{center}
   \vspace{-10 pt}
\end{figure*}

\begin{figure*}
 \begin{center}
 \includegraphics[width=160mm, height=74mm]{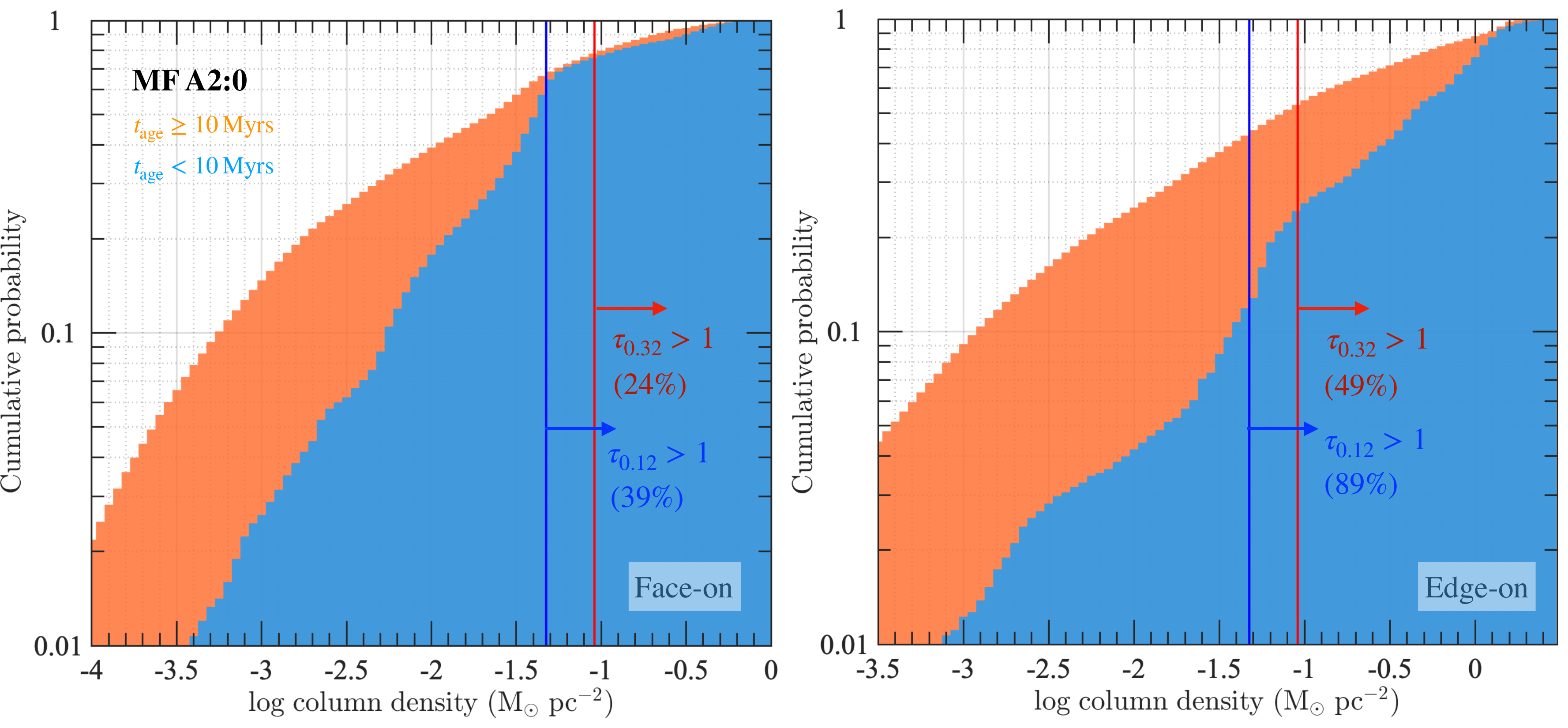}
 \caption{The cumulative probability distribution of dust column mass density (in units of $\rm kg\;m^{-2}$) of the stars of a selected $z=2$ \textsc{\small MassiveFIRE} galaxy ({same as in Fig.~\ref{fig.19}}) in the face-on (\textit{left} panel) and edge-on (\textit{right} panel) directions. In both panels, the blue and orange areas show the result of the young ($t_{\rm age}<10$ Myrs) and evolved ($t_{\rm age}\ge10$ Myrs) stars, respectively. The blue and red vertical lines indicate the column mass density above which the optical depth at $\lambda=1230\,\angstrom$ and $\lambda=3200\,\angstrom$ is over unity, respectively, assuming MW dust. The young stars dominate the stellar emission at $\lambda=1230\,\angstrom$, while the evolved stars dominate the emission $\lambda=3200\,\angstrom$ {(see \textit{left} panel of Fig.~\ref{fig.19})}. The mean dust column density of the stars is relatively higher in the edge-on direction than the face-on direction. The young stars, on the average,  have higher column density than the evolved stars. }
    \label{fig.20}
  \end{center}
   \vspace{-10 pt}
\end{figure*}

In Fig.~\ref{fig.19}, we explicitly show the attenuation curve of each of the 24 viewing angles of a selected disc-like galaxy (galaxy ID: MF A2:0) at $z=2$ (see Fig.~\ref{fig.1} for its visualisation). $\tilde{\tau}_{0.16}$ of the galaxy spans over the range of 0.81--1.73 amongst the viewing angles (Fig.~\ref{fig.18}), with the median value being 1.15. The figure clearly shows that the galaxy's edge-on (face-on) direction has roughly the highest (lowest) $\tilde{\tau}_{0.16}$ among all the sightlines. In Fig.~\ref{fig.20}, we show the cumulative probability distribution of dust column mass density (in units of {$\rm M_\odot\,pc^{-2}$}) of the star particles in both the face-on (\textit{left} panel) and edge-on (\textit{right} panel) directions. The mean column mass density in the face-on and edge-on directions are {$2.4\times10^{-2}$ and $7.6\times10^{-2}$ $\rm M_\odot\;pc^{-2}$}, respectively. The latter is higher by a factor of $\sim3$, thus explaining the larger $\tilde{\tau}_{0.16}$ for the edge-on viewing direction. 

There is also a clear trend that the attenuation curve becomes shallower (or `grayer') with increasing $\tilde{\tau}_{0.16}$ (\textit{right} panel of Fig.~\ref{fig.19}). The flattening of the attenuation curve leads to the decrease of $\mathcal{Y}$ with increasing $\tilde{\tau}_{0.16}$ (Eq.~\ref{eq.10}). This trend is driven by the variation of dust column density with stellar age --- younger stars reside in the more opaque regions where dust column density is higher.

To better illustrate this, we separately show in Fig.~\ref{fig.20} the column density distribution of the young stars ($t_{\rm age}<10$ Myrs) and the evolved stars ($t_{\rm age}\ge10$ Myrs) at the two viewing angles. The mean column density of the young stars in the face-on on edge-on directions are {$4.7\times10^{-2}$ and 0.47 $\rm M_\odot\;pc^{-1}$}, which are higher than that of the evolved stars by 0.3 and 0.8 dex (a factor of 1.9 and 6.3), respectively. 

The young stars dominate the emission at FUV and the attenuation of galaxy at FUV {strongly} depends on the obscuration of the young stars. We separately show the emission of the young and evolved stars of MF A2:0 in the \textit{left} panel of Fig.~\ref{fig.19} (black dashed lines). The young stars, which account for only $0.4\%$ of the total stellar mass of this galaxy, dominate the stellar emission of the galaxy at $\lambda\le2500\,\angstrom$, while the emission at longer wavelength is dominated by the evolved stars. A large fraction of these young stars are `highly obscured' at FUV (\ie,~$\tau_{\rm FUV}\simgreat1$). Specifically, $40\%$ ($90\%$) of the young stars have $\tau_{0.12}$ (measured at $\lambda_{\rm FUV}=1230\,\angstrom$) over unity in the face-on (edge-on) direction. The consequence of the high obscuration of the young stars is that the attenuation at FUV `responds' mildly to a change in the dust column density, as it turns from the face-on to the edge-on direction (the decline in $\rm e^{-\tau}$ `saturates' when $\tau\simgreat1$).

In contrast, the attenuation at NUV is more sensitive to a change in viewing direction because the evolved stars, which dominate the NUV emission, reside in the relatively diffuse environments. $76\%$ ($51\%$) of the evolved stars have $\tau_{0.32}$ (measured at $\lambda_{\rm NUV}=3200\,\angstrom$) less than unity in the face-on (edge-on) direction. As a result, the attenuation at NUV, increases more significantly with dust column density (${\rm e^{-\tau}}$ declines rapidly with increasing $\tau$ at $\tau<1$).  

Scattering plays a role in altering the shape of the attenuation curve, in particular, in the regime redwards of the `bump' ($\lambda>2175\,\angstrom$) (\textit{right} panel of Fig.~\ref{fig.19}). The attenuation curve becomes steeper by accounting for the {light scattered into the camera from dust} because {this component} compensates more for the loss of light {by extinction along the sightlines} at longer wavelength {in the UV-to-optical regime}. Furthermore, the scattered light accounts for a larger fraction of the total received light in the face-on direction due to the anisotropy of scattering (\textit{left} panel). The discrepancy in the steepness of the attenuation curve between the viewing angles is magnified due to the scattered light (\textit{right} panel).

\begin{figure*}
 \begin{center}
 \includegraphics[width=165mm]{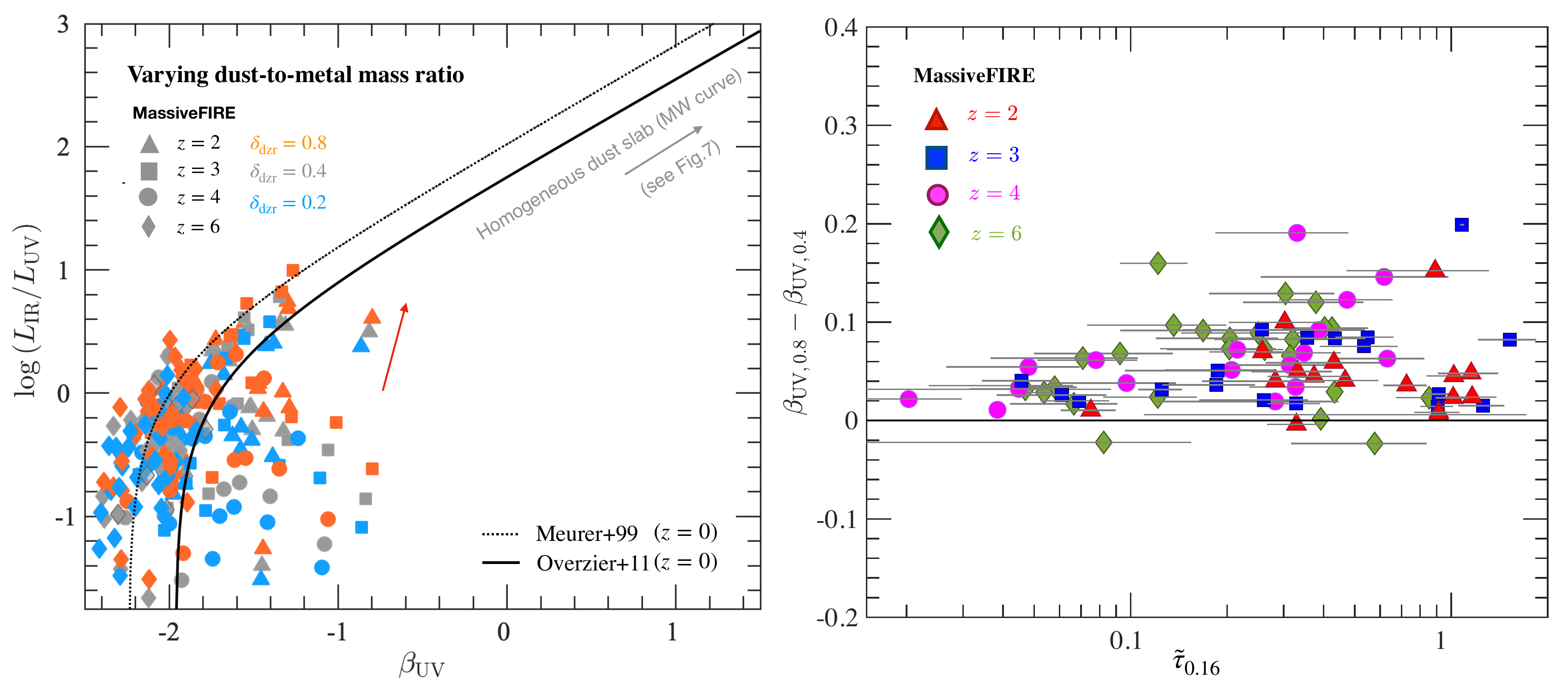}
 \caption{\textit{Left} panel: The IRX-$\beta_{\rm UV}$ relation of the \textsc{\small MassiveFIRE} galaxies with varying $\delta_{\rm dzr}$. The red, grey and blue symbols correspond to the cases of $\delta_{\rm dzr}=0.8$, $\delta_{\rm dzr}=0.4$ and $\delta_{\rm dzr}=0.2$, respectively. The dotted and solid black black lines indicate the local M99 and $\rm M99_{corr}$ relations, respectively. The red arrow indicates the direction of change in the IRX-$\beta_{\rm UV}$ relation by increasing $\delta_{\rm dzr}$. {The slope of the change is steeper than the analytic relation derived from the dust slab model with MW dust, indicating that $\tilde{\tau}_\lambda$ becomes shallower (`grayer') with increasing $\delta_{\rm dzr}$.} \textit{Right} panel: The difference in $\beta_{\rm UV}$ between the cases of $\delta_{\rm dzr}=0.8$ and $\delta_{\rm dzr}=0.4$ as a function of $\tilde{\tau}_{0.16}$ (calculated using $\delta_{\rm dzr}=0.4$). The horizontal error bars indicate the range of $\tilde{\tau}_{0.16}$ corresponding to $\delta_{\rm dzr}=0.2-0.8$.}
    \label{fig.21}
  \end{center}
  \vspace{-15pt}
\end{figure*}

Confirming the trend of shallower attenuation curve at higher $\tilde{\tau}_{0.16}$ (Fig.~\ref{fig.18}) observationally will be challenging. However, there may be \textit{indirect} evidence embedded in the IRX-$\beta_{\rm UV}$ relation. We have shown in Section~\ref{Sec:4e1} (Eq.~\ref{eq.22}, and see also in Section~\ref{Sec:4a}) that a steeper (more SMC-like) attenuation law can lead to a redder and shallower IRX-$\beta_{\rm UV}$ relation. This implies that the edge-on disc galaxies in a statistically large sample should appear to have redder $\beta_{\rm UV}$ than the face-on disc galaxies at given IRX since they on average have steeper attenuation curve. This trend has indeed been recently reported by \citet{Wang_2018}, who derive the result using a sample of UV-selected galaxies at $1.3<z<1.7$ extracted from the CANDELS field {(see also \citealt{Kriek_2013})}.

Note that although \citet{Wang_2018} has adopted a sample of disc galaxies, which is the most straightforward way of distinguishing the viewing directions of higher or lower $\tilde{\tau}_{0.16}$ observationally, we find that the trend of flattening attenuation curve with increasing  $\tilde{\tau}_{0.16}$ prevails among galaxies of varied morphology, as our sample also includes massive ellipticals, irregular galaxies and merging systems. That the attenuation curve is shallower in the direction of higher $\tilde{\tau}_{0.16}$ appears to be the general trend for galaxies of all types (Fig.~\ref{fig.18}). 

Finally, the inclination effect does not appear to be a major contributor to the scatter in the IRX-$\beta_{\rm UV}$ relation. Overall, it leads to a mean $3\sigma$ dispersion of $\beta_{\rm UV}$ of $\sim0.1$, {which is small} compared to the scatter driven by the variations {of} the intrinsic UV spectral slope of galaxies (Fig.~\ref{fig.16}) or the uncertainties in the underlying dust extinction law (Fig.~\ref{fig.17}).

\subsubsection{The effect of varying dust-to-metal mass ratio}
\label{Sec:4e3}

{Variations of the dust-to-metal ratio \citep[\eg][]{De_Cia_2013, De_Cia_2016, Wiseman_2017, De_Vis_2019, Li_2019} are another potential} source of scatter in the IRX-$\beta_{\rm UV}$ relation. So far, we have adopted a constant $\delta_{\rm dzr}=0.4$ in our analysis. In this subsection, we estimate the {impact of $\delta_{\rm dzr}$ variations on} the IRX-$\beta_{\rm UV}$ relation. 

We show in Fig.~\ref{fig.21} (\textit{left} panel) the IRX-$\beta_{\rm UV}$ relation of \textsc{\small MassvieFIRE} galaxies for the cases of $\delta_{\rm dzr}=0.2$, 0.4 and 0.8. Both IRX and $\beta_{\rm UV}$ increase with $\delta_{\rm dzr}$, {due to} the increase of dust column density. However, the changes of IRX and $\beta_{\rm UV}$ are small compared to the overall {galaxy-to-galaxy} scatter in the relation. Increasing (decreasing) $\delta_{\rm dzr}$ by a factor of 2 results in a {systematic} increase (decrease) of IRX and $\beta_{\rm UV}$ by 0.15 (0.20) dex and 0.05 (0.05) on average, respectively. This again shows that dust mass (in this case, {scaled to} $\delta_{\rm dzr}$) is not the key factor {that determines the location of galaxies on the IRX-$\beta_{\rm UV}$ diagram {(but rather, dust-to-star geometry)}.}


Interestingly, galaxies do not move parallel to the analytic curve of the dust slab model at large IRX (or $\beta_{\rm UV}$) {when $\delta_{\rm dzr}$ is varied}. {The change of the position in the IRX-$\beta_{\rm UV}$ plane with increasing $\delta_{\rm dzr}$ is almost parallel to the vertical direction and is steeper than the analytic relation predicted by the dust slab model for MW dust.} This indicates that the {attenuation curve of galaxies} becomes shallower (`grayer') with increasing $\delta_{\rm dzr}$. 

The attenuation curve {varies} with $\delta_{\rm dzr}$ {for the same reason it varies with viewing angle, see Section~\ref{Sec:4e2}.} Young stars, which dominate the stellar emission at shorter wavelength, reside in more dust obscured regions. As a consequence, UV light from a significant fraction of young stars is almost completely attenuated independent of the precise value of $\delta_{\rm dzr}$ (provided it is large enough). If every star or star-forming region had the same column density, the shape of the attenuation curve would be independent of $\delta_{\rm dzr}$, which follows from Eq.~\ref{eq.18}. The shift of the data in the IRX-$\beta_{\rm UV}$ plane {would then} be parallel to the analytic curve of the dust slab model with varying $\delta_{\rm dzr}$.

Finally, we note that the uncertainties in $\delta_{\rm dzr}$ we consider here can also be more generally viewed as equivalent to the uncertainties in the gas metallicities or the normalisation of the extinction curve, which are not well constrained observationally at high-$z$. A change of either of the two quantities will have the same effect on the attenuation curve as a change of $\delta_{\rm dzr}$ by the same factor. 

\begin{figure*}
 \begin{center}
 \includegraphics[width=165mm]{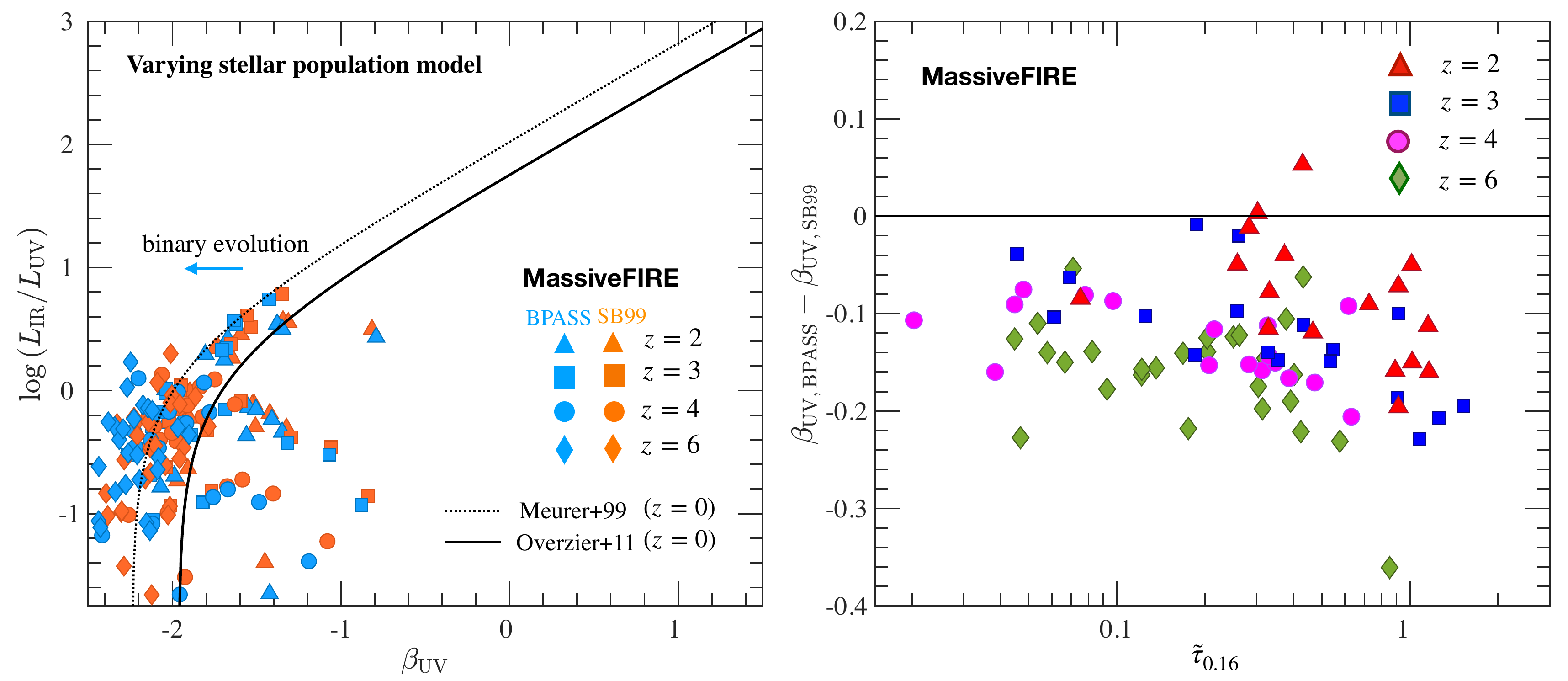}
 \caption{\textit{Left} panel: Comparison between the IRX-$\beta_{\rm UV}$ relation computed using the \textsc{bpass} (blue symbols) and the SB99 (orange symbols) stellar population models. The dotted and solid black lines indicate the local M99 and $\rm M99_{corr}$ relations.  \textit{Right} panel: The difference {in} $\beta_{\rm UV}$ produced by the two stellar population models as a function of $\tilde{\tau}_{0.16}$.} 
    \label{fig.22}
  \end{center}
  \vspace{-15pt}
\end{figure*}

\subsubsection{Binary evolution of stellar population}
\label{Sec:4e4}

\begin{figure}
 \begin{center}
 \includegraphics[width=83mm]{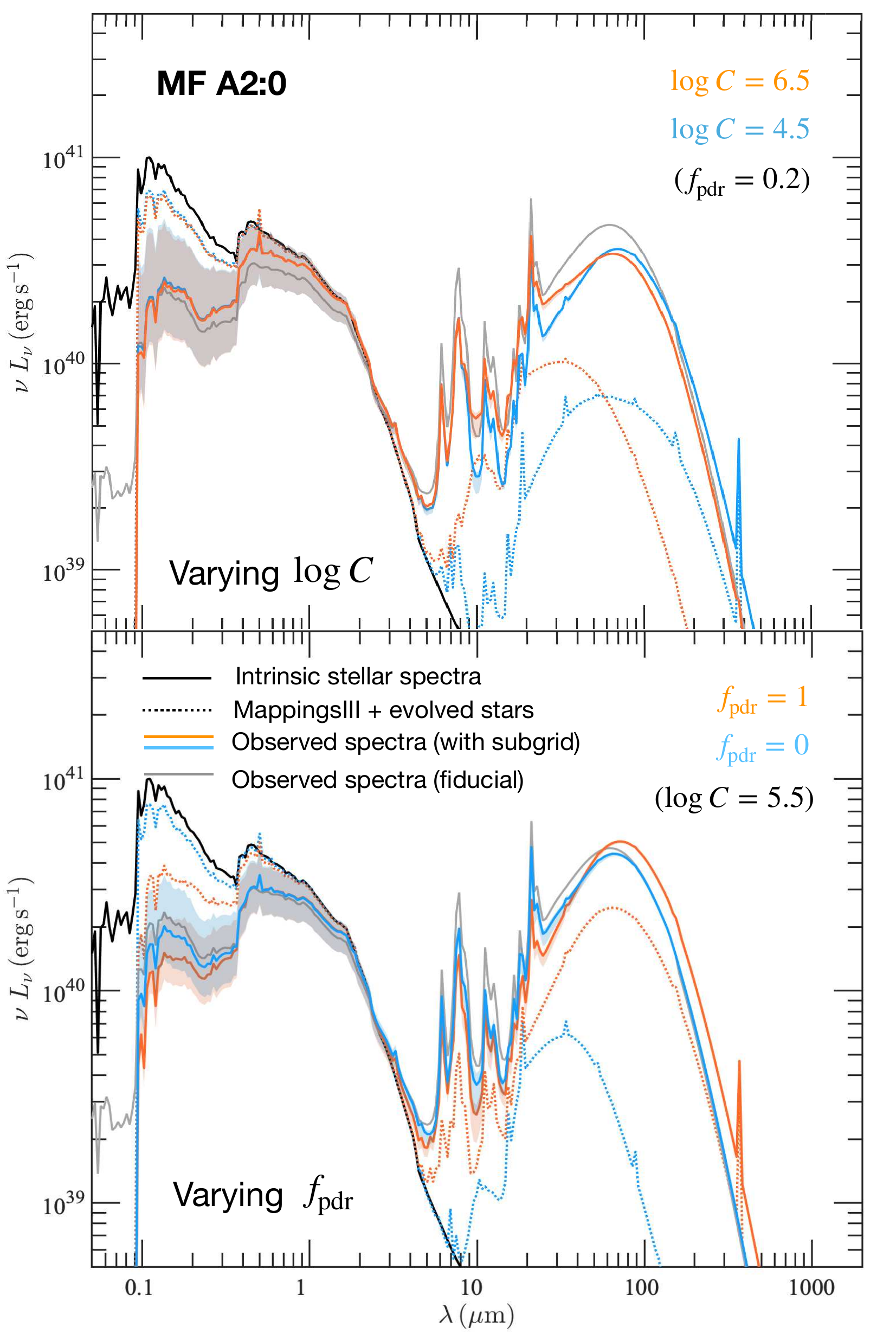}
 \caption{SEDs of a $z=2$ \textsc{\small MassiveFIRE} galaxy (galaxy ID: MF2:0) produced using different dust models. The coloured lines in the two panels show the results for the models where we adopt the subgrid prescription for the birth-clouds embedding the young ($t_{\rm age}\le 10$ Myrs) stars. {In the \textit{upper} panel, we show the result for ${\rm log}\,C=4.5$ (blue), ${\rm log}\,C=6.5$ (orange), with fixed $f_{\rm pdr}=0.2$. In the \textit{lower} panel, we show the observed spectra for $f_{\rm pdr}=0$ (blue), and $f_{\rm pdr}=1.0$ (orange) with fixed ${\rm log}\,C$ ($= 0.2$).} In each panel, black line shows the intrinsic stellar SED, while solid orange, blue and grey lines show the angle-averaged observed SEDs, each corresponding to a different dust model. Grey line corresponds to the fiducial RT model (without the subgrid prescription). Dotted lines (orange and blue) show the combined emission from the birth-clouds (pre-processed by the dust within the birth-clouds) and the evolved ($t_{\rm age}>10$ Myrs) stars (unattenuated by the diffuse dust component). The shaded orange and blue areas indicate the range of observed spectra spanning the different viewing directions for each model.}
    \label{fig.23}
  \end{center}
  \vspace{-10pt}
\end{figure}

\begin{figure}
 \begin{center}
 \includegraphics[width=85mm]{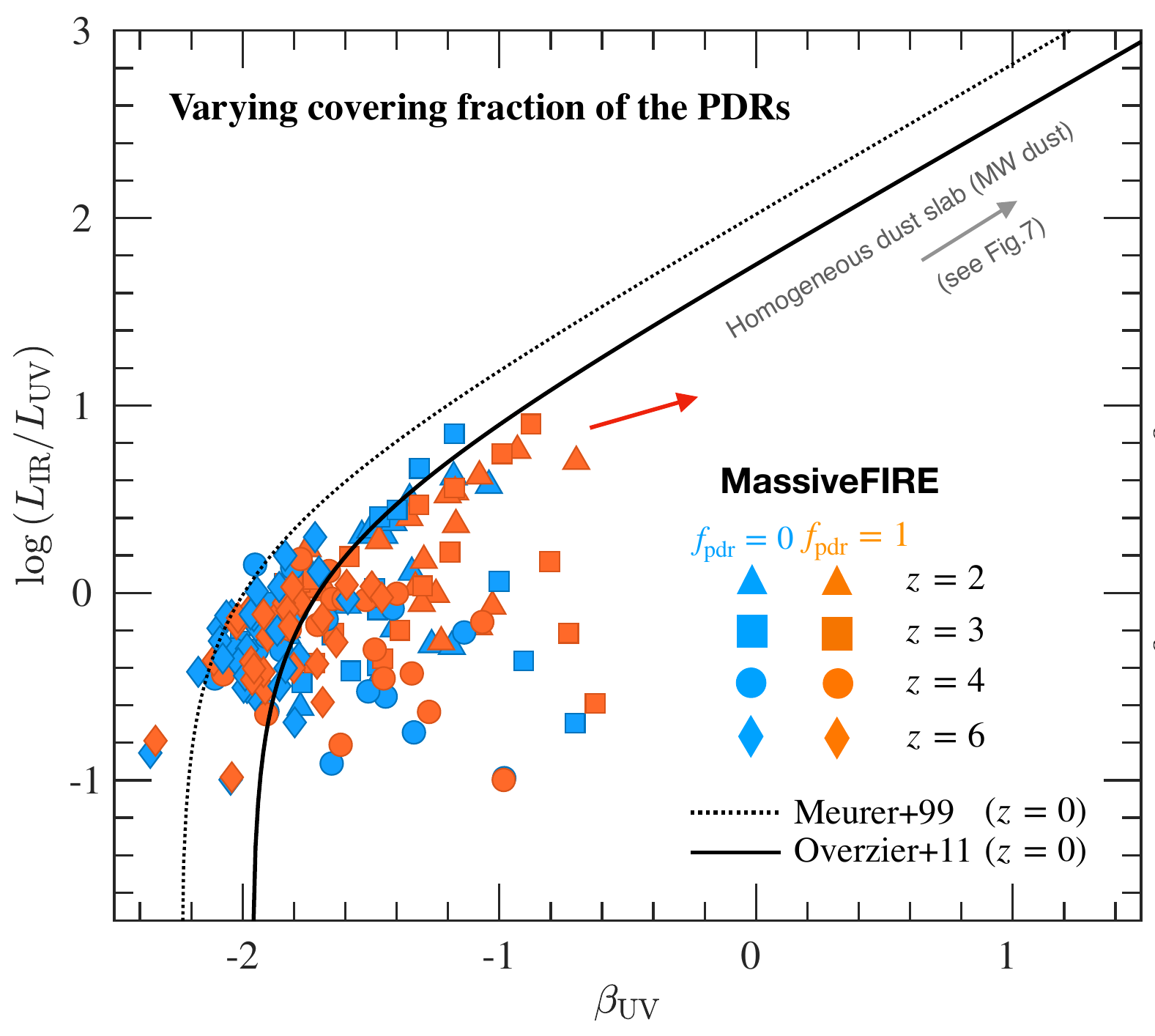}
 \caption{ The effect of the covering fraction of the PDR ($f_{\rm pdr}$) on the IRX-$\beta_{\rm UV}$ relation. Orange and blue symbols correspond to the cases of  $f_{\rm pdr}=1.0$ and $f_{\rm pdr}=0$, respectively.  The dotted and solid black black lines represent the local M99 and $\rm M99_{corr}$ relations, respectively. The red arrow marks the direction of change in the IRX-$\beta_{\rm UV}$ relation of galaxies with IRX$\simgreat1$ by increasing $f_{\rm pdr}$. The direction appears to be shallower than the analytic curve of the dust slab model, indicating the attenuation curve of galaxies become steeper (more SMC-like) with increasing $f_{\rm pdr}$.}
    \label{fig.24}
  \end{center}
  \vspace{-15pt}
\end{figure}

So far we have presented the results of the analysis using the SB99 stellar population model, which accounts for the evolution of single stellar {populations}. Recently there has also been growing attention to the effect of binary evolution of stellar populations on galaxy SED \citep[\eg][]{Stanway_2016, Ma_2016b, Reddy_2018}. Observations of the {stars in the solar neighbourhood} have shown that a considerable fraction of massive stars reside in binary systems \citep[\eg][]{Raghavan_2010, Sana_2012, Duch_ne_2013, El_Badry_2018}. Processes such as mass transfer between binary stars and binary mergers may increase the number of high-mass stars and effectively boost the UV part of the stellar emission. 

Here we provide an estimate of the change in the IRX-$\beta_{\rm UV}$ relation resulting from the binary evolution of stars. Specifically, we show the result derived from the recently developed ``Binary Population and Spectral Synthesis" (\textsc{bpass}) SED template libraries \citep[v2.2;][]{Eldridge_2012, Eldridge_2017, Stanway_2018} and compare it with the fiducial model. The \textsc{bpass} libraries are tabulated by stellar age and metallicity for a number of different IMFs. We adopt the \textsc{bpass} templates for the \citet{Chabrier_2003} IMF with a cut-off mass of $100\,M_\odot$, which is the closest available match in the libraries to the Kroupa IMF that has been implemented into \textsc{\small MassiveFIRE}.

In Fig.~\ref{fig.22}, we show the IRX-$\beta_{\rm UV}$ relation produced using the two stellar population models in the \textit{left} panel, and also the difference in $\beta_{\rm UV}$ between the two models as a function of $\tilde{\tau}_{0.16}$ in the \textit{right} panel. The \textsc{bpass} templates {predict} slightly bluer $\beta_{\rm UV}$ than the SB99 templates {as expected}. The mean difference of $\beta_{\rm UV}$ between the two stellar population models is 0.12 and the difference increases slightly with redshift {because} i) higher-$z$ galaxies are on average more bursty and thus contain a higher fraction of young OB stars {and ii)} they are metal-poorer \citep{Eldridge_2017}. The IRX is almost identical between the two {SED} models because {IRX} depends {primarily} on dust {properties} rather than {the intrinsic emission from stars}.

\subsubsection{The subresolution structure of the birth-clouds}
\label{Sec:4e5}

As mentioned before, our simulation may only marginally resolve the typical scale of the birth-clouds embedding the young star clusters \citep{Jonsson_2010}. Therefore, in order to check the uncertainty arising from small-scale ISM structures in the birth-clouds, we have performed additional RT analysis as in \citet{Liang_2019}, where we include a subgrid model for the birth-clouds. We summarise the details of the subgrid model and the resulting uncertainties in the IRX-$\beta_{\rm UV}$ relation in this subsection.

In brief, all the young ($t_{\rm age}\le10$ Myrs) star particles of a galaxy are assigned a \textsc{mappingsiii} source SED \citep{Groves_2008}. The \textsc{mappingsiii} SED templates are parametrized by the SFR and the metallicity of the star-forming regions, the pressure of the ambient ISM, the H \textsc{\small ii} region compactness (${\rm log}\,C$), and the covering fraction of the associated {photodissociation regions (PDR)}  ($f_{\rm pdr}$). The PDRs in the \textsc{mappingsiii} model are defined to have a hydrogen column depth of $10^{22}\,\rm cm^{-2}$ \citep{Jonsson_2010} based on both the observational and theoretical grounds \citep{Larson_1981, Solomon_1987, Rosolowsky_2003}.

In Fig.~\ref{fig.23} we show how the overall SED of galaxy depends on $f_{\rm pdr}$ (\textit{upper} panel) and ${\rm log}\,C$ (\textit{lower} panel). From the \textit{upper} panel, we can see that by changing ${\rm log}\,C$ alone barely affects the UV-to-optical part of the SED since the dust optical depth of the birth-clouds is unaffected. It does, however, affect the temperature distribution of dust within the birth-clouds and thus the dust re-emission at FIR \citep{Groves_2008}. A higher ${\rm log}\,C$ leads to a warmer dust SED shape of galaxy. The integrated IR luminosity as well as the overall IRX of galaxy, however, do not depend on ${\rm log}\,C$ since IRX well correlates with dust optical depth (Section~\ref{Sec:4b}), the latter being independent of ${\rm log}\,C$. Therefore, ${\rm log}\,C$ does not affect the IRX-$\beta_{\rm UV}$ relation of galaxies. 

On the other hand, a higher $f_{\rm pdr}$ leads to an increase in the global effective optical depth of galaxy (as is shown in the \textit{lower} panel of Fig.~\ref{fig.23}) and thus an increase in both $\beta_{\rm UV}$ and IRX. We show in Fig~\ref{fig.24} the IRX-$\beta_{\rm UV}$ relation of the \textsc{\small MassiveFIRE} galaxies for the cases of $f_{\rm pdr}=0$ (indicating that H \textsc{\small ii} regions are uncovered by the PDRs) and $f_{\rm pdr}=1.0$ (indicating that the PDRs entirely surround the H \textsc{\small ii} regions in the birth-clouds). The mean $\beta_{\rm UV}$ (IRX) of our sample is higher by 0.09 (0.02 dex) with $f_{\rm pdr}=1.0$.

We also note that the direction of change in the IRX-$\beta_{\rm UV}$ relation with $f_{\rm pdr}$ appears to be shallower than the slope of the analytic curve of the dust slab model, indicating a steeper (more SMC-like) attenuation curve with increasing $f_{\rm pdr}$. This is not surprising because a higher $f_{\rm pdr}$ means that a larger fraction of the ISM dust is associated to the birth-clouds. The hard UV photons emitted from the young stars therefore get more attenuated. This is in contrast with the scenario of increasing $\delta_{\rm dzr}$, where the dust mass evenly increases at any place within the galaxies but \textit{without} changing the dust-to-star geometry. In that case, the attenuation curve becomes shallower (`grayer') and the change of direction in the IRX-$\beta_{\rm UV}$ plane with increasing $\delta_{\rm dzr}$ becomes steeper than the analytic curve of the toy model (Section~\ref{Sec:4e3}).

\begin{figure*}
 \begin{center}
 \includegraphics[width=153mm]{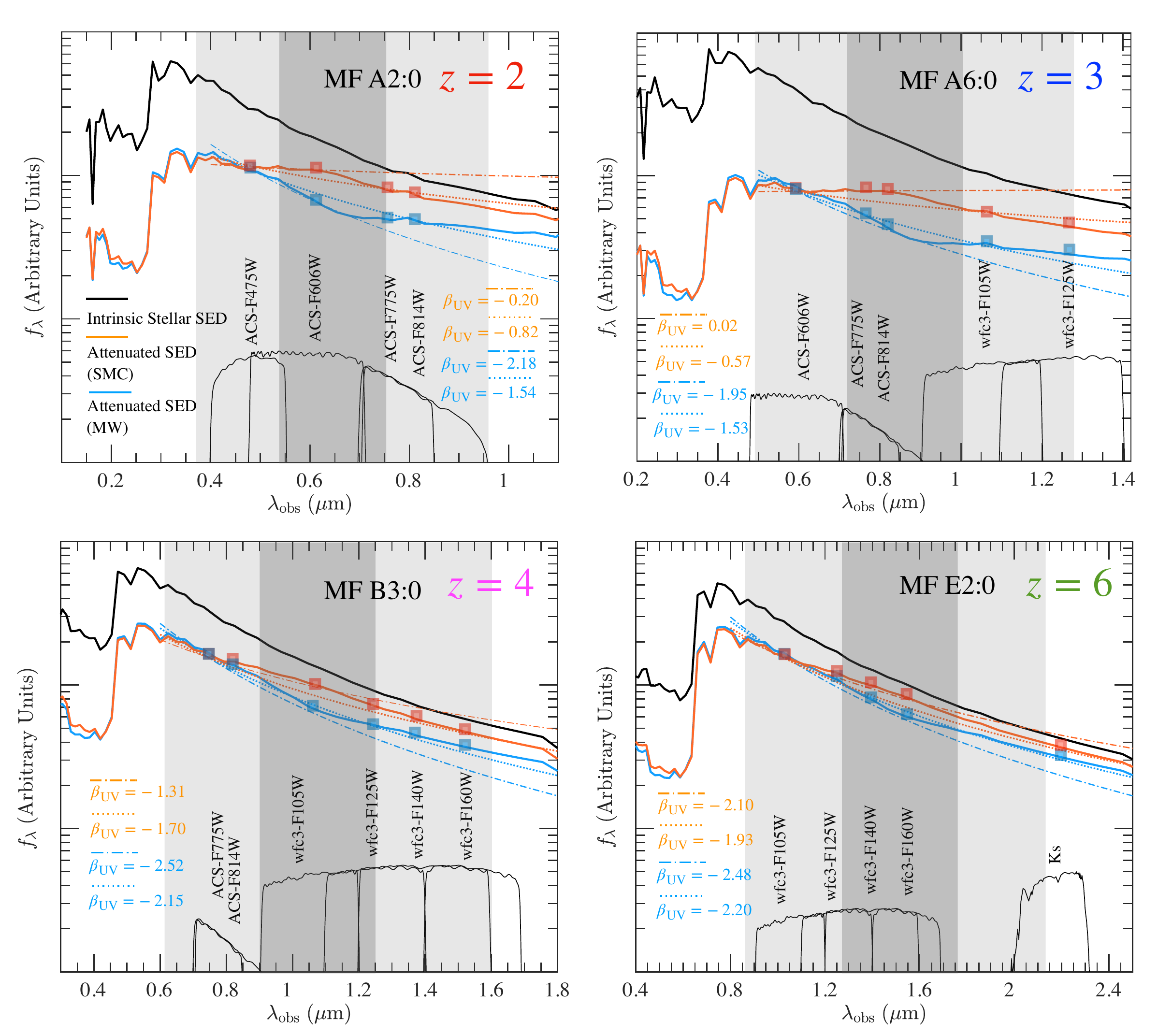}
 \caption{The coverage of the rest-frame UV wavelength regime by different filters for redshifts running from $z=2$ to $z=6$ (adapted from Fig. 9 of \citealt{Popping_2017}). The photometric sampling includes the \textit{HST} filters as part of the 3D-HST survey, on top of \textit{GALEX}, and a $\rm K_s$-band filter. In each panel, the thick solid black curve shows the intrinsic stellar SED at FUV-to-NUV of a \textsc{\small MassiveFIRE} galaxy at the corresponding redshift. The blue and orange lines in each panel show the attenuated spectra for the {MW and SMC dust}, respectively. The photometric data points of the attenuated spectra are mark by coloured squares. The relevant filters at each redshift and their transmission functions are plotted below the galaxy spectra. In each panel, the coloured dotted and dot-dashed lines show the best-fit power-law curves (for estimating $\beta_{\rm UV}$) to the two different combinations of photometric data points. From $z=2$ to $z=6$, the dotted (dot-dashed) line in the panel shows the result of the filter combinations of F475W+F775W (F606W), F606W+F105W (F814W), F775W+F125W (F105W) and F105W+$\rm K_s$ (F160W). The light grey areas indicate the wavelength range ($1230<\lambda<3200\,\angstrom$) where the photometric data points are used for estimating $\beta_{\rm UV}$ by the observations, while the dark grey areas show the regime ($1700<\lambda<2700\,\angstrom$) of the `bump' feature in the MW extinction curve.}
    \label{fig.25}
  \end{center}
  \vspace{-15pt}
\end{figure*}

\begin{figure}
 \begin{center}
 \includegraphics[width=85mm]{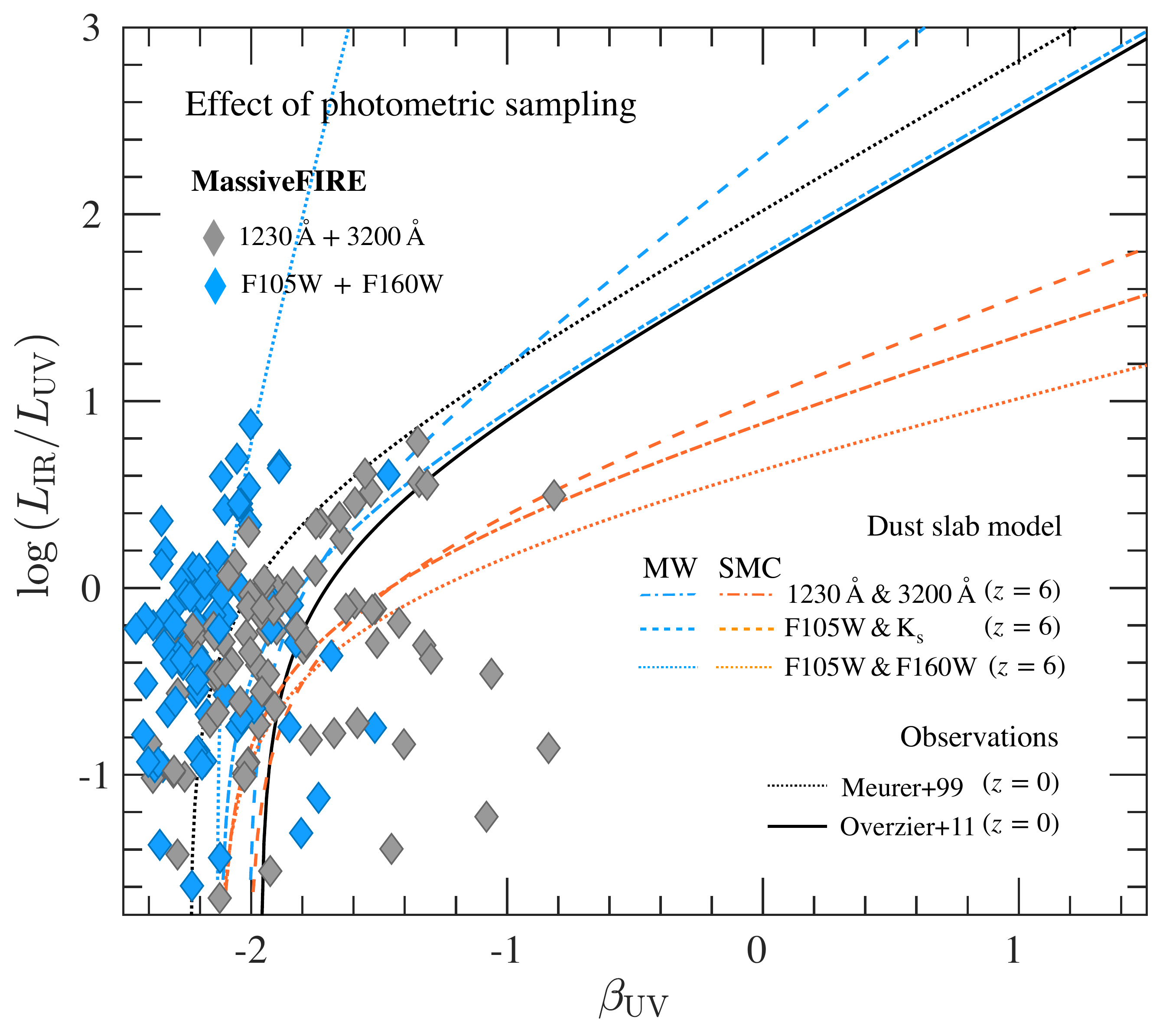}
 \caption{ The effect of photometric sampling on the derived IRX-$\beta_{\rm UV}$ relation. The grey symbols represent the result of the \textsc{\small MassiveFIRE} sample for MW dust with $\beta_{\rm UV}$ measured at $\lambda=1230$ and $3200\,\angstrom$. The blue symbols show the result with $\beta_{\rm UV}$ estimated using the filter combination of \textit{HST} F105W and F160W bands, as if the galaxies were all at $z=6$. The blue and orange lines show the analytic curves derived from the dust slab model for the MW and SMC dust of WD01, respectively. The dotted, dash-dotted and solid lines correspond to the cases where $\beta_{\rm UV}$ of the attenuated spectra (redshifted to $z=6$) is estimated using the filter combination of F105W+F160W bands, F105W+$\rm K_s$ bands and at rest-frame $1230\,\angstrom$ and $3200\,\angstrom$, respectively. All these curves are produced using the SB99 template SEDs for the case of $t_{\rm age}=50$ Myrs and $Z_*=0.1Z_\odot$. The black solid and dotted lines represent the M99 and $\rm M99_{corr}$ relations, respectively. Photometric sampling affects the measured IRX-$\beta_{\rm UV}$ relation of galaxies.}
   \vspace{-10pt}
    \label{fig.26}
  \end{center}
  \vspace{-10pt}
\end{figure}

\section{The deviation of the IRX-$\beta$ relation due to the observational effects}
\label{Sec:5}

In Section~\ref{Sec:4}, we explored the various sources of the \textit{intrinsic} scatter {of} the IRX-$\beta_{\rm UV}$ relation and quantified their relative contribution. However, measurements of $\beta_{\rm UV}$ and IRX of the distant galaxies can be uncertain due to different observational effects. Specifically, we will examine the uncertainties in the $\beta_{\rm UV}$ measurements due to different photometric samplings in Section~\ref{Sec:5a}. We will also discuss in Section~\ref{Sec:5b} the uncertainties in the `dust temperature' (or SED shape) used to infer $L_{\rm IR}$ of high-$z$ galaxies {by the observations, which results from the common dearth of photometric data points in the dust SEDs at high-$z$}. Finally, in Section~\ref{Sec:5c}, we review recent observational constraints {of} the IRX-$\beta_{\rm UV}$ relation for a sample of LBGs at $z\simgreat5$ in more detail.

\subsection{Measuring $\beta_{\rm UV}$ using broadband photometry}
\label{Sec:5a}

In the observational studies, $\beta_{\rm UV}$ are commonly estimated by fitting power-law SEDs to multi-band photometry within the wavelength range $1230<\lambda<3200\,\angstrom$. The true SED shape within this wavelength range, however,  may not be {well described by a} power law {(particularly, the 2175$\,\angstrom$ `bump' feature)}, and thus, the derived $\beta_{\rm UV}$ can depend on the {photometric} sampling. A poor sampling of UV photometry can result in {non-trivial offset of} the derived $\beta_{\rm UV}$ \citep[\eg][]{Popping_2017}. 

This issue is more important for intermediate- and high-$z$ observations, where galaxies often have few photometric data points at rest-UV \citep{Reddy_2018, Alvarez_Marquez_2019}. For instance, the $\beta_{\rm UV}$ of {the} intermediate-$z$ ($1.5\simless z\simless4$) samples {by \citet{Casey_2014} \& \citet{Wang_2018}} are estimated based on 3 photometric data points on average and the $\beta_{\rm UV}$ of the sources at higher-$z$ ($z\simgreat5$) are mostly derived based on two or three data points (see Table~\ref{T3} for a summary). 
 
In Fig.~\ref{fig.25}, we show an example of the coverage of the rest-UV wavelength range by different filters at different redshifts ($z=2$, 3, 4 and 6, corresponding to the four panels)\footnote{This figure is adapted from Fig. 9 of \citet{Popping_2017}. Note that the authors also present other filter combinations (see the appendix of that paper). In this work, we only use one combination as an example for the impact of photometric sampling {on} $\beta_{\rm UV}$ estimates.}. The photometric sampling we use here includes the \textit{HST} filters that were used for the 3D-HST survey \citep{Skelton_2014}, combined with \textit{GALEX}, and a $\rm K_s$-band filter. In each panel, we show the intrinsic (thick black line) and the dust-attenuated SEDs (solid blue and orange lines) of a selected \textsc{\small MassiveFIRE} galaxy at the corresponding redshift. We present the attenuated SEDs for both the MW (blue lines) and the SMC (orange lines) {dust models}. The photometric data points are calculated by convolving the attenuated SEDs with the transmission functions of the filters, which are shown below the galaxy spectra in each panel (thin black line). We also show in each panel the best-fit power-law curves obtained using different sampling of photometric data points (coloured dotted and dot-dashed lines). The dot-dashed curves correspond to the case where the two data points are both blueward of the `bump' at $\lambda=2175\,\angstrom$, whereas the dotted curves correspond to the case where the two data points are on either side of the `bump'.

Looking at the figure, it is clear that galaxy spectra are not pure power laws at rest-UV wavelengths. The spectra show a deficit of flux at $1700<\lambda<2700\,\angstrom$ (indicated by dark grey area) due to the enhanced extinction near the $2175\,\angstrom$ `bump'. $\beta_{\rm UV}$ derived with and without the data points in this regime have noticeable {differences}. We show in each panel of the figure the $\beta_{\rm UV}$ derived {using two photometric samplings} for each spectrum, one with two photometric data points lying both blueward of the `bump' and the other with the two data points lying on either side of the peak and outside the `bump' regime (dark grey area). For MW (SMC) dust, the former sampling yields noticeably bluer (redder) $\beta_{\rm UV}$.

The difference in $\beta_{\rm UV}$ {for the} different photometric {samplings increases} with larger $\tilde{\tau}_{0.16}$, {in line with the prediction} of the dust slab model. Specifically, we can rewrite Eq.~\ref{eq.15} for two different photometric {samplings}, \ie~

\begin{align}
\beta_{\rm UV,\,1} &= \beta_{\rm UV,\,0} + \mathcal{Z}_1\,\tilde{\tau}_{0.16}, \\
{\rm and}\;\;\;\;&\nonumber \\
\beta_{\rm UV,\,2} &= \beta_{\rm UV,\,0} + \mathcal{Z}_2\,\tilde{\tau}_{0.16}. 
\label{eq.24} 
\end{align}
where $ \mathcal{Z}_1$ and $ \mathcal{Z}_2$ represent the steepness of the \emph{same} attenuation curve and $\beta_{\rm UV,\,1}$ and $\beta_{\rm UV,\,2}$ represent the UV spectral slopes that are measured by two different combinations of bandpass filters. By subtracting one equation by the other, we obtain

\begin{equation}
\beta_{\rm UV,\,1} - \beta_{\rm UV,\,2} =  (\mathcal{Z}_1-\mathcal{Z}_2)\,\tilde{\tau}_{0.16},
\label{eq.25}
\end{equation}

\noindent {\ie.} the difference between the estimated $\beta_{\rm UV}$ {scales linearly with} $\tilde{\tau}_{0.16}$. {Since} IRX {increases with} $\tilde{\tau}_{0.16}$, $\beta_{\rm UV}$ measurements of galaxies of higher IRX are expected to be more influenced by the photometric sampling effect. 

In Fig.~\ref{fig.26}, we explicitly show the impact of the photometric {samplings} on the derived IRX-$\beta_{\rm UV}$ relation using both the analytic solutions of the dust slab model {and RT calculations} of the \textsc{\small MassiveFIRE} galaxies {for $z=6$}. {For the analytic solutions, we show the results for both the MW and SMC dust models and for three filter combinations, 1230+$3200\,\angstrom$, F105W+$\rm K_s$ and F105W+F160W. For the RT calculations, we only present the results for MW dust and for the filter combinations of 1230+$3200\,\angstrom$ and F105W+F160W. The latter has both filters blueward of the $2175\,\angstrom$ `bump', resulting in the lowest `measured' $\beta_{\rm UV}$ among the three samplings (see the \textit{lower right} panel of Fig.~\ref{fig.25}).}

The most significant uncertainty in the derived IRX-$\beta_{\rm UV}$ relation is the effect of the $2175\,\angstrom$ `bump' feature in the MW extinction curve. {With F105W+F160W}, the derived $\beta_{\rm UV}$ can be much `bluer' than the cases where both sides of the peak are covered {(\eg~1230+$3200\,\angstrom$)}. We will show in Section~\ref{Sec:5c} that {the location} of some \textsc{\small MassiveFIRE} galaxies {in the IRX-$\beta_{\rm UV}$ plane derived using the F105W+F160W filters (blue filled symbols) is in good agreement with a few detected LBGs at $z\simgreat5$, which appears to be significantly `bluer' than the M99 relation.} On the other hand, photometric sampling appears to be a less important issue for SMC dust due to the absence of the $2175\,\angstrom$ `bump' feature. Using the F105W+F160W filter combination leads to a slightly redder $\beta_{\rm UV}$ in this case.

The results we present here are derived using the local MW and SMC dust curves of the WD01 model. At higher redshifts, the shape of the dust extinction curve and, in particular, the strength of the $2175\,\angstrom$ `bump' feature are poorly constrained, although there has been preliminary evidence showing that high-$z$ galaxies tend to exhibit weaker `bump' than the MW curve \citep[\eg][]{Schady_2012, Kriek_2013, Zafar_2018}. Therefore, the significant offset of $\beta_{\rm UV}$ presented here {should} be viewed as a conservative maximum estimate. 

\begin{figure*}
 \begin{center}
 \includegraphics[width=170mm]{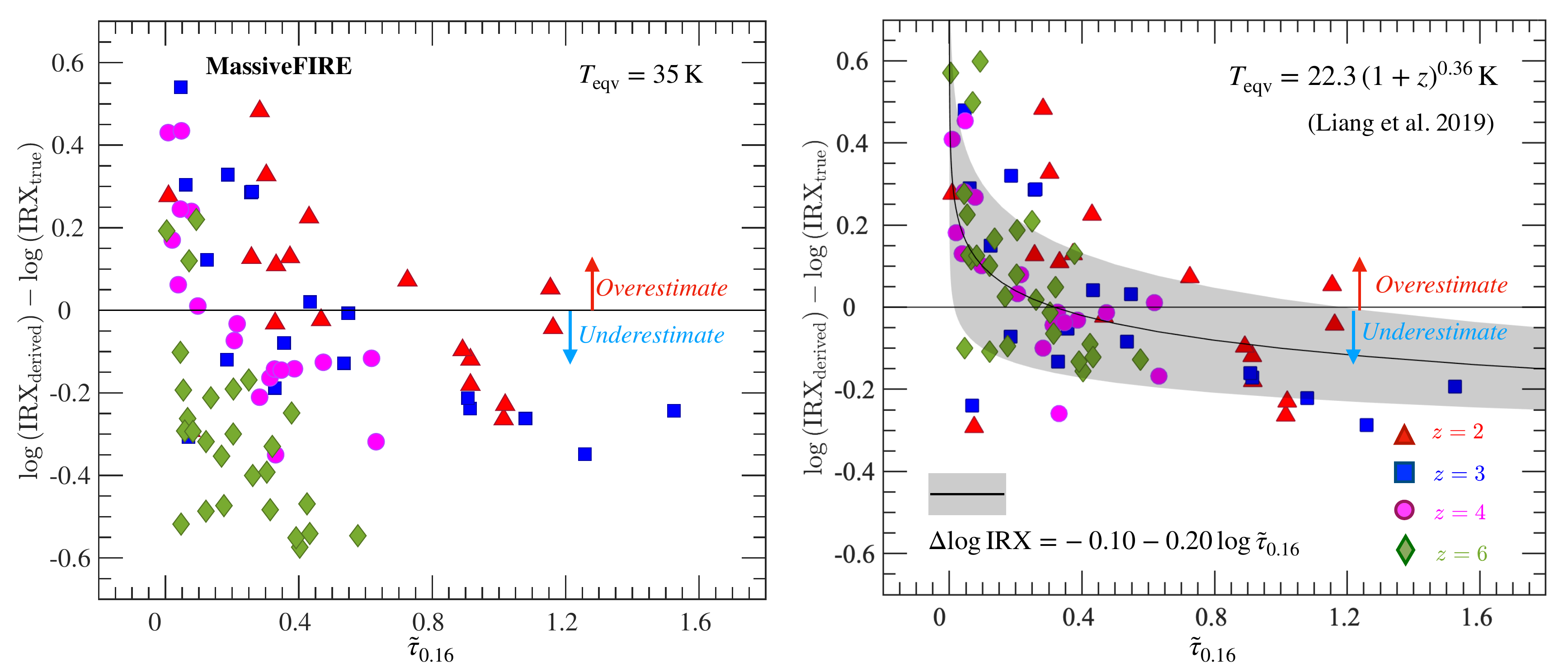}
 \caption{ The offset between IRX derived from converting ALMA band 6 ($\rm 1.1\,mm$) flux densities and the true IRX of galaxies as a function of $\tilde{\tau}_{0.16}$. In the \textit{left} panel, we show the result for the case where we adopt a MBB function (using $\beta_{\rm IR}=2.0$) with a constant $T_{\rm eqv}=35$ K submm-flux-to-IR-luminosity conversion. In the \textit{right} panel, we show the result when a redshift-dependent $T_{\rm eqv}$ following Eq.~\ref{eq.26} (using $\delta_{\rm dzr}=0.4$) is employed for the conversion. The grey semi-transparent area indicates the $3\sigma$ probability contour of the distribution of the data. The required $T_{\rm eqv}$ appears to show an additional positive correlation with $\tilde{\tau}_{0.16}$ at fixed redshift. }
    \label{fig.27}
  \end{center}
  \vspace{-15pt}
\end{figure*}

\subsection{The uncertainties in the `dust temperature'}
\label{Sec:5b}

Estimating $L_{\rm IR}$ (and hence IRX) of galaxies {at $z\simgreat5$ reliably} can be challenging \citep{Casey_2012, Liang_2019} because source detection at this epoch is difficult due to the high confusion {noise}. The majority of objects with detected dust emission at this epoch {have} only one or two photometric data points derived with ALMA (typically at band 6 or 7). To extrapolate $L_{\rm IR}$ of these galaxies, often {a specific} dust SED {shape is adopted and the `dust temperature' is set to a specific value (35 K)} \citep{Bouwens_2016, Liang_2019}. The derived IRX thus strongly depends on this assumed `dust temperature'. 

Recently, there has been growing evidence that galaxies at high-$z$ (\ie,~$z\simgreat5$) have higher `dust {temperatures}' than those at low and intermediate redshifts \citep[\eg,][]{Capak_2015, Bouwens_2016, Matthee_2017, Harikane_2020}. Specifically, it has been found that applying a constant `dust temperature' of 35 K together with a {MBB} function will lead to a significant IRX deficit of high-$z$ galaxies, resulting in galaxies far below the canonical M99 relation. Aside from having steeper dust attenuation curves, these galaxies may have a higher dust temperature so that the observed submm flux is lower for a given $L_{\rm IR}$.

Motivated by these observations, a number of recent theoretical studies have investigated the `dust temperature' of high-$z$ galaxies in detail. The dust SED shape of high-$z$ galaxies is found to {differ} noticeably from the template SEDs commonly adopted in the literature \citep[\eg][]{Liang_2019, Ma_2019}. Specifically, high-$z$ galaxies show more prominent emission on the Wien side of the dust SED compared to the low-$z$ counterparts in the cosmological `zoom' galaxy simulations. This part of the SED is associated with the warm dust component that is exposed to the hard UV radiation from the newly born young stars \citep{Casey_2012}. The more prominent emission of the warm dust component in high-$z$ galaxies can be attributed to enhanced star formation {activity} \citep{Safarzadeh_2017b, Liang_2019, Ma_2019} and/or a higher mean dust column density in vicinity of young stars \citep{Behrens_2018, Sommovigo_2020} in high-$z$ galaxies. Although the Wien side of the dust SED of galaxies at $z\simgreat5$ is typically not constrained by observations, a number of low- and intermediate-redshift observations with \textit{Herschel} have found a trend of enhanced mid-IR emission with increasing redshift based on stacking analysis \citep[\eg,][]{Bethermin_2015, Casey_2018b, Schreiber_2018}.

Hence, in order to account for this evolution of SED shape with redshift, an increase {in} `dust temperature' is needed {to convert} ALMA flux densities to $L_{\rm IR}$ with a {MBB} function. Using the \textsc{\small MassiveFIRE} galaxy sample at $z=2-6$, \citet{Liang_2019} derived the best-fit formula for this `equivalent' dust temperature {(see Eq.9 of \citealt{Liang_2019} for its definition)} using redshift and $\delta_{\rm dzr}$ as variables, \ie,

\begin{equation}
	T_{\rm eqv} = T_0\,(1+z)^\alpha (\delta_{\rm dzr}/0.4)^{\gamma}.\;\;\;\rm (L19)
	\label{eq.26}
\end{equation}

\begin{figure*}
 \begin{center}
 \includegraphics[width=170mm]{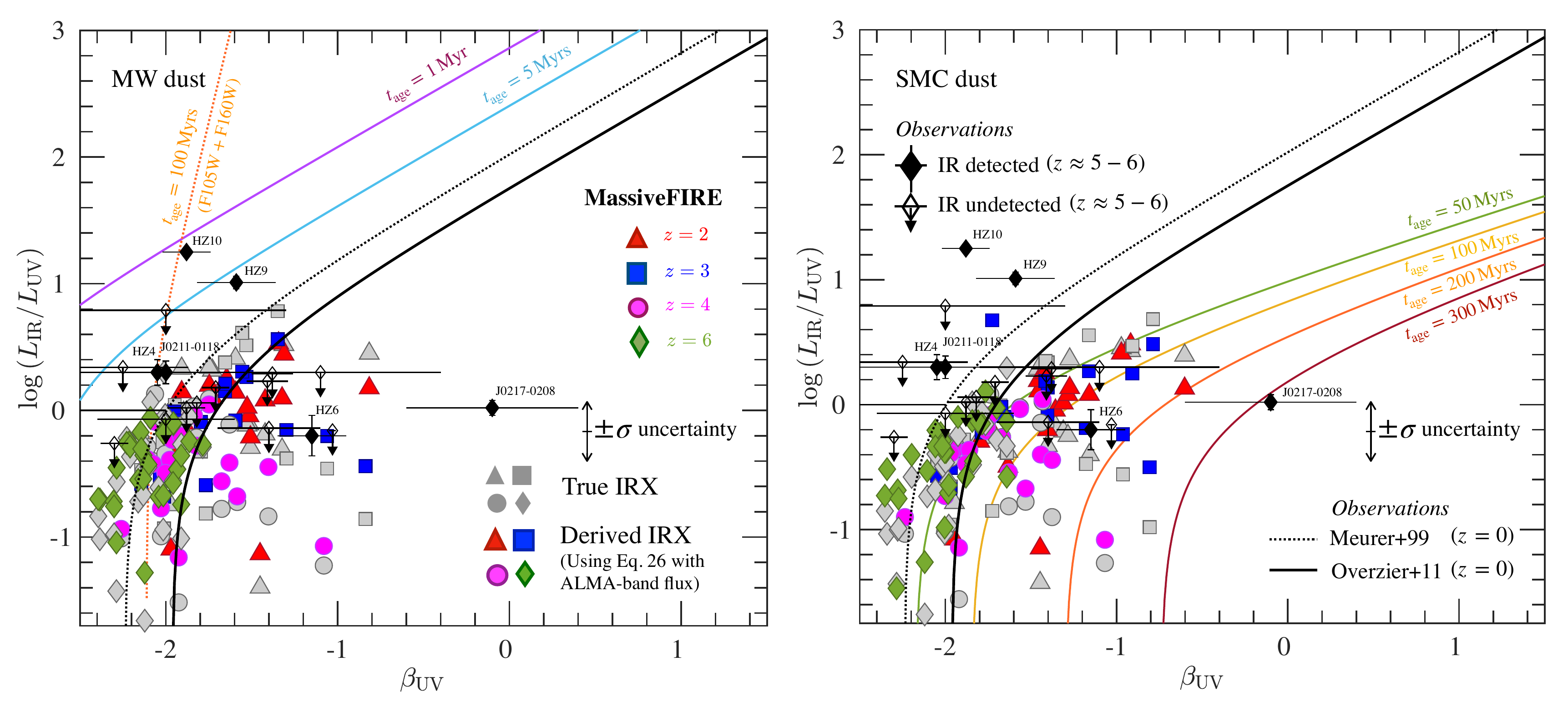}
 \caption{Comparison of the IRX-$\beta_{\rm UV}$ relation of the \textsc{\small MassiveFIRE} galaxies to the compilation of the recent observational data at $z\approx5-6.5$. The $L_{\rm IR}$ of the observational data are converted from their reported ALMA-band fluxes (either band 6 or 7) using a {standard MBB} function (assuming $\beta_{\rm IR}=2.0$) with $T_{\rm eqv}$ following Eq.~\ref{eq.26} (assuming $\delta_{\rm dzr}=0.4$). For undetected sources, downward arrows mark the $3\sigma$ upper limits of their IRX. The coloured filled symbols represent the result for \textsc{MassiveFIRE} derived using the same method as the observational data. For purpose of reference, the result for the true IRX is shown by grey symbols. In the \textit{left} panel, the violet and blue lines indicate the analytic curves of the dust slab model for {MW dust} and stellar age $t_{\rm age}=1$ and 5 Myrs, respectively. For these two curves, $\beta_{\rm UV}$ is measured at $\lambda=1230$ and $3200\,\angstrom$. The orange dotted line represents the case where $t_{\rm age}=100$ Myrs while $\beta_{\rm UV}$ is measured at the \textit{HST} F105W and F160W bands for the attenuated spectra (redshifted to $z=6$), both being blueward of the $2175\,\angstrom$ `bump' for $z=6$. In the \textit{right} panel, the green, brown, orange and dark red lines represent the analytic curves for SMC dust curve and $t_{\rm age}=50$, 100, 200 and 300 Myrs, respectively. $\beta_{\rm UV}$ is measured at $\lambda=1230$ and $3200\,\angstrom$. In both panels, the dotted and solid black lines indicate the local M99 and $\rm M99_{corr}$ relations, respectively.  }
    \label{fig.28}
  \end{center}
  \vspace{-15pt}
\end{figure*}

\noindent The `dust temperature' for inferring $L_{\rm IR}$ depends on the assumed functional shape of the dust SED and the observing bandpass, and it does not reflect the \textit{physical} temperature of the bulk of the ISM dust \citep[see][for the details]{Liang_2019}. For ALMA band 7 (6) fluxes, the best-fit parameter values are $T_0=24.4$ (22.3) K, $\alpha=0.31$ (0.36) and $\gamma=-0.13$ $(-0.15)$. The anti-correlation with $\delta_{\rm dzr}$ indicates that an increase {in} $\delta_{\rm dzr}$ leads to a reduced mass \textit{fraction} of the dust in the ISM {being} exposed to the hard UV photons from the young stars and hence less prominent emission on the Wien side {\citep{Scoville_2013, Faisst_2017, Liang_2019}}. The $\delta_{\rm dzr}$ of high-$z$ galaxies is not yet well constrained observationally. 

Using Eq.~\ref{eq.26} can mitigate the systematic underestimates of IRX at higher-$z$ when only ALMA data is available, considering that the redshift can often been determined through spectroscopy or photometry. We illustrate this in Fig.~\ref{fig.27}, where we show the difference between the IRX converted from the single-ALMA-band flux and the true value, $\rm \Delta\,(\rm log\, IRX)$, as a function of $\tilde{\tau}_{0.16}$ for the \textsc{\small MassiveFIRE} {sample} at different redshifts. We show the results for the cases where we adopt a constant $T_{\rm eqv}=35$ K and the redshift-dependent $T_{\rm eqv}$ following Eq.~\ref{eq.26} in the \textit{left} and \textit{right} panels, respectively. It can be seen that by using $T_{\rm eqv}=35$, the IRX {is systematically underestimated} {for} galaxies above $z=2$. The IRX of $z=6$ galaxies (green diamonds) is underestimated by $\sim0.4$ dex (a factor of $\sim3$). {$T_{\rm eqv}$ based on} Eq.~\ref{eq.26} can reduce the systematic error in the IRX estimates (\textit{right} panel). We note that {Eq.~\ref{eq.26}} is in good agreement with estimates based on source number counts in recent deep ALMA surveys \citep[\eg,][]{Bouwens_2016, Casey_2018b, Casey_2018}.

However, we can see from the \textit{right} panel that the scatter in $\Delta\,\rm log\,(IRX)$ is non-trivial, even after accounting for the systematic redshift evolution of $T_{\rm eqv}$. The $1\sigma$ dispersion of $\Delta\,\rm log\,(IRX)$ is 0.22 (corresponding to a factor of $\sim1.67$ in IRX). This non-trivial scatter is due to the variations in the dust SED shape at fixed redshift, which {are} not well accounted for by a redshift-dependent formula for $T_{\rm eqv}$. This is not surprising because galaxies at the same redshift have large dispersion in the starburstiness {\citep[\ie~sSFR, see \eg][]{Rodighiero_2011, Sparre_2017, Feldmann_2017b}} and very different spatial configuration of dust distribution near the UV-emitting OB stars {\citep[\eg][]{Lombardi_2014, Faisst_2017, Cochrane_2019, Sommovigo_2020}}. The prominence of the emission of the warm dust component at rest-MIR can thus have large {galaxy-to-galaxy} variations at a given redshift.

We expect that galaxies with higher $\tilde{\tau}_{0.16}$ show more prominent emission on the Wien side of the dust SED {when $\delta_{\rm dzr}$ is fixed}. Galaxies {with} higher $\tilde{\tau}_{0.16}$ tend to be more bursty {(Fig.~\ref{fig.13}, and see Section~\ref{Sec:4c})}, leading to a stronger emission {from} warm dust {component}. From the \textit{right} panel of Fig.~\ref{fig.27}, we {indeed} see a clear anti-correlation between $\rm \Delta\,(\rm log\, IRX)$ and $\tilde{\tau}_{0.16}$, indicating that a higher $T_{\rm eqv}$ is needed for recovering the $L_{\rm IR}$ of galaxies with higher $\tilde{\tau}_{0.16}$. Hence, if $\tilde{\tau}_{0.16}$ were known, it could be used to further improve the accuracy of the IRX estimate. 

Using the least-$\chi^2$ method, we derive the best-fit relation between $\Delta \rm log\,(IRX)$ and $\tilde{\tau}_{0.16}$ {for} the \textsc{\small MassiveFIRE} sample at $z=2-6$ over the range of $\tilde{\tau}_{0.16}=0.01-1.6$, 

\begin{equation}
\Delta \rm log\,(IRX) = -0.10\pm0.04-(0.20\pm0.04)\;{\rm log}\,\tilde{\tau}_{0.16}.
\label{eq.27}
\end{equation}

\noindent {Knowing that the derived IRX is proportional to $(T_{\rm eqv})^{3+\beta}$ for a given ALMA broadband flux (at band 6 or 7) in the Rayleigh-Jeans regime (see Eq. 10 of \citealt{Liang_2019}), we can then translate the above equation to a modified version of Eq.~\ref{eq.26}, including a correction term of $\tilde{\tau}_{0.16}$, \ie~}

\begin{align}
&T_{\rm eqv} = T_0\,(1+z)^\alpha (\delta_{\rm dzr}/0.4)^\gamma \,\Gamma, \nonumber \\
& \rm where\;\;\; \Gamma = 1.05\,(\tilde{\tau}_{0.16})^{0.04}.
\label{eq.29}
\end{align}

\vspace{-10pt}
\subsection{The IRX-$\beta$ relation at $z\simgreat5$}
\label{Sec:5c}

\begin{table*}
\caption{Observed properties of the galaxy compilation at $z\approx6$ that are are shown in Figure~\ref{fig.28}.}
\begin{threeparttable}
\begin{tabular}{ p{1.7 cm} p{0.8 cm}  p{1.4 cm} p{3.5 cm} p{1.4 cm}  p{1.5 cm}  p{1.6 cm} p{1.9 cm} }
 \hline
\multicolumn{1}{}{} \; ID \;\;\; & $z$ & $\beta_{\rm UV}$ & Rest-UV Photometry & ${\rm log}(L_{\rm UV}/L_\odot)$ & $S$ ($\rm \mu Jy$)\tnote{1}\; \tnote{2}  & ${\rm log}(L_{\rm IR}/L_\odot$)\tnote{1}\; \tnote{3}   & References\tnote{4}  \\
 \hline
 HZ8 &  5.1533 & $-1.41\pm0.12$ & F105W, F125W, F160W & $11.04\pm0.02$ & $<90$ (7)  &  $<11.26$ & C15, B17 \\
  \hline 
   HZ7 &  5.2532 & $-1.38\pm0.12$& F105W, F125W, F160W& $11.05\pm0.02$ & $<108$ (7) & $<11.33$ & C15, B17 \\
  \hline 
   HZ6 & 5.2928  & $-1.15\pm0.20$ & F105W, F125W, F160W  & $11.47\pm0.10$ & $129\pm36$ (7) & $11.07\pm0.12$ & C15, B17 \\
  \hline 
     HZ5 & 5.310  & $-1.03\pm0.02$ & F105W, F125W, F160W & $11.45\pm0.01$ & $<96$ (7) & $<11.28$ & C15, B17 \\
  \hline 
   HZ9 & 5.541 & $-1.59\pm0.23$ & F105W, F125W, F160W & $10.95\pm0.02$ & $516\pm42$ (7) & $11.96\pm0.04$ & C15, B17 \\
  \hline 
   HZ3 & 5.5416  & $-1.71\pm0.08$ & F105W, F125W, F160W & $11.08\pm0.01$ & $<93$ (7) & $<11.25$ & C15, B17 \\
  \hline 
   HZ4 &  5.544 & $-2.05\pm0.10$ & F105W, F125W, F160W & $11.28\pm0.01$ & $202\pm50$ (7)  & $11.58\pm0.10$ & C15, B17 \\
  \hline 
   HZ10 & 5.6566 & $-1.88\pm0.14$ & F105W, F125W, F160W & $11.14\pm0.02$ & $1261\pm44$ (7) & $12.39\pm0.02$ & C15, B17 \\
  \hline 
   HZ2 & 5.6597  & $-1.82\pm0.11$ & F105W, F125W, F160W & $11.15\pm0.01$ & $<87$ (7) & $<11.20$ & C15, B17 \\
  \hline 
   HZ1 &  5.6885 & $-1.88\pm0.11$ & F105W, F125W, F160W & $11.21\pm0.01$ & $<90$ (7) & $<11.22$ & C15, B17 \\
  \hline 
    A383-5.1 & 6.029  & $-2.0\pm0.7$ &  F814W, F110W, F125W, F160W, IRAC $3.6\rm\,\mu m$, $4.5\rm\,\mu m$ & $10.34\pm0.13$ & $<33$ (6) & $<11.12$ & R11, K16 \\
  \hline 
    J1211-0118 & 6.0293 & $-2.0\pm0.5$ & Subaru/HSC $z$ and $y$ bands & 11.43 & $348\pm72$ (7) & $11.73\pm0.09$ & H20 \\
  \hline 
    J0235-0532 &  6.0901  & $-2.6\pm0.6$ & Subaru/HSC $z$ and $y$ bands & 11.46 & $<162$ (7) & $<11.45$ & H20 \\
  \hline 
     J0217-0208 & 6.2037  & $-0.1\pm0.5$ & Subaru/HSC $z$ and $y$ bands & 11.63 & $310\pm42$ (7) & $11.65\pm0.06$ & H20 \\
  \hline 
    VR7 &  6.534 & $-1.4\pm0.3$ & F110W, F160W & $11.22\pm0.02$ & $<31.8$ (6) & $<11.07$ & M17, M19 \\
  \hline 
    MASOSA &  6.543 & $-1.1\pm0.7$ & F110W, F160W & $10.72\pm0.06$ & $<27.6$ (6)  & $<11.01$ & S15, M15, M17, M19 \\
  \hline 
   UVISTA-279127 &  6.58 & $-2.25\pm0.38$ & UVISTA $Y$, $J$, and $H$ bands & $11.37\pm0.06$ & $<138$ (6) & $<11.70$ & B18\\
  \hline 
    Himiko &  6.595 & $-2.0\pm0.4$ & F125W, F160W & 11.07 & $<27$ (6) & $<10.99$ & O13, C18\\
  \hline 
   CR7 & 6.604 & $-2.3\pm0.08$ &  UVISTA $Y$, $J$, $\rm K_s$-band & $11.15\pm0.04$ & $<21$ (6) & $<10.88$ & S15, M15, M17 \\
  \hline 
  \end{tabular}
  \begin{tablenotes}
 \footnotesize
   \item[1] \footnotesize{For the non-detections, we show the upper $3\sigma$ confidence limits.}
   \item[2] \footnotesize{{Bracketed numbers in this column (6 and 7) indicate the specific ALMA band at which the dust continuum was measured by the observations.}}
    \item[3] \footnotesize{$L_{\rm IR}$ is converted from $S$ using a {MBB} function (assuming $\beta_{\rm IR}=2.0$) with the `dust temperature' that follows Eq.~\ref{eq.26} (assuming $\delta_{\rm dzr}=0.4$).}
    \item[4]  \footnotesize{References: C15: \citep{Capak_2015}; B17: \citet{Barisic_2017}; R11: \citet{Richard_2011}; K16: \citet{Knudsen_2016}; H20: \citet{Harikane_2020}; M17, \citet{Matthee_2017}; M19: \citet{Matthee_2019}; S15: \citet{Sobral_2015}; M15: \citet{Matthee_2015}; B18, \citet{Bowler_2018}; O13: \citet{Ouchi_2013}; I16: \citet{Inoue_2016}; C18: \citet{Carniani_2018b}. }
\end{tablenotes}
\end{threeparttable}
   \label{T3}
\end{table*}

Fig.~\ref{fig.28} shows a compilation of ALMA-detected LBGs at $z\approx5-6.5$ (see Table~\ref{T3} for their observational properties). 6 out of 19 objects in this sample have $>3\sigma$ detection at either ALMA band 6 or 7. For the rest of the sample, we show in the figure their upper $3\sigma$ limits. We notice that $L_{\rm IR}$ of these galaxies {was originally} derived using different dust temperatures and dust {SEDs}. Hence, to make {a} fair comparison, we convert the reported ALMA-band fluxes to $L_{\rm IR}$ and hence IRX using the same {MBB} function (with $\beta_{\rm IR}=2.0$), constant $\delta_{\rm dzr}=0.4$ and {$T_{\rm eqv}$ from} Eq.~\ref{eq.26} (we still adopt Eq.~\ref{eq.26} instead of Eq.~\ref{eq.29} because $\tilde{\tau}_{0.16}$ is in practice very difficult to measure). We also derive $L_{\rm IR}$ and IRX of the \textsc{\small MassiveFIRE} {sample} by the same method. 

The observational data at $z\simgreat5$ {shows a} large scatter in the IRX-$\beta_{\rm UV}$ plane. While a few objects with ALMA detections (HZ10, HZ9, HZ4, and J0211-0118) appear to be much `bluer' than the canonical M99 relation (solid black line), there are also a few objects (HZ6 and J0217-0208) that {lie} below the relation. To better estimate the conditions of the different objects in the compiled sample, we also overplot a set of analytic curves for {different} stellar population ages and dust extinction curves.

In the \textit{left} panel, {we show} the analytic {solutions} for {MW dust} with {a} stellar population age of $t_{\rm age}=1$ Myr and 5 Myrs, {and} $\beta_{\rm UV}$ are measured at $\lambda=1230$ and $3200\,\angstrom$. The {galaxy} that shows the largest offset from the M99 relation on the `blue' side, HZ10, can be accounted for by a very young UV-weighted $t_{\rm age}$ of $\sim$1 Myr.  Alternatively, {its location in the IRX-$\beta_{\rm UV}$ plane} can also be {explained by} a strong $2175\,\angstrom$ `bump' in the attenuation curve, given that {$\beta_{\rm UV}$ of this object was} measured {using} only the photometry blueward of the `bump' (see Table~\ref{T3}, and also \citealt{Barisic_2017}). We show the analytic {IRX-$\beta_{\rm UV}$ relation} for $t_{\rm age}=100$ Myrs at $z=6$ and $\beta_{\rm UV}$ {being} measured at the \textit{HST} F105W and F160W bands. The location of HZ10 is {also} consistent with the analytic curve of this relatively old $t_{\rm age}$ because of the `bump' feature in the MW extinction curve. Therefore, to distinguish the potential contamination by the `bump' feature to the $\beta_{\rm UV}$ measurements of the $z\simgreat5$ galaxies, observations in the $\rm K_s$-band may be need \citep{Popping_2017}.

In the \textit{right} panel, we show the analytic {IRX-$\beta_{\rm UV}$ relation} for SMC dust curve {and a} much older stellar population. The location of the `reddest' galaxy, J0217-0208, is consistent with $t_{\rm age}=300$ Myrs, close to the oldest UV-weighted stellar age of the \textsc{\small MassiveFIRE} sample (Fig.~\ref{fig.16}). {However,} the location of J0217-0208 indicates a higher $\tilde{\tau}_{0.16}$ than the \textsc{\small MassiveFIRE} galaxy of similar age. J0217-0208 {may} also have younger UV-weighted $t_{\rm age}$ and an attenuation curve steeper than the SMC curve. This can either be due to a steep intrinsic extinction curve, or {a heavier dust obscuration of the young stars. Furthermore, the location of the detected source, HZ6, and the upper $3\sigma$ limit of the undetected source, HZ5, imply a $t_{\rm age}$ (or a lower limit of $t_{\rm age}$ for the undetected source) between 100 and 200 Myrs.} The data points or upper $3\sigma$ limits of the rest of the objects are all above the curve for $t_{\rm age}=100$ Myrs. 

The mean IRX of the undetected objects at $z\simgreat5$ derived from their stacked ALMA flux appeared to {lie} significantly below the canonical M99 relation \citep{Capak_2015}. This tension was alleviated by the work of \citet{Barisic_2017}, where the authors re-assessed the $\beta_{\rm UV}$ measurements using the \textit{HST}/Wide Field Camera 3 near-IR imaging and found a systematic bias of the previous ground-based data toward redder slopes. In this work, we {re-derive} the IRX of these objects using the best-fit formula for $T_{\rm eqv}$ (Eq.~\ref{eq.26}). 

We find that the majority of the upper $3\sigma$ limits of the undetected sources (10 out of 13) are {in agreement} with the canonical M99 relation and the {IRX-$\beta_{\rm UV}$} relation for SMC dust with $t_{\rm age}=50$ Myrs. The few objects that show {an} IRX deficit compared to these relations {can be explained by a relatively evolved stellar population, which is common in \textsc{\small MassiveFIRE}.} {Note that} IRX estimates of high-$z$ galaxies based on ALMA-band fluxes alone can have non-trivial uncertainties due to the variations in the dust SED shape of galaxies {at the wavelength range not covered by the ALMA bands}. We mark the $\pm 1\sigma$ uncertainty in the IRX estimate in both panels of Fig.~\ref{fig.28}. 

\vspace{-10pt}
\section{Summary and Conclusions}
\label{Sec:6}

The empirical relation between UV spectral slope ($\beta_{\rm UV}$) and infrared excess (IRX) of galaxies is frequently adopted for {estimating $L_{\rm IR}$ and the dust-obscured SFR of distant galaxies when only UV measurements are available; however, observations have shown evidence of non-trivial scatter of this relation among different galaxy populations.} In this work, we explore the nature of the IRX-$\beta_{\rm UV}$ relation and the different origins of the scatter. We adopt a sample of galaxies at $z=2-6$ {($M_*\approx 10^9-10^{12} \,M_\odot$)} that are extracted from the cosmological `zoom-in' simulations \textsc{\small MassiveFIRE} \citep{Feldmann_2016, Feldmann_2017}, which {are} part of the Feedback in Realistic Environments (\textsc{fire}) project \citep{Hopkins_2014}. Using {the} dust radiative transfer tool, \textsc{skirt} \citep{Baes_2011,Baes_2015}, we produce spatially resolved UV-to-mm {SEDs for} the \textsc{\small MassiveFIRE} sample, {and study their} observational properties, {in particular} $\beta_{\rm UV}$, integrated UV and IR luminosities. \\

The main findings of this work are:

\begin{itemize}[leftmargin=0.5cm]

\item Using the standard MW dust extinction law, the \textsc{\small MassiveFIRE} sample is in broad agreement with the canonical relation of \citet{Meurer_1999} (M99) derived using {a} local starburst population (Section~\ref{Sec:3a} \& Section~\ref{Sec:3b}). The deviation from the M99 relation correlates with the \textit{UV-luminosity}-weighted stellar age ($t_{\rm age,\,UV}$) (see Eq. 20), {but not} with the mass-weighted stellar age ($t_{\rm age,\,mw}$). 

\item The UV \textit{effective} optical depth of galaxies ($\tilde{\tau}_{0.16}$) is well correlated with IRX, specifically, $\rm IRX\propto{\rm e}^{\tilde{\tau}_{0.16}}-1$ (Eq.~\ref{eq.19}). This relation is {in agreement with} the analytic curve derived using the dust slab model (Section~\ref{Sec:4a}). In contrast, $\tilde{\tau}_{0.16}$ is weakly correlated to $\beta_{\rm UV}$ due to the large variations in the intrinsic UV spectral slope ($\beta_{\rm UV,\,0}$) of galaxies (Section~\ref{Sec:4b}). Thus, $\beta_{\rm UV}$ should not be used as reliable proxy for UV optical depth.

\item The increment (reddening) of UV spectral slope, $\beta_{\rm UV}$ - $\beta_{\rm UV,\,0}$, scales proportionally with $\tilde{\tau}_{0.16}$, broadly consistent with the expectation of the dust slab model. The slope of this linear relation depends on the steepness of the dust attenuation curve of galaxy. The shape of the attenuation curve and that of the underlying dust extinction curve are generally different {depending on the dust-to-star geometry} (Section~\ref{Sec:4b}).

\item $\tilde{\tau}_{0.16}$ (or equivalently, IRX) depends on the spatial configuration of dust {with respect to star-forming regions} in the galaxies and is not well correlated with $M_{\rm dust}$. $\tilde{\tau}_{0.16}$ increases during the starbursts as gas/dust configuration becomes more concentrated due to instabilities. Two galaxies of similar $M_{\rm dust}$ can have significantly different IRX (by over an order of magnitude) (Section~\ref{Sec:4c}).

\item A galaxy evolves in the IRX-$\beta_{\rm UV}$ plane over cosmic time. {The} evolutionary trajectory {consists of} counter-clockwise rotation on relatively short timescales of $\simless100$ Myrs. During {starbursts, galaxies move} upwards and leftwards {in} the plane (due to the increase of dust optical depth and the decrease of stellar age) while {during quiescent periods, galaxies move} downwards and rightwards (due to the decrease of optical depth and the increase of stellar age) (Section~\ref{Sec:4d}).  

\item The attenuation curve of galaxies varies with viewing direction, appearing to be shallower (or `grayer') in the direction of higher $\tilde{\tau}_{0.16}$, regardless of the morphology of galaxy. For disc galaxies, the edge-on (face-on) direction shows the highest (lowest) $\tilde{\tau}_{0.16}$ and the shallowest attenuation curve (Section~\ref{Sec:4e2}). {This is consistent with the recent observational finding reported by \citet{Wang_2018} (see also \citealt{Kriek_2013}).} 

\item Uncertainties in viewing direction, dust-to-metal mass ratio, stellar population model (single vs. binary star evolution) and {conditions of the star-forming birth-clouds} are secondary contributors to the scatter in the IRX-$\beta_{\rm UV}$ relation {of the \textsc{\small MassiveFIRE} sample} (Section~\ref{Sec:4e2}-\ref{Sec:4e4}). {For a given dust model}, the scatter can largely been accounted for by the variations in the intrinsic UV spectral slope of galaxies (Section~\ref{Sec:4d}). {The offset of $\beta_{\rm UV}$ resulting from the variation in the steepness of the extinction curve scales linearly with $\tilde{\tau}_{0.16}$} (Section~\ref{Sec:4e1}). 
\end{itemize}

Estimating {observationally} the relatively contributions of the various sources to the scatter in the {IRX-$\beta_{\rm UV}$} relation is challenging. Most of the current observational constraints in the intermediate redshift range ($2\simless z\simless 4$) are derived based on a stacking method, which {does} not {capture} the variations of individual sources (Section~\ref{Sec:3b}). A few studies using DSFGs have complete detections. These samples are biased to the IR-luminous objects, and show a clear `secondary dependence' of the relation on $L_{\rm IR}$ (Section~\ref{Sec:3a}). The overall dispersion of this relation at high-$z$ among a general, unbiased galaxy sample is still uncertain.

Measurements of $\beta_{\rm UV}$ and $L_{\rm IR}$ carry additional uncertainties mainly because of the common dearth of photometric data points in the SEDs at high-$z$. $\beta_{\rm UV}$ estimates may depend on the photometric sampling at rest-UV and be susceptible to the `contamination' by the $2175\,\angstrom$ `bump' feature in the dust extinction curve (Section~\ref{Sec:5a}). {The `contamination' can be more severe in the systems of high $\tilde{\tau}_{0.16}$ (or IRX), which is indicated by the analytic solution of the dust slab model (Eq.~\ref{eq.25}).} Additionally, $L_{\rm IR}$ estimates may also be uncertain due to the variations in the dust SED shape in the wavelength range {(\ie~the Wien side of the dust SED)} that is not well covered by ALMA bands (Section~\ref{Sec:5b}). 

We have also assessed the IRX-$\beta_{\rm UV}$ relation of a LBG sample at $z\approx5-6.5$ that had previously been reported to show significant IRX deficit \citep[\eg,][]{Capak_2015}. Using the recently updated $\beta_{\rm UV}$ measurements \citep[\eg,][]{Barisic_2017} and the IRXs derived using our best-fit formula for the `equivalent dust temperature' ($T_{\rm eqv}$) based on \textsc{\small MassiveFIRE} (Eq.~\ref{eq.26}, and see \citealt{Liang_2019}), the location of these objects {in the IRX-$\beta_{\rm UV}$ plane} shows no clear tension with the locally derived dust attenuation laws (MW and SMC). The objects in the sample that show redder $\beta_{\rm UV}$ compared to the M99 relation can be accounted for by SMC dust together with a relatively evolved stellar population. We also note that the $L_{\rm IR}$ estimates of these galaxies based on ALMA fluxes will typically not be better than $\pm0.22$ dex due to a `secondary dependences' of $T_{\rm eqv}$. The next-generation space IR telescope \textsc{\small SPICA} \citep{Spinoglio_2017, Egami_2018}, {which covers the spectral range of $\rm 12-230\,\mu m$ with much improved sensitivity compared to the \textit{Spitzer} and \textit{Herschel} telescopes}, may improve our constraints on the dust SED shape and thus the $L_{\rm IR}$ estimates of the high-$z$ galaxies. 

\vspace{-10pt}
\section*{Acknowledgements}

This manuscript has benefited from discussions with Pascal Oesch (Geneva), Nick Z. Scoville (Caltech), Xuejian (Jacob) Shen (Caltech), Marcel Neeleman (MPIA), Laura Sommovigo (Scuola Normale Superiore) and Andrea Ferrara (Scuola Normale Superiore). We thank Caitlin Casey for providing us with the data that are not publicly available for producing Fig. 4. LL would like to thank the hospitality of the Department of Astronomy of the University of Florida (UF), where part of this manuscript was improved. His research stay at UF was supported by the GRC Grant awarded by the University of Zurich. RF acknowledges financial support from the Swiss National Science Foundation (grant no. 157591). Simulations were run with resources provided by the NASA High-End Computing (HEC) Programme. Additional computing support was provided by HEC allocations SMD-14-5189, SMD-15-5950, SMD-16-7561, SMD-17-1204, by NSF XSEDE allocations AST120025, AST140023, AST150045, by allocations s697, s698 at the Swiss National Supercomputing center (CSCS), and by S3IT resources at the University of Zurich. DN was supported by NSF grants AST-1715206, AST-1908137 and AST-1909153, as well as HST-AR-15043.001. DK acknowledges support from the NSF grant AST-1715101 and the Cottrell Scholar Award from the Research Corporation for Science Advancement. CAFG was supported by NSF through grants AST-1517491, AST-1715216, and CAREER award AST-1652522; by NASA through grant 17-ATP17-0067; by STScI through grant HST-AR-14562.001; and by a Cottrell Scholar Award from the Research Corporation for Science Advancement. PFH was supported by an Alfred P. Sloan Research Fellowship, NASA ATP Grant NNX14AH35G, and NSF Collaborative Research Grant \#1411920 and CAREER grant \#1455342. This research was supported by the Munich Institute for Astro- and Particle Physics (MIAPP) of the Deutsche Forschungsgemeinschaft (DFG) cluster of excellence ``Origin and Structure of the Universe". The Flatiron Institute is supported by the Simons Foundation. 

\vspace{-10pt}

\bibliographystyle{mnras}
\bibliography{irxbeta}

\begin{thebibliography}{}
\makeatletter
\relax
\def\mn@urlcharsother{\let\do\@makeother \do\$\do\&\do\#\do\^\do\_\do\%\do\~}
\def\mn@doi{\begingroup\mn@urlcharsother \@ifnextchar [ {\mn@doi@}
  {\mn@doi@[]}}
\def\mn@doi@[#1]#2{\def\@tempa{#1}\ifx\@tempa\@empty \href
  {http://dx.doi.org/#2} {doi:#2}\else \href {http://dx.doi.org/#2} {#1}\fi
  \endgroup}
\def\mn@eprint#1#2{\mn@eprint@#1:#2::\@nil}
\def\mn@eprint@arXiv#1{\href {http://arxiv.org/abs/#1} {{\tt arXiv:#1}}}
\def\mn@eprint@dblp#1{\href {http://dblp.uni-trier.de/rec/bibtex/#1.xml}
  {dblp:#1}}
\def\mn@eprint@#1:#2:#3:#4\@nil{\def\@tempa {#1}\def\@tempb {#2}\def\@tempc
  {#3}\ifx \@tempc \@empty \let \@tempc \@tempb \let \@tempb \@tempa \fi \ifx
  \@tempb \@empty \def\@tempb {arXiv}\fi \@ifundefined
  {mn@eprint@\@tempb}{\@tempb:\@tempc}{\expandafter \expandafter \csname
  mn@eprint@\@tempb\endcsname \expandafter{\@tempc}}}

\bibitem[\protect\citeauthoryear{Adelberger \& Steidel}{Adelberger \&
  Steidel}{2000}]{Adelberger_2000}
Adelberger K.~L.,  Steidel C.~C.,  2000, \mn@doi [\apj] {10.1086/317183}, 544,
  218

\bibitem[\protect\citeauthoryear{{\'A}lvarez-M{\'a}rquez
  et~al.,}{{\'A}lvarez-M{\'a}rquez et~al.}{2016}]{Alvarez_Marquez_2016}
{\'A}lvarez-M{\'a}rquez J.,  et~al., 2016, \mn@doi [A\&A]
  {10.1051/0004-6361/201527190}, 587, A122

\bibitem[\protect\citeauthoryear{{\'A}lvarez-M{\'a}rquez, Burgarella, Buat,
  Ilbert  \& P{\'e}rez-Gonz{\'a}lez}{{\'A}lvarez-M{\'a}rquez
  et~al.}{2019}]{Alvarez_Marquez_2019}
{\'A}lvarez-M{\'a}rquez J.,  Burgarella D.,  Buat V.,  Ilbert O.,
  P{\'e}rez-Gonz{\'a}lez P.~G.,  2019, \mn@doi [A\&A]
  {10.1051/0004-6361/201935719}, 630, A153

\bibitem[\protect\citeauthoryear{{Angl{\'e}s-Alc{\'a}zar},
  {Faucher-Gigu{\`e}re}, {Kere{\v{s}}}, {Hopkins}, {Quataert}  \&
  {Murray}}{{Angl{\'e}s-Alc{\'a}zar} et~al.}{2017}]{Angles-Alcazar_2017}
{Angl{\'e}s-Alc{\'a}zar} D.,  {Faucher-Gigu{\`e}re} C.-A.,  {Kere{\v{s}}} D.,
  {Hopkins} P.~F.,  {Quataert} E.,   {Murray} N.,  2017, \mn@doi [\mnras]
  {10.1093/mnras/stx1517}, \href
  {https://ui.adsabs.harvard.edu/abs/2017MNRAS.470.4698A} {470, 4698}

\bibitem[\protect\citeauthoryear{Baes \& Camps}{Baes \&
  Camps}{2015}]{Baes_2015}
Baes M.,  Camps P.,  2015, \mn@doi [A&C] {10.1016/j.ascom.2015.05.006}, 12, 33

\bibitem[\protect\citeauthoryear{Baes, Verstappen, De~Looze, Fritz, Saftly,
  Vidal~P{\'e}rez, Stalevski  \& Valcke}{Baes et~al.}{2011}]{Baes_2011}
Baes M.,  Verstappen J.,  De~Looze I.,  Fritz J.,  Saftly W.,  Vidal~P{\'e}rez
  E.,  Stalevski M.,   Valcke S.,  2011, \mn@doi [\apjs]
  {10.1088/0067-0049/196/2/22}, 196, 22

\bibitem[\protect\citeauthoryear{Bakx et~al.,}{Bakx et~al.}{2020}]{Bakx_2020}
Bakx T. J. L.~C.,  et~al., 2020, \mn@doi [\mnras] {10.1093/mnras/staa509}, 493,
  4294

\bibitem[\protect\citeauthoryear{Ba{\~n}ados et~al.,}{Ba{\~n}ados
  et~al.}{2019}]{Banados_2019}
Ba{\~n}ados E.,  et~al., 2019, \mn@doi [\apj] {10.3847/2041-8213/ab3659}, 881,
  L23

\bibitem[\protect\citeauthoryear{Barisic et~al.,}{Barisic
  et~al.}{2017}]{Barisic_2017}
Barisic I.,  et~al., 2017, \mn@doi [\apj] {10.3847/1538-4357/aa7eda}, 845, 41

\bibitem[\protect\citeauthoryear{{Behrens}, {Pallottini}, {Ferrara},
  {Gallerani}  \& {Vallini}}{{Behrens} et~al.}{2018}]{Behrens_2018}
{Behrens} C.,  {Pallottini} A.,  {Ferrara} A.,  {Gallerani} S.,   {Vallini} L.,
   2018, \mn@doi [\mnras] {10.1093/mnras/sty552}, \href
  {http://adsabs.harvard.edu/abs/2018MNRAS.477..552B} {477, 552}

\bibitem[\protect\citeauthoryear{Bell}{Bell}{2002}]{Bell_2002}
Bell E.~F.,  2002, \mn@doi [\apj] {10.1086/342127}, 577, 150

\bibitem[\protect\citeauthoryear{Bell}{Bell}{2003}]{Bell_2003}
Bell E.~F.,  2003, \mn@doi [\apj] {10.1086/367829}, 586, 794

\bibitem[\protect\citeauthoryear{Berta et~al.,}{Berta
  et~al.}{2011}]{Berta_2011}
Berta S.,  et~al., 2011, \mn@doi [A\&A] {10.1051/0004-6361/201116844}, 532, A49

\bibitem[\protect\citeauthoryear{B{\'e}thermin et~al.,}{B{\'e}thermin
  et~al.}{2015}]{Bethermin_2015}
B{\'e}thermin M.,  et~al., 2015, \mn@doi [A\&A] {10.1051/0004-6361/201425031},
  573, A113

\bibitem[\protect\citeauthoryear{Boquien et~al.,}{Boquien
  et~al.}{2012}]{Boquien_2012}
Boquien M.,  et~al., 2012, \mn@doi [A\&A] {10.1051/0004-6361/201118624}, 539,
  A145

\bibitem[\protect\citeauthoryear{Bourne et~al.,}{Bourne
  et~al.}{2017}]{Bourne_2017}
Bourne N.,  et~al., 2017, \mn@doi [\mnras] {10.1093/mnras/stx031}, p. stx031

\bibitem[\protect\citeauthoryear{Bouwens et~al.,}{Bouwens
  et~al.}{2009}]{Bouwens_2009}
Bouwens R.~J.,  et~al., 2009, \mn@doi [\apj] {10.1088/0004-637x/705/1/936},
  705, 936

\bibitem[\protect\citeauthoryear{Bouwens et~al.,}{Bouwens
  et~al.}{2014}]{Bouwens_2014}
Bouwens R.~J.,  et~al., 2014, \mn@doi [\apj] {10.1088/0004-637x/795/2/126},
  795, 126

\bibitem[\protect\citeauthoryear{Bouwens et~al.,}{Bouwens
  et~al.}{2016}]{Bouwens_2016}
Bouwens R.,  et~al., 2016, \mn@doi [\apj] {10.3847/1538-4357/833/1/72}, 833, 72

\bibitem[\protect\citeauthoryear{{Bouwens} et~al.,}{{Bouwens}
  et~al.}{2020}]{Bouwens_2020}
{Bouwens} R.,  et~al., 2020, arXiv e-prints, \href
  {https://ui.adsabs.harvard.edu/abs/2020arXiv200910727B} {2009.10727}

\bibitem[\protect\citeauthoryear{Bowler, Bourne, Dunlop, McLure  \&
  McLeod}{Bowler et~al.}{2018}]{Bowler_2018}
Bowler R. A.~A.,  Bourne N.,  Dunlop J.~S.,  McLure R.~J.,   McLeod D.~J.,
  2018, \mn@doi [\mnras] {10.1093/mnras/sty2368}, 481, 1631

\bibitem[\protect\citeauthoryear{Buat, Boselli, Gavazzi  \& Bonfanti}{Buat
  et~al.}{2002}]{Buat_2002}
Buat V.,  Boselli A.,  Gavazzi G.,   Bonfanti C.,  2002, \mn@doi [A\&A]
  {10.1051/0004-6361:20011832}, 383, 801

\bibitem[\protect\citeauthoryear{Buat et~al.,}{Buat et~al.}{2005}]{Buat_2005}
Buat V.,  et~al., 2005, \mn@doi [\apj] {10.1086/423241}, 619, L51

\bibitem[\protect\citeauthoryear{Buat, Marcillac, Burgarella, Le~Floc'h,
  Takeuchi, Iglesias-Par{\'a}mo  \& Xu}{Buat et~al.}{2007}]{Buat_2007}
Buat V.,  Marcillac D.,  Burgarella D.,  Le~Floc'h E.,  Takeuchi T.~T.,
  Iglesias-Par{\'a}mo J.,   Xu C.~K.,  2007, \mn@doi [A\&A]
  {10.1051/0004-6361:20066685}, 469, 19

\bibitem[\protect\citeauthoryear{Buat, Takeuchi, Burgarella, Giovannoli  \&
  Murata}{Buat et~al.}{2009}]{Buat_2009}
Buat V.,  Takeuchi T.~T.,  Burgarella D.,  Giovannoli E.,   Murata K.~L.,
  2009, \mn@doi [A\&A] {10.1051/0004-6361/200912024}, 507, 693

\bibitem[\protect\citeauthoryear{Burgarella et~al.,}{Burgarella
  et~al.}{2013}]{Burgarella_2013}
Burgarella D.,  et~al., 2013, \mn@doi [A\&A] {10.1051/0004-6361/201321651},
  554, A70

\bibitem[\protect\citeauthoryear{Calzetti}{Calzetti}{1997}]{Calzetti_1997}
Calzetti D.,  1997, \mn@doi [\apj] {10.1086/118242}, 113, 162

\bibitem[\protect\citeauthoryear{Calzetti, Kinney  \&
  Storchi-Bergmann}{Calzetti et~al.}{1994}]{Calzetti_1994}
Calzetti D.,  Kinney A.~L.,   Storchi-Bergmann T.,  1994, \mn@doi [\apj]
  {10.1086/174346}, 429, 582

\bibitem[\protect\citeauthoryear{Calzetti, Armus, Bohlin, Kinney, Koornneef  \&
  Storchi‐Bergmann}{Calzetti et~al.}{2000}]{Calzetti_2000}
Calzetti D.,  Armus L.,  Bohlin R.~C.,  Kinney A.~L.,  Koornneef J.,
  Storchi‐Bergmann T.,  2000, \mn@doi [\apj] {10.1086/308692}, 533, 682

\bibitem[\protect\citeauthoryear{Camps \& Baes}{Camps \&
  Baes}{2015}]{Camps_2015a}
Camps P.,  Baes M.,  2015, \mn@doi [A&C] {10.1016/j.ascom.2014.10.004}, 9, 20

\bibitem[\protect\citeauthoryear{Camps et~al.,}{Camps
  et~al.}{2015}]{Camps_2015b}
Camps P.,  et~al., 2015, \mn@doi [A\&A] {10.1051/0004-6361/201525998}, 580, A87

\bibitem[\protect\citeauthoryear{{Camps}, {Trayford}, {Baes}, {Theuns},
  {Schaller}  \& {Schaye}}{{Camps} et~al.}{2016}]{Camps_2016}
{Camps} P.,  {Trayford} J.~W.,  {Baes} M.,  {Theuns} T.,  {Schaller} M.,
  {Schaye} J.,  2016, \mn@doi [\mnras] {10.1093/mnras/stw1735}, \href
  {http://adsabs.harvard.edu/abs/2016MNRAS.462.1057C} {462, 1057}

\bibitem[\protect\citeauthoryear{Camps et~al.,}{Camps
  et~al.}{2018}]{Camps_2018}
Camps P.,  et~al., 2018, \mn@doi [\apjs] {10.3847/1538-4365/aaa24c}, 234, 20

\bibitem[\protect\citeauthoryear{Capak et~al.,}{Capak
  et~al.}{2004}]{Capak_2004}
Capak P.,  et~al., 2004, \mn@doi [\apj] {10.1086/380611}, 127, 180

\bibitem[\protect\citeauthoryear{Capak et~al.,}{Capak
  et~al.}{2007}]{Capak_2007}
Capak P.,  et~al., 2007, \mn@doi [\apjs] {10.1086/519081}, 172, 99

\bibitem[\protect\citeauthoryear{Capak et~al.,}{Capak
  et~al.}{2015}]{Capak_2015}
Capak P.~L.,  et~al., 2015, \mn@doi [Nature] {10.1038/nature14500}, 522, 455

\bibitem[\protect\citeauthoryear{Carniani et~al.,}{Carniani
  et~al.}{2018a}]{Carniani_2018}
Carniani S.,  et~al., 2018a, \mn@doi [\mnras] {10.1093/mnras/sty1088}, 478,
  1170

\bibitem[\protect\citeauthoryear{Carniani, Maiolino, Smit  \&
  Amor{\'\i}n}{Carniani et~al.}{2018b}]{Carniani_2018b}
Carniani S.,  Maiolino R.,  Smit R.,   Amor{\'\i}n R.,  2018b, \mn@doi [\apj]
  {10.3847/2041-8213/aaab45}, 854, L7

\bibitem[\protect\citeauthoryear{Casey}{Casey}{2012}]{Casey_2012}
Casey C.~M.,  2012, \mn@doi [\mnras] {10.1111/j.1365-2966.2012.21455.x}, 425,
  3094

\bibitem[\protect\citeauthoryear{Casey, Narayanan  \& Cooray}{Casey
  et~al.}{2014a}]{Casey_2014P}
Casey C.~M.,  Narayanan D.,   Cooray A.,  2014a, \mn@doi [Physics Reports]
  {10.1016/j.physrep.2014.02.009}, 541, 45

\bibitem[\protect\citeauthoryear{Casey et~al.,}{Casey
  et~al.}{2014b}]{Casey_2014}
Casey C.~M.,  et~al., 2014b, \mn@doi [\apj] {10.1088/0004-637x/796/2/95}, 796,
  95

\bibitem[\protect\citeauthoryear{Casey et~al.,}{Casey
  et~al.}{2018a}]{Casey_2018b}
Casey C.~M.,  et~al., 2018a, \mn@doi [\apj] {10.3847/1538-4357/aac82d}, 862, 77

\bibitem[\protect\citeauthoryear{Casey, Hodge, Zavala, Spilker, da Cunha,
  Staguhn, Finkelstein  \& Drew}{Casey et~al.}{2018b}]{Casey_2018}
Casey C.~M.,  Hodge J.,  Zavala J.~A.,  Spilker J.,  da Cunha E.,  Staguhn J.,
  Finkelstein S.~L.,   Drew P.,  2018b, \mn@doi [\apj]
  {10.3847/1538-4357/aacd11}, 862, 78

\bibitem[\protect\citeauthoryear{Chabrier}{Chabrier}{2003}]{Chabrier_2003}
Chabrier G.,  2003, \mn@doi [\pasp] {10.1086/376392}, 115, 763

\bibitem[\protect\citeauthoryear{Charlot \& Fall}{Charlot \&
  Fall}{2000}]{Charlot_2000}
Charlot S.,  Fall S.~M.,  2000, \mn@doi [\apj] {10.1086/309250}, 539, 718

\bibitem[\protect\citeauthoryear{Chary \& Elbaz}{Chary \&
  Elbaz}{2001}]{Chary_2001}
Chary R.,  Elbaz D.,  2001, \mn@doi [\apj] {10.1086/321609}, 556, 562

\bibitem[\protect\citeauthoryear{Cochrane et~al.,}{Cochrane
  et~al.}{2019}]{Cochrane_2019}
Cochrane R.~K.,  et~al., 2019, \mn@doi [\mnras] {10.1093/mnras/stz1736}, 488,
  1779

\bibitem[\protect\citeauthoryear{Conroy}{Conroy}{2013}]{Conroy_2013}
Conroy C.,  2013, \mn@doi [ARA\&A] {10.1146/annurev-astro-082812-141017}, 51,
  393

\bibitem[\protect\citeauthoryear{{Dale} \& {Helou}}{{Dale} \&
  {Helou}}{2002}]{Dale_2002}
{Dale} D.~A.,  {Helou} G.,  2002, \mn@doi [\apj] {10.1086/341632}, \href
  {http://adsabs.harvard.edu/abs/2002ApJ...576..159D} {576, 159}

\bibitem[\protect\citeauthoryear{Dale, Helou, Magdis, Armus, D{\'\i}az-Santos
  \& Shi}{Dale et~al.}{2014}]{Dale_2014}
Dale D.~A.,  Helou G.,  Magdis G.~E.,  Armus L.,  D{\'\i}az-Santos T.,   Shi
  Y.,  2014, \mn@doi [\apj] {10.1088/0004-637x/784/1/83}, 784, 83

\bibitem[\protect\citeauthoryear{De~Cia, Ledoux, Savaglio, Schady  \&
  Vreeswijk}{De~Cia et~al.}{2013}]{De_Cia_2013}
De~Cia A.,  Ledoux C.,  Savaglio S.,  Schady P.,   Vreeswijk P.~M.,  2013,
  \mn@doi [A\&A] {10.1051/0004-6361/201321834}, 560, A88

\bibitem[\protect\citeauthoryear{De~Cia, Ledoux, Mattsson, Petitjean, Srianand,
  Gavignaud  \& Jenkins}{De~Cia et~al.}{2016}]{De_Cia_2016}
De~Cia A.,  Ledoux C.,  Mattsson L.,  Petitjean P.,  Srianand R.,  Gavignaud
  I.,   Jenkins E.~B.,  2016, \mn@doi [A\&A] {10.1051/0004-6361/201527895},
  596, A97

\bibitem[\protect\citeauthoryear{De~Looze et~al.,}{De~Looze
  et~al.}{2014}]{De_Looze_2014}
De~Looze I.,  et~al., 2014, \mn@doi [A\&A] {10.1051/0004-6361/201424747}, 571,
  A69

\bibitem[\protect\citeauthoryear{De~Vis et~al.,}{De~Vis
  et~al.}{2019}]{De_Vis_2019}
De~Vis P.,  et~al., 2019, \mn@doi [A\&A] {10.1051/0004-6361/201834444}, 623, A5

\bibitem[\protect\citeauthoryear{Dey et~al.,}{Dey et~al.}{2008}]{Dey_2008}
Dey A.,  et~al., 2008, \mn@doi [\apj] {10.1086/529516}, 677, 943

\bibitem[\protect\citeauthoryear{Dole et~al.,}{Dole et~al.}{2004}]{Dole_2004}
Dole H.,  et~al., 2004, \mn@doi [\apjs] {10.1086/422690}, 154, 93

\bibitem[\protect\citeauthoryear{{Draine} et~al.,}{{Draine}
  et~al.}{2007}]{Draine_2007}
{Draine} B.~T.,  et~al., 2007, \mn@doi [ApJ] {10.1086/518306}, \href
  {http://adsabs.harvard.edu/abs/2007ApJ...663..866D} {663, 866}

\bibitem[\protect\citeauthoryear{Duch{\^e}ne \& Kraus}{Duch{\^e}ne \&
  Kraus}{2013}]{Duch_ne_2013}
Duch{\^e}ne G.,  Kraus A.,  2013, \mn@doi [ARA\&A]
  {10.1146/annurev-astro-081710-102602}, 51, 269

\bibitem[\protect\citeauthoryear{Dunlop et~al.,}{Dunlop
  et~al.}{2017}]{Dunlop_2017}
Dunlop J.~S.,  et~al., 2017, \mn@doi [\mnras] {10.1093/mnras/stw3088}, 466, 861

\bibitem[\protect\citeauthoryear{{Dwek}}{{Dwek}}{1998}]{Dwek_1998}
{Dwek} E.,  1998, \mn@doi [\apj] {10.1086/305829}, \href
  {http://adsabs.harvard.edu/abs/1998ApJ...501..643D} {501, 643}

\bibitem[\protect\citeauthoryear{Egami et~al.,}{Egami
  et~al.}{2018}]{Egami_2018}
Egami E.,  et~al., 2018, \mn@doi [PASA] {10.1017/pasa.2018.41}, 35

\bibitem[\protect\citeauthoryear{El-Badry \& Rix}{El-Badry \&
  Rix}{2018}]{El_Badry_2018}
El-Badry K.,  Rix H.-W.,  2018, \mn@doi [\mnras: Letters]
  {10.1093/mnrasl/sly206}

\bibitem[\protect\citeauthoryear{Elbaz et~al.,}{Elbaz
  et~al.}{2011}]{Elbaz_2011}
Elbaz D.,  et~al., 2011, \mn@doi [A\&A] {10.1051/0004-6361/201117239}, 533,
  A119

\bibitem[\protect\citeauthoryear{Eldridge \& Stanway}{Eldridge \&
  Stanway}{2012}]{Eldridge_2012}
Eldridge J.~J.,  Stanway E.~R.,  2012, \mn@doi [\mnras]
  {10.1111/j.1365-2966.2011.19713.x}, 419, 479

\bibitem[\protect\citeauthoryear{Eldridge, Stanway, Xiao, McClelland, Taylor,
  Ng, Greis  \& Bray}{Eldridge et~al.}{2017}]{Eldridge_2017}
Eldridge J.~J.,  Stanway E.~R.,  Xiao L.,  McClelland L. A.~S.,  Taylor G.,  Ng
  M.,  Greis S. M.~L.,   Bray J.~C.,  2017, \mn@doi [\pasp]
  {10.1017/pasa.2017.51}, 34

\bibitem[\protect\citeauthoryear{Ellis et~al.,}{Ellis
  et~al.}{2013}]{Ellis_2013}
Ellis R.~S.,  et~al., 2013, \mn@doi [\apj] {10.1088/2041-8205/763/1/l7}, 763,
  L7

\bibitem[\protect\citeauthoryear{Faisst et~al.,}{Faisst
  et~al.}{2017}]{Faisst_2017}
Faisst A.~L.,  et~al., 2017, \mn@doi [\apj] {10.3847/1538-4357/aa886c}, 847, 21

\bibitem[\protect\citeauthoryear{{Faisst}, {Fudamoto}, {Oesch}, {Scoville},
  {Riechers}, {Pavesi}  \& {Capak}}{{Faisst} et~al.}{2020}]{Faisst_2020}
{Faisst} A.~L.,  {Fudamoto} Y.,  {Oesch} P.~A.,  {Scoville} N.,  {Riechers}
  D.~A.,  {Pavesi} R.,   {Capak} P.,  2020, arXiv e-prints, \href
  {https://ui.adsabs.harvard.edu/abs/2020arXiv200507716F} {2005.07716}

\bibitem[\protect\citeauthoryear{Faucher-Gigu{\`e}re}{Faucher-Gigu{\`e}re}{2017}]{Faucher_Giguere_2017}
Faucher-Gigu{\`e}re C.-A.,  2017, \mn@doi [\mnras] {10.1093/mnras/stx2595},
  473, 3717

\bibitem[\protect\citeauthoryear{Faucher-Gigu{\`e}re, Lidz, Zaldarriaga  \&
  Hernquist}{Faucher-Gigu{\`e}re et~al.}{2009}]{Faucher_Giguere_2009}
Faucher-Gigu{\`e}re C.-A.,  Lidz A.,  Zaldarriaga M.,   Hernquist L.,  2009,
  \mn@doi [\apj] {10.1088/0004-637x/703/2/1416}, 703, 1416

\bibitem[\protect\citeauthoryear{Feldmann}{Feldmann}{2017}]{Feldmann_2017b}
Feldmann R.,  2017, \mn@doi [\mnras: Letters] {10.1093/mnrasl/slx073}, 470, L59

\bibitem[\protect\citeauthoryear{Feldmann, Hopkins, Quataert,
  Faucher-Gigu{\`e}re  \& Kere{\v s}}{Feldmann et~al.}{2016}]{Feldmann_2016}
Feldmann R.,  Hopkins P.~F.,  Quataert E.,  Faucher-Gigu{\`e}re C.-A.,
  Kere{\v s} D.,  2016, \mn@doi [\mnras: Letters] {10.1093/mnrasl/slw014}, 458,
  L14

\bibitem[\protect\citeauthoryear{Feldmann, Quataert, Hopkins,
  Faucher-Gigu{\`e}re  \& Kere{\v s}}{Feldmann et~al.}{2017}]{Feldmann_2017}
Feldmann R.,  Quataert E.,  Hopkins P.~F.,  Faucher-Gigu{\`e}re C.-A.,
  Kere{\v s} D.,  2017, \mn@doi [\mnras] {10.1093/mnras/stx1120}, 470, 1050

\bibitem[\protect\citeauthoryear{Ferrara, Hirashita, Ouchi  \&
  Fujimoto}{Ferrara et~al.}{2017}]{Ferrara_2017}
Ferrara A.,  Hirashita H.,  Ouchi M.,   Fujimoto S.,  2017, \mn@doi [\mnras]
  {10.1093/mnras/stx1898}, 471, 5018

\bibitem[\protect\citeauthoryear{Fischera, Dopita  \& Sutherland}{Fischera
  et~al.}{2003}]{Fischera_2003}
Fischera J.,  Dopita M.~A.,   Sutherland R.~S.,  2003, \mn@doi [\apj]
  {10.1086/381190}, 599, L21

\bibitem[\protect\citeauthoryear{Fitzpatrick}{Fitzpatrick}{1999}]{Fitzpatrick_1999}
Fitzpatrick E.~L.,  1999, \mn@doi [\pasp] {10.1086/316293}, 111, 63

\bibitem[\protect\citeauthoryear{{Flores Vel{\'a}zquez} et~al.,}{{Flores
  Vel{\'a}zquez} et~al.}{2020}]{Flores_2020}
{Flores Vel{\'a}zquez} J.~A.,  et~al., 2020, arXiv e-prints, \href
  {https://ui.adsabs.harvard.edu/abs/2020arXiv200808582F} {2008.08582}

\bibitem[\protect\citeauthoryear{Fudamoto et~al.,}{Fudamoto
  et~al.}{2017}]{Fudamoto_2017}
Fudamoto Y.,  et~al., 2017, \mn@doi [\mnras] {10.1093/mnras/stx1948}, 472, 483

\bibitem[\protect\citeauthoryear{Fudamoto et~al.,}{Fudamoto
  et~al.}{2020}]{Fudamoto_2020}
Fudamoto Y.,  et~al., 2020, \mn@doi [\mnras] {10.1093/mnras/stz3248}, 491, 4724

\bibitem[\protect\citeauthoryear{Galametz et~al.,}{Galametz
  et~al.}{2013}]{Galametz_2013}
Galametz A.,  et~al., 2013, \mn@doi [\apjs] {10.1088/0067-0049/206/2/10}, 206,
  10

\bibitem[\protect\citeauthoryear{Gill, Knebe  \& Gibson}{Gill
  et~al.}{2004}]{Gill_2004}
Gill S. P.~D.,  Knebe A.,   Gibson B.~K.,  2004, \mn@doi [\mnras]
  {10.1111/j.1365-2966.2004.07786.x}, 351, 399

\bibitem[\protect\citeauthoryear{Goldader, Meurer, Heckman, Seibert, Sanders,
  Calzetti  \& Steidel}{Goldader et~al.}{2002}]{Goldader_2002}
Goldader J.~D.,  Meurer G.,  Heckman T.~M.,  Seibert M.,  Sanders D.~B.,
  Calzetti D.,   Steidel C.~C.,  2002, \mn@doi [\apj] {10.1086/339165}, 568,
  651

\bibitem[\protect\citeauthoryear{Gordon, Calzetti  \& Witt}{Gordon
  et~al.}{1997}]{Gordon_1997}
Gordon K.~D.,  Calzetti D.,   Witt A.~N.,  1997, \mn@doi [\apj]
  {10.1086/304654}, 487, 625

\bibitem[\protect\citeauthoryear{Gordon, Clayton, Misselt, Landolt  \&
  Wolff}{Gordon et~al.}{2003}]{Gordon_2003}
Gordon K.~D.,  Clayton G.~C.,  Misselt K.~A.,  Landolt A.~U.,   Wolff M.~J.,
  2003, \mn@doi [\apj] {10.1086/376774}, 594, 279

\bibitem[\protect\citeauthoryear{Granato, Lacey, Silva, Bressan, Baugh, Cole
  \& Frenk}{Granato et~al.}{2000}]{Granato_2000}
Granato G.~L.,  Lacey C.~G.,  Silva L.,  Bressan A.,  Baugh C.~M.,  Cole S.,
  Frenk C.~S.,  2000, \mn@doi [\apj] {10.1086/317032}, 542, 710

\bibitem[\protect\citeauthoryear{Grasha, Calzetti, Andrews, Lee  \&
  Dale}{Grasha et~al.}{2013}]{Grasha_2013}
Grasha K.,  Calzetti D.,  Andrews J.~E.,  Lee J.~C.,   Dale D.~A.,  2013,
  \mn@doi [\apj] {10.1088/0004-637x/773/2/174}, 773, 174

\bibitem[\protect\citeauthoryear{Griffin et~al.,}{Griffin
  et~al.}{2010}]{Griffin_2010}
Griffin M.~J.,  et~al., 2010, \mn@doi [A\&A] {10.1051/0004-6361/201014519},
  518, L3

\bibitem[\protect\citeauthoryear{Groves, Dopita, Sutherland, Kewley, Fischera,
  Leitherer, Brandl  \& van Breugel}{Groves et~al.}{2008}]{Groves_2008}
Groves B.,  Dopita M.~A.,  Sutherland R.~S.,  Kewley L.~J.,  Fischera J.,
  Leitherer C.,  Brandl B.,   van Breugel W.,  2008, \mn@doi [\apjs]
  {10.1086/528711}, 176, 438

\bibitem[\protect\citeauthoryear{Gruppioni et~al.,}{Gruppioni
  et~al.}{2013}]{Gruppioni_2013}
Gruppioni C.,  et~al., 2013, \mn@doi [\mnras] {10.1093/mnras/stt308}, 432, 23

\bibitem[\protect\citeauthoryear{Guo et~al.,}{Guo et~al.}{2013}]{Guo_2013}
Guo Y.,  et~al., 2013, \mn@doi [\apjs] {10.1088/0067-0049/207/2/24}, 207, 24

\bibitem[\protect\citeauthoryear{Hahn \& Abel}{Hahn \& Abel}{2011}]{Hahn_2011}
Hahn O.,  Abel T.,  2011, \mn@doi [\mnras] {10.1111/j.1365-2966.2011.18820.x},
  415, 2101

\bibitem[\protect\citeauthoryear{Hao, Kennicutt, Johnson, Calzetti, Dale  \&
  Moustakas}{Hao et~al.}{2011}]{Hao_2011}
Hao C.-N.,  Kennicutt R.~C.,  Johnson B.~D.,  Calzetti D.,  Dale D.~A.,
  Moustakas J.,  2011, \mn@doi [\apj] {10.1088/0004-637x/741/2/124}, 741, 124

\bibitem[\protect\citeauthoryear{Harikane et~al.,}{Harikane
  et~al.}{2020}]{Harikane_2020}
Harikane Y.,  et~al., 2020, \mn@doi [\apj] {10.3847/1538-4357/ab94bd}, 896, 93

\bibitem[\protect\citeauthoryear{Harvey et~al.,}{Harvey
  et~al.}{2013}]{Harvey_2013}
Harvey P.~M.,  et~al., 2013, \mn@doi [\apj] {10.1088/0004-637x/764/2/133}, 764,
  133

\bibitem[\protect\citeauthoryear{Hayward \& Smith}{Hayward \&
  Smith}{2015}]{Hayward_2015}
Hayward C.~C.,  Smith D. J.~B.,  2015, \mn@doi [\mnras]
  {10.1093/mnras/stu2195}, 446, 1512

\bibitem[\protect\citeauthoryear{Hayward, Kere{\v s}, Jonsson, Narayanan, Cox
  \& Hernquist}{Hayward et~al.}{2011}]{Hayward_2011}
Hayward C.~C.,  Kere{\v s} D.,  Jonsson P.,  Narayanan D.,  Cox T.~J.,
  Hernquist L.,  2011, \mn@doi [\apj] {10.1088/0004-637x/743/2/159}, 743, 159

\bibitem[\protect\citeauthoryear{Hayward, Jonsson, Kere{\v s}, Magnelli,
  Hernquist  \& Cox}{Hayward et~al.}{2012}]{Hayward_2012}
Hayward C.~C.,  Jonsson P.,  Kere{\v s} D.,  Magnelli B.,  Hernquist L.,   Cox
  T.~J.,  2012, \mn@doi [\mnras] {10.1111/j.1365-2966.2012.21254.x}, 424, 951

\bibitem[\protect\citeauthoryear{Heinis et~al.,}{Heinis
  et~al.}{2013}]{Heinis_2013}
Heinis S.,  et~al., 2013, \mn@doi [\mnras] {10.1093/mnras/sts397}, 429, 1113

\bibitem[\protect\citeauthoryear{{Hildebrand}}{{Hildebrand}}{1983}]{Hildebrand_1983}
{Hildebrand} R.~H.,  1983, \qjras, \href
  {https://ui.adsabs.harvard.edu/abs/1983QJRAS..24..267H} {24, 267}

\bibitem[\protect\citeauthoryear{Hinshaw et~al.,}{Hinshaw
  et~al.}{2013}]{Hinshaw_2013}
Hinshaw G.,  et~al., 2013, \mn@doi [ApJS] {10.1088/0067-0049/208/2/19}, 208, 19

\bibitem[\protect\citeauthoryear{Hirashita, Nozawa, Villaume  \&
  Srinivasan}{Hirashita et~al.}{2015}]{Hirashita_2015}
Hirashita H.,  Nozawa T.,  Villaume A.,   Srinivasan S.,  2015, \mn@doi
  [\mnras] {10.1093/mnras/stv2095}, 454, 1620

\bibitem[\protect\citeauthoryear{Hopkins}{Hopkins}{2013}]{Hopkins_2013}
Hopkins P.~F.,  2013, \mn@doi [\mnras] {10.1093/mnras/sts210}, 428, 2840

\bibitem[\protect\citeauthoryear{{Hopkins}}{{Hopkins}}{2015}]{Hopkins_2015}
{Hopkins} P.~F.,  2015, \mn@doi [\mnras] {10.1093/mnras/stv195}, \href
  {http://adsabs.harvard.edu/abs/2015MNRAS.450...53H} {450, 53}

\bibitem[\protect\citeauthoryear{Hopkins, Kere{\v s}, O{\~n}orbe,
  Faucher-Gigu{\`e}re, Quataert, Murray  \& Bullock}{Hopkins
  et~al.}{2014}]{Hopkins_2014}
Hopkins P.~F.,  Kere{\v s} D.,  O{\~n}orbe J.,  Faucher-Gigu{\`e}re C.-A.,
  Quataert E.,  Murray N.,   Bullock J.~S.,  2014, \mn@doi [\mnras]
  {10.1093/mnras/stu1738}, 445, 581

\bibitem[\protect\citeauthoryear{Howell et~al.,}{Howell
  et~al.}{2010}]{Howell_2010}
Howell J.~H.,  et~al., 2010, \mn@doi [\apj] {10.1088/0004-637x/715/1/572}, 715,
  572

\bibitem[\protect\citeauthoryear{Ilbert et~al.,}{Ilbert
  et~al.}{2009}]{Ilbert_2009}
Ilbert O.,  et~al., 2009, \mn@doi [\apj] {10.1088/0004-637x/690/2/1236}, 690,
  1236

\bibitem[\protect\citeauthoryear{Imara, Loeb, Johnson, Conroy  \&
  Behroozi}{Imara et~al.}{2018}]{Imara_2018}
Imara N.,  Loeb A.,  Johnson B.~D.,  Conroy C.,   Behroozi P.,  2018, \mn@doi
  [\apj] {10.3847/1538-4357/aaa3f0}, 854, 36

\bibitem[\protect\citeauthoryear{Inoue et~al.,}{Inoue
  et~al.}{2016}]{Inoue_2016}
Inoue A.~K.,  et~al., 2016, \mn@doi [Science] {10.1126/science.aaf0714}, 352,
  1559

\bibitem[\protect\citeauthoryear{Jin et~al.,}{Jin et~al.}{2019}]{Jin_2019}
Jin S.,  et~al., 2019, \mn@doi [\apj] {10.3847/1538-4357/ab55d6}, 887, 144

\bibitem[\protect\citeauthoryear{Jonsson, Cox, Primack  \& Somerville}{Jonsson
  et~al.}{2006}]{Jonsson_2006}
Jonsson P.,  Cox T.~J.,  Primack J.~R.,   Somerville R.~S.,  2006, \mn@doi
  [\apj] {10.1086/497567}, 637, 255

\bibitem[\protect\citeauthoryear{{Jonsson}, {Groves}  \& {Cox}}{{Jonsson}
  et~al.}{2010}]{Jonsson_2010}
{Jonsson} P.,  {Groves} B.~A.,   {Cox} T.~J.,  2010, \mn@doi [\mnras]
  {10.1111/j.1365-2966.2009.16087.x}, \href
  {http://adsabs.harvard.edu/abs/2010MNRAS.403...17J} {403, 17}

\bibitem[\protect\citeauthoryear{Kennicutt}{Kennicutt}{1998}]{Kennicutt_1998}
Kennicutt R.~C.,  1998, \mn@doi [ARA\&A] {10.1146/annurev.astro.36.1.189}, 36,
  189

\bibitem[\protect\citeauthoryear{Kennicutt \& Evans}{Kennicutt \&
  Evans}{2012}]{Kennicutt_2012}
Kennicutt R.~C.,  Evans N.~J.,  2012, \mn@doi [ARA&A]
  {10.1146/annurev-astro-081811-125610}, 50, 531

\bibitem[\protect\citeauthoryear{Kinney, Bohlin, Calzetti, Panagia  \&
  Wyse}{Kinney et~al.}{1993}]{Kinney_1993}
Kinney A.~L.,  Bohlin R.~C.,  Calzetti D.,  Panagia N.,   Wyse R. F.~G.,  1993,
  \mn@doi [\apjs] {10.1086/191771}, 86, 5

\bibitem[\protect\citeauthoryear{Knollmann \& Knebe}{Knollmann \&
  Knebe}{2009}]{Knollmann_2009}
Knollmann S.~R.,  Knebe A.,  2009, \mn@doi [\apjs]
  {10.1088/0067-0049/182/2/608}, 182, 608

\bibitem[\protect\citeauthoryear{Knudsen, Richard, Kneib, Jauzac, Cl{\'e}ment,
  Drouart, Egami  \& Lindroos}{Knudsen et~al.}{2016}]{Knudsen_2016}
Knudsen K.~K.,  Richard J.,  Kneib J.-P.,  Jauzac M.,  Cl{\'e}ment B.,  Drouart
  G.,  Egami E.,   Lindroos L.,  2016, \mn@doi [\mnras: Letters]
  {10.1093/mnrasl/slw114}, 462, L6

\bibitem[\protect\citeauthoryear{Kong, Charlot, Brinchmann  \& Fall}{Kong
  et~al.}{2004}]{Kong_2004}
Kong X.,  Charlot S.,  Brinchmann J.,   Fall S.~M.,  2004, \mn@doi [\mnras]
  {10.1111/j.1365-2966.2004.07556.x}, 349, 769

\bibitem[\protect\citeauthoryear{Koprowski et~al.,}{Koprowski
  et~al.}{2018}]{Koprowski_2018}
Koprowski M.~P.,  et~al., 2018, \mn@doi [\mnras] {10.1093/mnras/sty1527}, 479,
  4355

\bibitem[\protect\citeauthoryear{Kriek \& Conroy}{Kriek \&
  Conroy}{2013}]{Kriek_2013}
Kriek M.,  Conroy C.,  2013, \mn@doi [\apj] {10.1088/2041-8205/775/1/l16}, 775,
  L16

\bibitem[\protect\citeauthoryear{Krumholz \& Gnedin}{Krumholz \&
  Gnedin}{2011}]{Krumholz_2011}
Krumholz M.~R.,  Gnedin N.~Y.,  2011, \mn@doi [\apj]
  {10.1088/0004-637x/729/1/36}, 729, 36

\bibitem[\protect\citeauthoryear{Laigle et~al.,}{Laigle
  et~al.}{2016}]{Laigle_2016}
Laigle C.,  et~al., 2016, \mn@doi [\apjs] {10.3847/0067-0049/224/2/24}, 224, 24

\bibitem[\protect\citeauthoryear{Laporte et~al.,}{Laporte
  et~al.}{2016}]{Laporte_2016}
Laporte N.,  et~al., 2016, \mn@doi [\apj] {10.3847/0004-637x/820/2/98}, 820, 98

\bibitem[\protect\citeauthoryear{Laporte et~al.,}{Laporte
  et~al.}{2017}]{Laporte_2017}
Laporte N.,  et~al., 2017, \mn@doi [\apj] {10.3847/2041-8213/aa62aa}, 837, L21

\bibitem[\protect\citeauthoryear{Larson}{Larson}{1981}]{Larson_1981}
Larson R.~B.,  1981, \mn@doi [\mnras] {10.1093/mnras/194.4.809}, 194, 809

\bibitem[\protect\citeauthoryear{Leitherer \& Heckman}{Leitherer \&
  Heckman}{1995}]{Leitherer_1995}
Leitherer C.,  Heckman T.~M.,  1995, \mn@doi [\apjs] {10.1086/192112}, 96, 9

\bibitem[\protect\citeauthoryear{Leitherer et~al.,}{Leitherer
  et~al.}{1999}]{Leitherer_1999}
Leitherer C.,  et~al., 1999, \mn@doi [\apjs] {10.1086/313233}, 123, 3

\bibitem[\protect\citeauthoryear{Li, Narayanan  \& Dav{\'e}}{Li
  et~al.}{2019}]{Li_2019}
Li Q.,  Narayanan D.,   Dav{\'e} R.,  2019, \mn@doi [\mnras]
  {10.1093/mnras/stz2684}, 490, 1425

\bibitem[\protect\citeauthoryear{Liang, Feldmann, Faucher-Gigu{\`e}re, Kere{\v
  s}, Hopkins, Hayward, Quataert  \& Scoville}{Liang et~al.}{2018}]{Liang_2018}
Liang L.,  Feldmann R.,  Faucher-Gigu{\`e}re C.-A.,  Kere{\v s} D.,  Hopkins
  P.~F.,  Hayward C.~C.,  Quataert E.,   Scoville N.~Z.,  2018, \mn@doi
  [\mnras: Letters] {10.1093/mnrasl/sly071}, 478, L83

\bibitem[\protect\citeauthoryear{{Liang} et~al.,}{{Liang}
  et~al.}{2019}]{Liang_2019}
{Liang} L.,  et~al., 2019, \mn@doi [\mnras] {10.1093/mnras/stz2134}, \href
  {https://ui.adsabs.harvard.edu/abs/2019MNRAS.489.1397L} {489, 1397}

\bibitem[\protect\citeauthoryear{Lombardi, Bouy, Alves  \& Lada}{Lombardi
  et~al.}{2014}]{Lombardi_2014}
Lombardi M.,  Bouy H.,  Alves J.,   Lada C.~J.,  2014, \mn@doi [A\&A]
  {10.1051/0004-6361/201323293}, 566, A45

\bibitem[\protect\citeauthoryear{Lutz}{Lutz}{2014}]{Lutz_2014}
Lutz D.,  2014, \mn@doi [ARA\&A] {10.1146/annurev-astro-081913-035953}, 52, 373

\bibitem[\protect\citeauthoryear{Ma et~al.,}{Ma et~al.}{2015}]{Ma_2015}
Ma J.,  et~al., 2015, \mn@doi [\mnras] {10.1093/mnras/stv2073}, 454, 1751

\bibitem[\protect\citeauthoryear{Ma, Hopkins, Faucher-Gigu{\`e}re, Zolman,
  Muratov, Kere{\v s}  \& Quataert}{Ma et~al.}{2016a}]{Ma_2016a}
Ma X.,  Hopkins P.~F.,  Faucher-Gigu{\`e}re C.-A.,  Zolman N.,  Muratov A.~L.,
  Kere{\v s} D.,   Quataert E.,  2016a, \mn@doi [\mnras]
  {10.1093/mnras/stv2659}, 456, 2140

\bibitem[\protect\citeauthoryear{Ma, Hopkins, Kasen, Quataert,
  Faucher-Gigu{\`e}re, Kere{\v s}, Murray  \& Strom}{Ma
  et~al.}{2016b}]{Ma_2016b}
Ma X.,  Hopkins P.~F.,  Kasen D.,  Quataert E.,  Faucher-Gigu{\`e}re C.-A.,
  Kere{\v s} D.,  Murray N.,   Strom A.,  2016b, \mn@doi [\mnras]
  {10.1093/mnras/stw941}, 459, 3614

\bibitem[\protect\citeauthoryear{Ma, Ge, Zhao, Prochaska, Zhang, Ji  \&
  Schneider}{Ma et~al.}{2017}]{Ma_2017}
Ma J.,  Ge J.,  Zhao Y.,  Prochaska J.~X.,  Zhang S.,  Ji T.,   Schneider
  D.~P.,  2017, \mn@doi [\mnras] {10.1093/mnras/stx2117}, 472, 2196

\bibitem[\protect\citeauthoryear{Ma et~al.,}{Ma et~al.}{2019}]{Ma_2019}
Ma X.,  et~al., 2019, \mn@doi [\mnras] {10.1093/mnras/stz1324}, 487, 1844

\bibitem[\protect\citeauthoryear{Madau \& Dickinson}{Madau \&
  Dickinson}{2014}]{Madau_2014}
Madau P.,  Dickinson M.,  2014, \mn@doi [ARA\&A]
  {10.1146/annurev-astro-081811-125615}, 52, 415

\bibitem[\protect\citeauthoryear{Magnelli, Elbaz, Chary, Dickinson, Le~Borgne,
  Frayer  \& Willmer}{Magnelli et~al.}{2009}]{Magnelli_2009}
Magnelli B.,  Elbaz D.,  Chary R.~R.,  Dickinson M.,  Le~Borgne D.,  Frayer
  D.~T.,   Willmer C. N.~A.,  2009, \mn@doi [A\&A]
  {10.1051/0004-6361:200811443}, 496, 57

\bibitem[\protect\citeauthoryear{Magnelli, Elbaz, Chary, Dickinson, Le~Borgne,
  Frayer  \& Willmer}{Magnelli et~al.}{2011}]{Magnelli_2011}
Magnelli B.,  Elbaz D.,  Chary R.~R.,  Dickinson M.,  Le~Borgne D.,  Frayer
  D.~T.,   Willmer C. N.~A.,  2011, \mn@doi [A\&A]
  {10.1051/0004-6361/200913941}, 528, A35

\bibitem[\protect\citeauthoryear{Matthee, Sobral, Santos, R{\"o}ttgering,
  Darvish  \& Mobasher}{Matthee et~al.}{2015}]{Matthee_2015}
Matthee J.,  Sobral D.,  Santos S.,  R{\"o}ttgering H.,  Darvish B.,   Mobasher
  B.,  2015, \mn@doi [\mnras] {10.1093/mnras/stv947}, 451, 400

\bibitem[\protect\citeauthoryear{Matthee et~al.,}{Matthee
  et~al.}{2017}]{Matthee_2017}
Matthee J.,  et~al., 2017, \mn@doi [\apj] {10.3847/1538-4357/aa9931}, 851, 145

\bibitem[\protect\citeauthoryear{Matthee et~al.,}{Matthee
  et~al.}{2019}]{Matthee_2019}
Matthee J.,  et~al., 2019, \mn@doi [\apj] {10.3847/1538-4357/ab2f81}, 881, 124

\bibitem[\protect\citeauthoryear{McCracken et~al.,}{McCracken
  et~al.}{2012}]{McCracken_2012}
McCracken H.~J.,  et~al., 2012, \mn@doi [A\&A] {10.1051/0004-6361/201219507},
  544, A156

\bibitem[\protect\citeauthoryear{McLeod, McLure, Dunlop, Robertson, Ellis  \&
  Targett}{McLeod et~al.}{2015}]{McLeod_2015}
McLeod D.~J.,  McLure R.~J.,  Dunlop J.~S.,  Robertson B.~E.,  Ellis R.~S.,
  Targett T.~A.,  2015, \mn@doi [\mnras] {10.1093/mnras/stv780}, 450, 3032

\bibitem[\protect\citeauthoryear{McLure et~al.,}{McLure
  et~al.}{2013}]{McLure_2013}
McLure R.~J.,  et~al., 2013, \mn@doi [\mnras] {10.1093/mnras/stt627}, 432, 2696

\bibitem[\protect\citeauthoryear{McLure et~al.,}{McLure
  et~al.}{2018}]{McLure_2018}
McLure R.~J.,  et~al., 2018, \mn@doi [\mnras] {10.1093/mnras/sty522}, 476, 3991

\bibitem[\protect\citeauthoryear{Meurer, Heckman, Leitherer, Kinney, Robert  \&
  Garnett}{Meurer et~al.}{1995}]{Meurer_1995}
Meurer G.~R.,  Heckman T.~M.,  Leitherer C.,  Kinney A.,  Robert C.,   Garnett
  D.~R.,  1995, \mn@doi [\apj] {10.1086/117721}, 110, 2665

\bibitem[\protect\citeauthoryear{Meurer, Heckman  \& Calzetti}{Meurer
  et~al.}{1999}]{Meurer_1999}
Meurer G.~R.,  Heckman T.~M.,   Calzetti D.,  1999, \mn@doi [\apj]
  {10.1086/307523}, 521, 64

\bibitem[\protect\citeauthoryear{Morrissey et~al.,}{Morrissey
  et~al.}{2007}]{Morrissey_2007}
Morrissey P.,  et~al., 2007, \mn@doi [\apjs] {10.1086/520512}, 173, 682

\bibitem[\protect\citeauthoryear{Mortlock, McLure, Bowler, McLeod,
  M{\'a}rmol-Queralt{\'o}, Parsa, Dunlop  \& Bruce}{Mortlock
  et~al.}{2017}]{Mortlock_2017}
Mortlock A.,  McLure R.~J.,  Bowler R. A.~A.,  McLeod D.~J.,
  M{\'a}rmol-Queralt{\'o} E.,  Parsa S.,  Dunlop J.~S.,   Bruce V.~A.,  2017,
  \mn@doi [\mnras] {10.1093/mnras/stw2728}, 465, 672

\bibitem[\protect\citeauthoryear{Muratov, Kere{\v s}, Faucher-Gigu{\`e}re,
  Hopkins, Quataert  \& Murray}{Muratov et~al.}{2015}]{Muratov_2015}
Muratov A.~L.,  Kere{\v s} D.,  Faucher-Gigu{\`e}re C.-A.,  Hopkins P.~F.,
  Quataert E.,   Murray N.,  2015, \mn@doi [\mnras] {10.1093/mnras/stv2126},
  454, 2691

\bibitem[\protect\citeauthoryear{Narayanan et~al.,}{Narayanan
  et~al.}{2010}]{Narayanan_2010}
Narayanan D.,  et~al., 2010, \mn@doi [\mnras]
  {10.1111/j.1365-2966.2010.16997.x}, 407, 1701

\bibitem[\protect\citeauthoryear{Narayanan, Dav{\'e}, Johnson, Thompson, Conroy
   \& Geach}{Narayanan et~al.}{2018a}]{Narayanan_2018}
Narayanan D.,  Dav{\'e} R.,  Johnson B.~D.,  Thompson R.,  Conroy C.,   Geach
  J.,  2018a, \mn@doi [\mnras] {10.1093/mnras/stx2860}, 474, 1718

\bibitem[\protect\citeauthoryear{Narayanan, Conroy, Dav{\'e}, Johnson  \&
  Popping}{Narayanan et~al.}{2018b}]{Narayanan_2018b}
Narayanan D.,  Conroy C.,  Dav{\'e} R.,  Johnson B.~D.,   Popping G.,  2018b,
  \mn@doi [\apj] {10.3847/1538-4357/aaed25}, 869, 70

\bibitem[\protect\citeauthoryear{{Narayanan} et~al.,}{{Narayanan}
  et~al.}{2020}]{Narayanan_2020}
{Narayanan} D.,  et~al., 2020, arXiv e-prints, \href
  {https://ui.adsabs.harvard.edu/abs/2020arXiv200610757N} {2006.10757}

\bibitem[\protect\citeauthoryear{Natta \& Panagia}{Natta \&
  Panagia}{1984}]{Natta_1984}
Natta A.,  Panagia N.,  1984, \mn@doi [\apj] {10.1086/162681}, 287, 228

\bibitem[\protect\citeauthoryear{Neeleman, Prochaska, Kanekar  \&
  Rafelski}{Neeleman et~al.}{2020}]{Neeleman_2020}
Neeleman M.,  Prochaska J.~X.,  Kanekar N.,   Rafelski M.,  2020, \mn@doi
  [Nature] {10.1038/s41586-020-2276-y}, 581, 269

\bibitem[\protect\citeauthoryear{Nguyen et~al.,}{Nguyen
  et~al.}{2010}]{Nguyen_2010}
Nguyen H.~T.,  et~al., 2010, \mn@doi [A\&A] {10.1051/0004-6361/201014680}, 518,
  L5

\bibitem[\protect\citeauthoryear{Noll, Burgarella, Giovannoli, Buat, Marcillac
  \& Mu{\~n}oz-Mateos}{Noll et~al.}{2009}]{Noll_2009}
Noll S.,  Burgarella D.,  Giovannoli E.,  Buat V.,  Marcillac D.,
  Mu{\~n}oz-Mateos J.~C.,  2009, \mn@doi [A\&A] {10.1051/0004-6361/200912497},
  507, 1793

\bibitem[\protect\citeauthoryear{Novak et~al.,}{Novak
  et~al.}{2019}]{Novak_2019}
Novak M.,  et~al., 2019, \mn@doi [\apj] {10.3847/1538-4357/ab2beb}, 881, 63

\bibitem[\protect\citeauthoryear{Oesch, Bouwens, Illingworth, Franx, Ammons,
  Dokkum, Trenti  \& Labb{\'e}}{Oesch et~al.}{2015}]{Oesch_2015}
Oesch P.~A.,  Bouwens R.~J.,  Illingworth G.~D.,  Franx M.,  Ammons S.~M.,
  Dokkum P. G.~v.,  Trenti M.,   Labb{\'e} I.,  2015, \mn@doi [\apj]
  {10.1088/0004-637x/808/1/104}, 808, 104

\bibitem[\protect\citeauthoryear{Oesch, Bouwens, Illingworth, Labb{\'e}  \&
  Stefanon}{Oesch et~al.}{2018}]{Oesch_2018}
Oesch P.~A.,  Bouwens R.~J.,  Illingworth G.~D.,  Labb{\'e} I.,   Stefanon M.,
  2018, \mn@doi [\apj] {10.3847/1538-4357/aab03f}, 855, 105

\bibitem[\protect\citeauthoryear{Oteo et~al.,}{Oteo et~al.}{2013}]{Oteo_2013}
Oteo I.,  et~al., 2013, \mn@doi [A\&A] {10.1051/0004-6361/201321478}, 554, L3

\bibitem[\protect\citeauthoryear{Ouchi et~al.,}{Ouchi
  et~al.}{2013}]{Ouchi_2013}
Ouchi M.,  et~al., 2013, \mn@doi [\apj] {10.1088/0004-637x/778/2/102}, 778, 102

\bibitem[\protect\citeauthoryear{Overzier et~al.,}{Overzier
  et~al.}{2011}]{Overzier_2011}
Overzier R.~A.,  et~al., 2011, \mn@doi [\apj] {10.1088/2041-8205/726/1/l7},
  726, L7

\bibitem[\protect\citeauthoryear{Pei}{Pei}{1992}]{Pei_1992}
Pei Y.~C.,  1992, \mn@doi [\apj] {10.1086/171637}, 395, 130

\bibitem[\protect\citeauthoryear{Penner et~al.,}{Penner
  et~al.}{2012}]{Penner_2012}
Penner K.,  et~al., 2012, \mn@doi [\apj] {10.1088/0004-637x/759/1/28}, 759, 28

\bibitem[\protect\citeauthoryear{Poglitsch et~al.,}{Poglitsch
  et~al.}{2010}]{Poglitsch_2010}
Poglitsch A.,  et~al., 2010, \mn@doi [A\&A] {10.1051/0004-6361/201014535}, 518,
  L2

\bibitem[\protect\citeauthoryear{Popping, Puglisi  \& Norman}{Popping
  et~al.}{2017}]{Popping_2017}
Popping G.,  Puglisi A.,   Norman C.~A.,  2017, \mn@doi [\mnras]
  {10.1093/mnras/stx2202}, 472, 2315

\bibitem[\protect\citeauthoryear{Qiu, Mutch, da Cunha, Poole  \& Wyithe}{Qiu
  et~al.}{2019}]{Qiu_2019}
Qiu Y.,  Mutch S.~J.,  da Cunha E.,  Poole G.~B.,   Wyithe J. S.~B.,  2019,
  \mn@doi [\mnras] {10.1093/mnras/stz2233}, 489, 1357

\bibitem[\protect\citeauthoryear{Raghavan et~al.,}{Raghavan
  et~al.}{2010}]{Raghavan_2010}
Raghavan D.,  et~al., 2010, \mn@doi [\apjs] {10.1088/0067-0049/190/1/1}, 190, 1

\bibitem[\protect\citeauthoryear{Reddy, Steidel, Fadda, Yan, Pettini, Shapley,
  Erb  \& Adelberger}{Reddy et~al.}{2006}]{Reddy_2006}
Reddy N.~A.,  Steidel C.~C.,  Fadda D.,  Yan L.,  Pettini M.,  Shapley A.~E.,
  Erb D.~K.,   Adelberger K.~L.,  2006, \mn@doi [\apj] {10.1086/503739}, 644,
  792

\bibitem[\protect\citeauthoryear{Reddy, Erb, Pettini, Steidel  \&
  Shapley}{Reddy et~al.}{2010}]{Reddy_2010}
Reddy N.~A.,  Erb D.~K.,  Pettini M.,  Steidel C.~C.,   Shapley A.~E.,  2010,
  \mn@doi [\apj] {10.1088/0004-637x/712/2/1070}, 712, 1070

\bibitem[\protect\citeauthoryear{Reddy et~al.,}{Reddy
  et~al.}{2018}]{Reddy_2018}
Reddy N.~A.,  et~al., 2018, \mn@doi [\apj] {10.3847/1538-4357/aaa3e7}, 853, 56

\bibitem[\protect\citeauthoryear{Richard, Kneib, Ebeling, Stark, Egami  \&
  Fiedler}{Richard et~al.}{2011}]{Richard_2011}
Richard J.,  Kneib J.-P.,  Ebeling H.,  Stark D.~P.,  Egami E.,   Fiedler
  A.~K.,  2011, \mn@doi [\mnras: Letters] {10.1111/j.1745-3933.2011.01050.x},
  414, L31

\bibitem[\protect\citeauthoryear{Rodighiero et~al.,}{Rodighiero
  et~al.}{2011}]{Rodighiero_2011}
Rodighiero G.,  et~al., 2011, \mn@doi [\apj] {10.1088/2041-8205/739/2/l40},
  739, L40

\bibitem[\protect\citeauthoryear{Rosolowsky, Engargiola, Plambeck  \&
  Blitz}{Rosolowsky et~al.}{2003}]{Rosolowsky_2003}
Rosolowsky E.,  Engargiola G.,  Plambeck R.,   Blitz L.,  2003, \mn@doi [\apj]
  {10.1086/379166}, 599, 258

\bibitem[\protect\citeauthoryear{{Safarzadeh}, {Hayward}, {Ferguson}  \&
  {Somerville}}{{Safarzadeh} et~al.}{2017a}]{Safarzadeh_2017b}
{Safarzadeh} M.,  {Hayward} C.~C.,  {Ferguson} H.~C.,   {Somerville} R.~S.,
  2017a, \mn@doi [\apj] {10.3847/0004-637X/818/1/62}, \href
  {http://adsabs.harvard.edu/abs/2016ApJ...818...62S} {818, 62}

\bibitem[\protect\citeauthoryear{Safarzadeh, Hayward  \& Ferguson}{Safarzadeh
  et~al.}{2017b}]{Safarzadeh_2017}
Safarzadeh M.,  Hayward C.~C.,   Ferguson H.~C.,  2017b, \mn@doi [\apj]
  {10.3847/1538-4357/aa6c5b}, 840, 15

\bibitem[\protect\citeauthoryear{Salim \& Boquien}{Salim \&
  Boquien}{2019}]{Salim_2019}
Salim S.,  Boquien M.,  2019, \mn@doi [\apj] {10.3847/1538-4357/aaf88a}, 872,
  23

\bibitem[\protect\citeauthoryear{{Salim} \& {Narayanan}}{{Salim} \&
  {Narayanan}}{2020}]{Salim_2020}
{Salim} S.,  {Narayanan} D.,  2020, arXiv e-prints, \href
  {https://ui.adsabs.harvard.edu/abs/2020arXiv200103181S} {2001.03181}

\bibitem[\protect\citeauthoryear{Sana et~al.,}{Sana et~al.}{2012}]{Sana_2012}
Sana H.,  et~al., 2012, \mn@doi [Science] {10.1126/science.1223344}, 337, 444

\bibitem[\protect\citeauthoryear{Schady et~al.,}{Schady
  et~al.}{2012}]{Schady_2012}
Schady P.,  et~al., 2012, \mn@doi [A\&A] {10.1051/0004-6361/201117414}, 537,
  A15

\bibitem[\protect\citeauthoryear{{Schreiber}, {Elbaz}, {Pannella}, {Ciesla},
  {Wang}  \& {Franco}}{{Schreiber} et~al.}{2018}]{Schreiber_2018}
{Schreiber} C.,  {Elbaz} D.,  {Pannella} M.,  {Ciesla} L.,  {Wang} T.,
  {Franco} M.,  2018, \mn@doi [A\&A] {10.1051/0004-6361/201731506}, \href
  {http://adsabs.harvard.edu/abs/2018A%26A...609A..30S} {609, A30}

\bibitem[\protect\citeauthoryear{Schulz, Popping, Pillepich, Nelson,
  Vogelsberger, Marinacci  \& Hernquist}{Schulz et~al.}{2020}]{Schulz_2020}
Schulz S.,  Popping G.,  Pillepich A.,  Nelson D.,  Vogelsberger M.,  Marinacci
  F.,   Hernquist L.,  2020, \mn@doi [\mnras] {10.1093/mnras/staa1900}

\bibitem[\protect\citeauthoryear{{Scoville}}{{Scoville}}{2013}]{Scoville_2013}
{Scoville} N.~Z.,  2013, in Falc{\'o}n-Barroso J., Knapen J. H., eds, Secular
  Evolution of Galaxies.
Cambridge University Press, Cambridge, UK, p.~491

\bibitem[\protect\citeauthoryear{Seon \& Draine}{Seon \&
  Draine}{2016}]{Seon_2016}
Seon K.-I.,  Draine B.~T.,  2016, \mn@doi [\apj] {10.3847/1538-4357/833/2/201},
  833, 201

\bibitem[\protect\citeauthoryear{Shao et~al.,}{Shao et~al.}{2010}]{Shao_2010}
Shao L.,  et~al., 2010, \mn@doi [A\&A] {10.1051/0004-6361/201014606}, 518, L26

\bibitem[\protect\citeauthoryear{Shen et~al.,}{Shen et~al.}{2020}]{Shen_2020}
Shen X.,  et~al., 2020, \mn@doi [\mnras] {10.1093/mnras/staa1423}, 495, 4747

\bibitem[\protect\citeauthoryear{Skelton et~al.,}{Skelton
  et~al.}{2014}]{Skelton_2014}
Skelton R.~E.,  et~al., 2014, \mn@doi [\apjs] {10.1088/0067-0049/214/2/24},
  214, 24

\bibitem[\protect\citeauthoryear{Sklias et~al.,}{Sklias
  et~al.}{2014}]{Sklias_2014}
Sklias P.,  et~al., 2014, \mn@doi [A\&A] {10.1051/0004-6361/201322424}, 561,
  A149

\bibitem[\protect\citeauthoryear{Snyder et~al.,}{Snyder
  et~al.}{2015}]{Snyder_2015}
Snyder G.~F.,  et~al., 2015, \mn@doi [\mnras] {10.1093/mnras/stv2078}, 454,
  1886

\bibitem[\protect\citeauthoryear{Sobral, Matthee, Darvish, Schaerer, Mobasher,
  R{\"o}ttgering, Santos  \& Hemmati}{Sobral et~al.}{2015}]{Sobral_2015}
Sobral D.,  Matthee J.,  Darvish B.,  Schaerer D.,  Mobasher B.,
  R{\"o}ttgering H. J.~A.,  Santos S.,   Hemmati S.,  2015, \mn@doi [\apj]
  {10.1088/0004-637x/808/2/139}, 808, 139

\bibitem[\protect\citeauthoryear{Solomon, Rivolo, Barrett  \& Yahil}{Solomon
  et~al.}{1987}]{Solomon_1987}
Solomon P.~M.,  Rivolo A.~R.,  Barrett J.,   Yahil A.,  1987, \mn@doi [\apj]
  {10.1086/165493}, 319, 730

\bibitem[\protect\citeauthoryear{Somerville \& Dav{\'e}}{Somerville \&
  Dav{\'e}}{2015}]{Somerville_2015}
Somerville R.~S.,  Dav{\'e} R.,  2015, \mn@doi [ARA\&A]
  {10.1146/annurev-astro-082812-140951}, 53, 51

\bibitem[\protect\citeauthoryear{Sommovigo, Ferrara, Pallottini, Carniani,
  Gallerani  \& Decataldo}{Sommovigo et~al.}{2020}]{Sommovigo_2020}
Sommovigo L.,  Ferrara A.,  Pallottini A.,  Carniani S.,  Gallerani S.,
  Decataldo D.,  2020, \mn@doi [\mnras] {10.1093/mnras/staa1959}, 497, 956

\bibitem[\protect\citeauthoryear{Sparre, Hayward, Feldmann,
  Faucher-Gigu{\`e}re, Muratov, Kere{\v s}  \& Hopkins}{Sparre
  et~al.}{2017}]{Sparre_2017}
Sparre M.,  Hayward C.~C.,  Feldmann R.,  Faucher-Gigu{\`e}re C.-A.,  Muratov
  A.~L.,  Kere{\v s} D.,   Hopkins P.~F.,  2017, \mn@doi [\mnras]
  {10.1093/mnras/stw3011}, 466, 88

\bibitem[\protect\citeauthoryear{Spinoglio et~al.,}{Spinoglio
  et~al.}{2017}]{Spinoglio_2017}
Spinoglio L.,  et~al., 2017, \mn@doi [PASA] {10.1017/pasa.2017.48}, 34

\bibitem[\protect\citeauthoryear{Stanway \& Eldridge}{Stanway \&
  Eldridge}{2018}]{Stanway_2018}
Stanway E.~R.,  Eldridge J.~J.,  2018, \mn@doi [\mnras]
  {10.1093/mnras/sty1353}, 479, 75

\bibitem[\protect\citeauthoryear{Stanway, Eldridge  \& Becker}{Stanway
  et~al.}{2016}]{Stanway_2016}
Stanway E.~R.,  Eldridge J.~J.,   Becker G.~D.,  2016, \mn@doi [\mnras]
  {10.1093/mnras/stv2661}, 456, 485

\bibitem[\protect\citeauthoryear{Steidel, Giavalisco, Dickinson  \&
  Adelberger}{Steidel et~al.}{1996}]{Steidel_1996}
Steidel C.~C.,  Giavalisco M.,  Dickinson M.,   Adelberger K.~L.,  1996,
  \mn@doi [\apj] {10.1086/118019}, 112, 352

\bibitem[\protect\citeauthoryear{Stratta, Maiolino, Fiore  \& D'Elia}{Stratta
  et~al.}{2007}]{Stratta_2007}
Stratta G.,  Maiolino R.,  Fiore F.,   D'Elia V.,  2007, \mn@doi [\apj]
  {10.1086/518502}, 661, L9

\bibitem[\protect\citeauthoryear{Swinyard et~al.,}{Swinyard
  et~al.}{2010}]{Swinyard_2010}
Swinyard B.~M.,  et~al., 2010, \mn@doi [A\&A] {10.1051/0004-6361/201014605},
  518, L4

\bibitem[\protect\citeauthoryear{Takeuchi, Yuan, Ikeyama, Murata  \&
  Inoue}{Takeuchi et~al.}{2012}]{Takeuchi_2012}
Takeuchi T.~T.,  Yuan F.-T.,  Ikeyama A.,  Murata K.~L.,   Inoue A.~K.,  2012,
  \mn@doi [\apj] {10.1088/0004-637x/755/2/144}, 755, 144

\bibitem[\protect\citeauthoryear{Torrey et~al.,}{Torrey
  et~al.}{2015}]{Torrey_2015}
Torrey P.,  et~al., 2015, \mn@doi [\mnras] {10.1093/mnras/stu2592}, 447, 2753

\bibitem[\protect\citeauthoryear{{Trayford} et~al.,}{{Trayford}
  et~al.}{2017}]{Trayford_2017}
{Trayford} J.~W.,  et~al., 2017, \mn@doi [\mnras] {10.1093/mnras/stx1051},
  \href {http://adsabs.harvard.edu/abs/2017MNRAS.470..771T} {470, 771}

\bibitem[\protect\citeauthoryear{Trayford, Lagos, Robotham  \&
  Obreschkow}{Trayford et~al.}{2019}]{Trayford_2019}
Trayford J.~W.,  Lagos C. d.~P.,  Robotham A. S.~G.,   Obreschkow D.,  2019,
  \mn@doi [\mnras] {10.1093/mnras/stz3234}

\bibitem[\protect\citeauthoryear{Tremonti et~al.,}{Tremonti
  et~al.}{2004}]{Tremonti_2004}
Tremonti C.~A.,  et~al., 2004, \mn@doi [\apj] {10.1086/423264}, 613, 898

\bibitem[\protect\citeauthoryear{Venemans et~al.,}{Venemans
  et~al.}{2017}]{Venemans_2017}
Venemans B.~P.,  et~al., 2017, \mn@doi [\apj] {10.3847/2041-8213/aa943a}, 851,
  L8

\bibitem[\protect\citeauthoryear{Vogelsberger, Marinacci, Torrey  \&
  Puchwein}{Vogelsberger et~al.}{2020}]{Vogelsberger_2020}
Vogelsberger M.,  Marinacci F.,  Torrey P.,   Puchwein E.,  2020, \mn@doi
  [Nature Reviews Physics] {10.1038/s42254-019-0127-2}, 2, 42

\bibitem[\protect\citeauthoryear{Walter et~al.,}{Walter
  et~al.}{2016}]{Walter_2016}
Walter F.,  et~al., 2016, \mn@doi [\apj] {10.3847/1538-4357/833/1/67}, 833, 67

\bibitem[\protect\citeauthoryear{Wang \& Heckman}{Wang \&
  Heckman}{1996}]{Wang_1996}
Wang B.,  Heckman T.~M.,  1996, \mn@doi [\apj] {10.1086/176760}, 457, 645

\bibitem[\protect\citeauthoryear{Wang et~al.,}{Wang et~al.}{2018}]{Wang_2018}
Wang W.,  et~al., 2018, \mn@doi [\apj] {10.3847/1538-4357/aaef79}, 869, 161

\bibitem[\protect\citeauthoryear{Watson, Christensen, Knudsen, Richard,
  Gallazzi  \& Micha{\l}owski}{Watson et~al.}{2015}]{Watson_2015}
Watson D.,  Christensen L.,  Knudsen K.~K.,  Richard J.,  Gallazzi A.,
  Micha{\l}owski M.~J.,  2015, \mn@doi [Nature] {10.1038/nature14164}, 519, 327

\bibitem[\protect\citeauthoryear{{Weingartner} \& {Draine}}{{Weingartner} \&
  {Draine}}{2001}]{Weingartner_2001}
{Weingartner} J.~C.,  {Draine} B.~T.,  2001, \mn@doi [ApJ] {10.1086/318651},
  \href {http://adsabs.harvard.edu/abs/2001ApJ...548..296W} {548, 296}

\bibitem[\protect\citeauthoryear{Whitaker, Pope, Cybulski, Casey, Popping  \&
  Yun}{Whitaker et~al.}{2017}]{Whitaker_2017}
Whitaker K.~E.,  Pope A.,  Cybulski R.,  Casey C.~M.,  Popping G.,   Yun M.~S.,
   2017, \mn@doi [\apj] {10.3847/1538-4357/aa94ce}, 850, 208

\bibitem[\protect\citeauthoryear{Williams, Quadri, Franx, van Dokkum  \&
  Labb{\'e}}{Williams et~al.}{2009}]{Williams_2009}
Williams R.~J.,  Quadri R.~F.,  Franx M.,  van Dokkum P.,   Labb{\'e} I.,
  2009, \mn@doi [\apj] {10.1088/0004-637x/691/2/1879}, 691, 1879

\bibitem[\protect\citeauthoryear{Wiseman, Schady, Bolmer, Kr{\"u}hler, Yates,
  Greiner  \& Fynbo}{Wiseman et~al.}{2017}]{Wiseman_2017}
Wiseman P.,  Schady P.,  Bolmer J.,  Kr{\"u}hler T.,  Yates R.~M.,  Greiner J.,
    Fynbo J. P.~U.,  2017, \mn@doi [A\&A] {10.1051/0004-6361/201629228}, 599,
  A24

\bibitem[\protect\citeauthoryear{Witt \& Gordon}{Witt \&
  Gordon}{1996}]{Witt_1996}
Witt A.~N.,  Gordon K.~D.,  1996, \mn@doi [\apj] {10.1086/177282}, 463, 681

\bibitem[\protect\citeauthoryear{Witt \& Gordon}{Witt \&
  Gordon}{2000}]{Witt_2000}
Witt A.~N.,  Gordon K.~D.,  2000, \mn@doi [\apj] {10.1086/308197}, 528, 799

\bibitem[\protect\citeauthoryear{Xue et~al.,}{Xue et~al.}{2011}]{Xue_2011}
Xue Y.~Q.,  et~al., 2011, \mn@doi [\apjs] {10.1088/0067-0049/195/1/10}, 195, 10

\bibitem[\protect\citeauthoryear{Zafar, Watson, Fynbo, Malesani, Jakobsson  \&
  de Ugarte~Postigo}{Zafar et~al.}{2011}]{Zafar_2011}
Zafar T.,  Watson D.,  Fynbo J. P.~U.,  Malesani D.,  Jakobsson P.,   de
  Ugarte~Postigo A.,  2011, \mn@doi [A\&A] {10.1051/0004-6361/201116663}, 532,
  A143

\bibitem[\protect\citeauthoryear{Zafar et~al.,}{Zafar
  et~al.}{2018}]{Zafar_2018}
Zafar T.,  et~al., 2018, \mn@doi [\mnras] {10.1093/mnras/sty1380}, 479, 1542

\bibitem[\protect\citeauthoryear{da Cunha et~al.,}{da~Cunha
  et~al.}{2013}]{Cunha_2013}
da Cunha E.,  et~al., 2013, \mn@doi [\apj] {10.1088/0004-637x/766/1/13}, 766,
  13

\makeatother
\end{thebibliography}
\vspace{-10pt}
\bsp


\label{lastpage}
\end{document}